\documentclass[paper]{aa} 

\usepackage{times} 
\usepackage{graphicx} 
\usepackage{xspace} 
\usepackage{epsfig} 
\usepackage{natbib} 
\usepackage{rotating} 
\usepackage{dcolumn} 
\usepackage{amssymb,amsmath}

\begin{document} 
 
\title{Results from DROXO. II.}

\subtitle{[Ne\,II] and X-ray emission from $\rho$ Ophiuchi young stellar
objects}
 
\author{E. Flaccomio\inst{1} \and B. Stelzer\inst{1} \and S.
Sciortino\inst{1} \and G. Micela\inst{1} \and I. Pillitteri\inst{1,2}
\and L. Testi\inst{3,4}} 

\offprints{E. Flaccomio} 
 
\institute{INAF - Osservatorio Astronomico di Palermo, 
  Piazza del Parlamento 1, 
  I-90134 Palermo, Italy \\ \email{E. Flaccomio, ettoref@astropa.unipa.it}
 \and
Dip. di Scienze Fisiche e Astronomiche - Sez. di Astronomia - Universit\`a di Palermo, Piazza del Parlamento 1, 90134 Palermo, Italy
  \and
INAF - Osservatorio Astrofisico di Arcetri, Largo E. Fermi 5, 50125 Firenze, Italy  
 \and
 ESO, Karl Schwarzschild Str. 2 - D-85748, Garching bei M\"unchen, Germany
}  
 
\titlerunning{[Ne\,II] and X-ray emission from $\rho$ Oph YSOs} 
 
\date{Received $<$date$>$ / Accepted $<$date$>$}

\abstract 
% context heading (optional) 
{
The infrared [Ne\,II] and [Ne\,III] fine structure lines at
12.81$\mu$m and 15.55$\mu$m have recently been theoretically predicted
to trace the circumstellar disk gas subject to X-ray heating and
ionization.
} 
% aims heading (mandatory) 
{
We observationally investigate the origin of the neon fine structure
line emission by comparing observations with models of X-ray
irradiated disks and by searching for empirical correlations between
the line luminosities and stellar and circumstellar parameters.
} 
% methods heading (mandatory) 
{ 
We select a sample of 28 young stellar objects in the $\rho$ Ophiuchi
star formation region for which good quality infrared spectra and
X-ray data have been obtained, the former with the Spitzer IRS and the
latter with the Deep Rho Ophiuchi XMM-Newton Observation. We measure
neon line fluxes and X-ray luminosities; we complement these data with
stellar/circumstellar parameters obtained by fitting the spectral energy
distributions of our objects (from optical to millimeter wavelengths)
with star/disk/envelope models.
}
% results heading (mandatory) 
{
We detect the [Ne\,II] and the [Ne\,III] lines in 10 and 1 cases,
respectively. Line luminosities show no correlation with X-ray
emission. The luminosity of the [Ne\,II] line for one star, and that
of both the [Ne\,II] and [Ne\,III] lines for a second star, match the
predictions of published models of X-ray irradiated disks; for the
remaining 8 objects the [Ne\,II] emission is 1-3\,dex higher than
predicted on the basis of their $L_{\rm X}$. However, the
stellar/circumstellar characteristics assumed in published models do
not match those of most of the stars in our sample. Class\,I objects
show significantly stronger [Ne\,II] lines than Class\,II and
Class\,III ones. A correlation is moreover found between the [Ne\,II]
line emission and the disk mass accretion rates estimated from the
spectral energy distributions. This might point
toward a role of accretion-generated UV emission in the generation of
the line or to other mechanisms related to mass inflows from
circumstellar disks and envelopes and/or to the associated mass
outflows (winds and jets).
} 
% conclusions heading (optional) 
{
The X-ray luminosity is clearly not the only parameter that determines
the [Ne\,II] emission. For more exacting tests of X-ray irradiated
disk models, these must be computed for the stellar and circumstellar
characteristics of the observed objects. Explaining the strong
[Ne\,II] emission of Class\,I objects likely requires the inclusion
in the models of additional physical components such as the envelope,
inflows, and outflows. 
} 
 
\keywords{Stars: activity -- Stars: pre-main sequence -- Stars:
formation -- circumstellar matter -- planetary systems: protoplanetary
disks} 

\maketitle

\section{Introduction}\label{sect:intro} 

The first million years in the formation of a low-mass star are
characterized by several complex and still not fully understood
phenomena involving the circumstellar envelope, the disk, and the
central protostar, e.g., envelope and disk mass accretion, outflows,
disk evolution including planet formation, star/disk magnetic
interactions, and other manifestations of the magnetic field such as
the intense X-ray emission from hot magnetically confined plasma. The
first studies of Young Stellar Objects (YSOs) have often neglected
important interactions between these phenomena. A noteworthy example
is the effect of X-rays from the central object on the surrounding
molecular cloud, the accretion envelope, and the disk. YSOs, indeed,
have stronger X-ray emission than main sequence stars \citep{fei99}.

The origin of magnetic phenomena in YSOs, and of their X-ray emission
in particular, is an intriguing and still poorly understood problem.
Renewed interest in YSO X-ray emission comes from the recent
recognition that X-rays may ionize and modify in several other ways
the thermal and chemical structure of star forming clouds
\citep{lor01}, circumstellar disks 
\citep[e.g.][]{gla97,ilg06c,mei08,gor08,erc08}, and planetary
atmospheres \citep{cec06}. We here focus on the response of YSO disks
to X-ray irradiation by the central (proto)star. 

The structure and temporal evolution of circumstellar disks is of
paramount importance for the understanding of the star- and
planet-formation process.  The structure of the gas-phase component,
by far more massive than the dust component, is particularly uncertain
being the most affected by high energy radiation (far/extreme
ultraviolet and X-ray). \citet{gla97} have shown that, in a
protostellar disk illuminated by a central X-ray source, X-ray
ionization dominates over that of galactic cosmic rays, giving rise to
a vertically layered ionization structure with an outer {\em active}
surface and a mostly neutral inner {\em dead zone}
\citep{gla97,gam96}. \citet{ilg06c} calculated the disk ionization
structure as a function of the X-ray luminosity and emitting plasma
temperature, and found that the disk is divided into three distinct
radial zones: an inner active region, a central region where the depth
of the dead-zone depends on the X-ray spectral and temporal
characteristics, and an outer region with non-variable dead-zone. In
addition to ionization, X-rays can heat the external layers of
disk atmospheres, as shown by \citet{gla04} who predict temperatures
of the order of 5000\,K.

Theoretical calculations depend critically on several ingredients: the
disk model, the chemical network, the spectral and temporal
characteristics of the X-ray source and its assumed spatial location
with respect to the disk. Observational tests are therefore highly
desirable and could help constrain the model assumptions. The lines of
ionized neon are particularly useful as a proxy of the effect of high
energy radiation, as its 1$^{st}$ and 2$^{nd}$ ionization potential
are 21.56 and 41\,eV, respectively and therefore ionization can occur
only by photons in the EUV and X-ray range. Moreover, due to its
closed shell configuration, the Ne chemistry is particularly simple.
\citet{gla07}, \citet[][hereafter MGN\,08]{mei08}, \citet[][hereafter
GH\,08]{gor08}, and \citet{erc08} have recently calculated the
strength of fine structure emission lines from ionized neon
originating in a disk exposed to stellar X-rays. \citet{gla07}
estimate that in low-mass YSOs the ionization of neon is dominated by
X-rays, because the photospheres of these stars emit few UV photons
and cosmic rays are removed by the strong winds. Ne\,II and Ne\,III
ions, predominantly resulting from K-shell photoionization of neutral
neon by X-rays with energy E$>$0.87\,KeV, give rise to magnetic dipole
transitions at 12.81$\mu$m and 15.55$\mu$m, respectively. The
predicted line luminosities are, for the reference disk/star models of
MGN\,08 and GH\,08, and for X-ray luminosities of $\sim2\cdot
10^{30}$\,erg~s$^{-1}$, of the order of 10$^{28}$ and 10$^{27}$\,erg
s$^{-1}$ for [Ne\,II] and [Ne\,III], respectively. MGN\,08 predict
that the line luminosities increase with X-ray luminosity following a
steeper-than-linear relation. \citet{erc08} predict, with respect to
MGN\,08 and GH\,08, lower luminosities by a factor of 3-5. The
theoretical models are far from unique: other recent calculations of
[Ne\,II] line emission from a EUV-induced photoevaporative disk wind,
that neglect X-ray irradiation, yield luminosities similar to those
obtained by  MGN\,08 and GH\,08 for their reference models
\citep{ale08}. The observation of line shifts and broadenings,
accessible through ground-based high resolution spectrographs, may help
to discriminate among the different proposed emission mechanisms
\citep[e.g.][]{herc07,van09}.

Detection of the [Ne\,II] 12.81$\mu$m line was first reported by
\citet{pas07} for four stars out of a sample of 6 transition-disk
systems. \citet{lah07} detected the line in 15 more T-Tauri stars and
\citet{esp07} added three more detections. The relation of the line
strengths with X-ray luminosity and with other system parameters has
remained unclear: \citet{pas07} report a correlation of the [Ne\,II]
line luminosity with $L_{\rm X}$ and an anticorrelation with mass
accretion rate\footnote{Note, however, that the four [Ne\,II]
detections of \citet{pas07} span very small ranges of [Ne\,II] line
and X-ray luminosities, both $\sim$0.2\,dex, comparable with 
uncertainties. The same applies to $\dot{M}$, for which the range is
0.5\,dex.}, $\dot{M}$; \citet{esp07}, complementing the \citet{pas07}
data with their own, fail to confirm the correlation with $L_{\rm X}$
and find a possible {\em direct} correlation with $\dot{M}$; the
sample of \citet{lah07} had only sparse X-ray data; in all cases the
samples are small and inhomogeneous, comprising stars in different
star-forming regions with different ages and distances.

In this contribution we investigate the connections between the neon
fine structure line emission and the stellar/circumstellar properties
with particular reference to the X-ray luminosity. We choose to focus
on $\rho$ Ophiuchi, one of the closest, youngest and most studied 
low-mass star forming regions in the solar neighborhood \citep[for a
recent review see][]{wil08}. This is motivated by the fact that (i)
the region has been extensively observed with {\em Spitzer} IRS, (ii)
high quality X-ray data have recently been obtained by ourselves with
the {\em Deep Rho Ophiuchi XMM-Newton Observation} \citep[DROXO,
][]{sci06}, allowing the cross-correlation of {\em Spitzer} sources
with well-characterized X-ray emitters. Moreover, the young
($\lesssim$1\,Myr) $\rho$ Oph members have hard and luminous X-ray
emission, characteristics that are expected to favor an observable
disk response. YSOs in $\rho$ Oph cover a range of evolutionary phases
and include a significant number of Class\,I protostars. While on the
one hand this fact makes our sample inhomogeneous, it also results in
a better coverage of the star/disk/envelope parameter space. We here
assume that $\rho$ Oph is at a distance of 120\,pc, the most recent
value derived by \citet{lom08}.

This paper is organized as follows: we first introduce, in
\S\,\ref{sect:sample} the main X-ray and NIR data, as well as
additional data both original and from the literature; in
\S\,\ref{sect:analysis} we derive the quantities used for the
subsequent analysis. We then compare the theoretical prediction for
the X-ray ionization proxies with the observations
(\S\,\ref{sect:results}) and look for physically meaningful
correlations between these and other stellar/circumstellar properties.
Section \ref{sect:conclusions} summarizes our results and presents our
conclusions. An Appendix describes our method to characterize the YSOs
in our sample by fitting theoretical models to their spectral energy
distributions.

\section{DROXO/{\em Spitzer} sample}
\label{sect:sample}

In order to correlate the [Ne\,II] and [Ne\,III] line strengths with
the stellar X-ray emission and with the properties of the
circumstellar material, we decided to focus on a physically
homogeneous and well characterized sample of YSOs. Our sample includes
YSOs that: $i$) belong to the $\rho$ Ophiuchi star forming region, and
are therefore both young and relatively coeval, $ii$) have been
observed with the {\em Spitzer}/IRS in high resolution mode, $iii$)
fall in the field of view of DROXO, and have therefore well
characterized X-ray emission. 

\subsection{{\em Spitzer}/IRS data}
\label{sect:data_spitzer}

We started by searching the {\em Spitzer} archive for IRS observations
of $\rho$ Ophiuchi members in the {\em XMM-Newton} field (cf. \S
\ref{sect:data_xmm}) performed with the short-high module (SH:
$\lambda=9.9-16.9$\,$\mu$m; $R\sim$600; slit
size\,=\,4.7\arcsec$\times$11.3\arcsec). Excluding GY\,65, which was
identified by \citet{luh99b} as a background star, our sample consists
of 28 YSOs. Table\,\ref{tab:target_obspar} lists these objects, the
{\em Spitzer} program(s) under which they were observed, the total IRS
(SH) exposure times and the details of the observing strategy: the
number of exposures, the number of data collecting events (DCE) per
exposure,  and the integration time for each DCE. Note that four
objects were targeted by two separate programs and have therefore been
observed twice.

\begin{table*}\begin{center}
\caption{Observation parameters for the {\em Spitzer}/IRS
and DROXO data.}
\label{tab:target_obspar}
\begin{tabular}{lrrr|lrr}
\hline
       Object  &   Prog.Id.  & Exp.Time   &   $\rm n_{exp}\times n_{dce}\times T_{dce}$     & DROXO\#& Offax&Exp.T         \\  
               &             &      [s]     &         [s] 				        & &[$^\prime$]  &[Ks]\\
\hline											       
 DoAr25/GY17   &     179     &       975.2  &  4$\times$2$\times$121.9  		        &   3 &  12.4   & 140/130/165  \\
  IRS14/GY54   &     179     &        62.9  &  2$\times$1$\times$31.5			        &     &  13.7   & 127/  -/  -  \\
  WL12/GY111   &     172     &        62.9  &  2$\times$1$\times$31.5			        &   8 &   9.1   & 195/179/238  \\
  WL22/GY174   &      93     &      2265.0  & 12$\times$6$\times$31.5			        &  27 &   6.5   & 251/234/285  \\
  WL16/GY182   &      93     &       251.7  &  4$\times$2$\times$31.5			        &     &   4.2   & 308/285/400  \\
  WL17/GY205   &       2     &        12.6  &  2$\times$1$\times$6.3			        &  34 &   2.9   & 326/317/440  \\
  WL10/GY211   &    3303     &      2265.0  & 12$\times$6$\times$31.5			        &  35 &   6.2   & 263/ 17/337  \\
  EL29/GY214   &    93+2     &       125.8  &  6$\times$3$\times$6.3+2$\times$1$\times$ 6.3     &  38 &   3.2   & 309/309/434  \\
       GY224   &     172     &       251.7  &  4$\times$2$\times$31.5			        &  39 &   1.4   & 357/368/490  \\
  WL19/GY227   &     172     &       251.7  &  4$\times$2$\times$31.5			        &  40 &   2.0   & 359/341/469  \\
  WL11/GY229   &    3303     &     15414.1  &  8$\times$4$\times$481.7  		        &     &   5.4   & 283/ 19/376  \\
  WL20/GY240   &     172     &       251.7  &  4$\times$2$\times$31.5			        &  46 &   1.4   & 376/357/486  \\
 IRS37/GY244   &     172     &        62.9  &  2$\times$1$\times$31.5			        &     &  11.2   &   -/ 11/ 55  \\
   WL5/GY246   &    3303     &      8776.6  & 12$\times$6$\times$121.9  		        &  49 &  11.2   &   -/ 11/206  \\
 IRS42/GY252   &   172+2     &        75.5  &  2$\times$1$\times$31.5+2$\times$1$\times$ 6.3    &  54 &   1.9   & 415/401/495  \\
       GY253   &    3303     &     15414.1  &  8$\times$4$\times$481.7  		        &  55 &   3.7   & 367/390/478  \\
   WL6/GY254   &     172     &        62.9  &  2$\times$1$\times$31.5			        &  56 &  10.3   & 183/ 12/247  \\
      CRBR85   &   172+2     &       306.7  &  2$\times$2$\times$121.9+2$\times$1$\times$31.5   &     &   2.0   & 415/401/511  \\
 IRS43/GY265   &       2     &        12.6  &  2$\times$1$\times$6.3			        &  62 &   2.5   & 412/396/504  \\
 IRS44/GY269   &       2     &        12.6  &  2$\times$1$\times$6.3			        &  64 &   2.6   & 395/377/494  \\
 IRS45/GY273   &     179     &       251.7  &  4$\times$2$\times$31.5			        &  65 &  13.0   &   -/  -/184  \\
 IRS46/GY274   &     172     &       251.7  &  4$\times$2$\times$31.5			        &  67 &   3.0   & 388/374/487  \\
 IRS47/GY279   &     172     &        62.9  &  2$\times$1$\times$31.5			        &  68 &  12.8   &   -/ 10/196  \\
       GY289   &    3303     &     15414.1  &  8$\times$4$\times$481.7  		        &  75 &   7.6   & 256/ 17/346  \\
       GY291   &    3303     &      3900.6  &  8$\times$4$\times$121.9  		        &  76 &   8.4   & 237/ 16/323  \\
 IRS48/GY304   &       2     &        12.6  &  2$\times$1$\times$6.3			        &     &  10.6   & 196/ 13/263  \\
 IRS51/GY315   &   172+2     &       264.3  &  4$\times$2$\times$31.5+2$\times$1$\times$ 6.3    &  87 &   6.1   & 345/356/432  \\
 IRS54/GY378   &       2     &        12.6  &  2$\times$1$\times$6.3			        &  97 &  11.6   & 187/ 13/257  \\
\hline
\end{tabular}\end{center}
Column description. (1): Object names. (2) {\em Spitzer} programs
from which the IRS data was taken. The identification numbers
correspond to the following programs: 2={\em Spectroscopy of
protostellar, protoplanetary and debris disks}  (P.I.: J.R. Houck);
93={\em Survey of PAH Emission, 10-19.5$\mu$m} (P.I.: D. Cruikshank);
172/179={\em From Molecular Cores to Planet-Forming Disks}, (P.I.: N.
Evans);  3303={\em The Evolution of Astrophysical Ices: The Carbon
Dioxide Diagnostic} (P.I.: D. Whittet). (3) Total IRS exposure time
accumulated for each object in the short-high configuration. (4)
Observing strategy, i.e. the number of exposures, $\rm n_{exp}$, the
number of {\em data collecting events} (DCE) per exposure, $\rm
n_{dce}$, and the exposure time of each DCE, $\rm T_{dce}$. For
targets observed by multiple programs these figures are reported for
each program. (5)  DROXO source number from Pillitteri et al. (2009)
for objects detected in the DROXO data. (6)  Distance, in arcmin, from
the {\em XMM-Newton} optical axis during the DROXO exposure. (7)
Exposure times for the three EPIC detectors (PN/MOS1/MOS2); missing
values indicate that the source was outside the detector FOV.
\end{table*}

We downloaded the short-wavelength, high-resolution Basic Calibrated
Data (BCD) for the 28 stars in our sample from the {\em Spitzer}
archive. In order to produce final spectra we used the tools suggested
on the {\em Spitzer} Science Center web pages\footnote{ 
http://ssc.spitzer.caltech.edu/postbcd/irs\_reduction.html}.
Specifically, after removing bad pixels with {\sl IRSCLEAN} v\,1.9 we
extracted the spectra of each DCE in the {\em Spitzer IRS Custom
Extraction (SPICE)} v\,2.0.4 environment. We then added up all the
spectra from a given observation of a given target, from a minimum of
two (the two nod positions) up to 72. This leaves us with 32 spectra
(28 + 4 for the objects that were observed twice). Finally, we used
{\sl IRSFRINGE} v\,1.1 for the defringing. The background (sky)
emission as a function of wavelength was estimated using {\em
SPOT}\footnote{http://ssc.spitzer.caltech.edu/propkit/spot}, in steps
of 0.5$\mu$m for $\lambda$=10-19$\mu$m and 0.01$\mu$m for
$\lambda$=12.76-12.87$\mu$m (the region of the [Ne\,II] line). These 
values, computed for the sky coordinates of the objects and the
observation date, include the expected contributions from the Zodiacal
light, the interstellar medium, and the cosmic infrared
background\footnote{http://ssc.spitzer.caltech.edu/documents/background}. 
They do not consider any eventual extended emission in the target
proximity. Note, however, that this should not affect our main
purpose, i.e. measuring the [Ne\,II] and [Ne\,III] line fluxes, since
emission from these lines is not expected from the cold molecular
cloud in the absence of hot ionizing stars. Previous observations
of YSOs have moreover indicated that the emission of these lines is
spatially unresolved at the {\em Spitzer} resolution \citep{lah07}.
Any continuum emission from the molecular cloud, if present, will thus
be subtracted along with the stellar continuum when measuring line
fluxes.  However, multiple emission components eventually present
within the {\em Spitzer} beam (4-5\arcsec) will obviously be included
in the [Ne\,II] and [Ne\,III] fluxes, including e.g. those that may be
associated with outflows as shown by \citet{van09} for the T Tauri
system.

\subsection{X-ray data}
\label{sect:data_xmm}

The {\em Deep Rho Ophiuchi XMM-Newton Observation} (DROXO) is the most
sensitive X-ray exposure of the $\rho$\,Oph  star forming region
performed so far \citep{sci06}. It consists of an observation of
Core\,F performed with the {\em XMM-Newton} X-ray telescope
\citep{jan01}. The nominal pointing position was  $\alpha_{\rm 2000} =
16$:$27$:$16.5$, $\delta_{\rm 2000} = -24$:$40$:$06.8$.  The
observation, interrupted only by the constraints of the satellite
orbit, was carried out in five subsequent revolutions
(\#\,0961...\#\,0965), for a total exposure time of $515$\,ksec. We
use here data from the European Photon Imaging Camera
\citep[EPIC;][]{str01,tur01}, consisting of three almost co-pointed
imaging detectors (MOS1, MOS2, and pn) sensitive to 0.3-10.0\,KeV
photons and with a combined field of view of $\sim$0.2 square degrees.
Source detection resulted in a list of 111 X-ray emitters,  60 of
which are identified with a mid-infrared object detected by
\citet{bon01} with ISOCAM at 6.7 and/or 14.3$\mu$m. Details of the
data reduction and general results from DROXO are found in Pillitteri
et al. (2009, submitted). The present study is limited to the $28$
YSOs with {\em Spitzer}/IRS coverage (see \S
\ref{sect:data_spitzer}). 

Twenty-two of our 28 YSOs are positionally matched with a DROXO source
using a 5$\arcsec$ identification radius. All the identifications are
unambiguous. Cols.\,5-7 of Table \ref{tab:droxo_irs_results} list the
DROXO source numbers from Pillitteri et al. (2009) and, for all 28
objects, off-axis angles and effective exposure times at the YSO
position for the three EPIC detectors\footnote{The off-axis angle
gives an indication of the quality of the point spread function, which
is sharpest on the optical axis; the effective exposure times are
normalized values taking into account the vignetting of the optical
system and  the detector efficiency at the source position.}. Six YSOs
are not detected in DROXO (see \S\,\ref{sect_ana_X_noDROXO}). In
three cases we used {\em Chandra} ACIS data from the literature. In
the remaining three cases we have computed upper limits for the count
rate as described in Pillitteri et al. (2009). {\em Chandra} ACIS data
were also used for one of the DROXO-detected sources, IRS42/GY252. In
the DROXO data the photon extraction region for this object is
contaminated by a nearby bright source. The higher spatial resolution
{\em Chandra} data is not affected by this problem.

\subsection{Ancillary data and SED fits}
\label{sect:anc_SED}

We have collected additional data for our targets from which we derive
relevant physical parameters. A summary of the results is given in
Table~\ref{tab:target_litdata}. Col.~2 lists the ISOCAM source number
from \citet{bon01}; col.~3 the YSO class derived both from the
spectral slope between $2\,\mu$m and $14\,\mu$m \citep[reported from
][]{bon01} and from our own model fitting of the SED (see below and
Appendix~\ref{ap:sed_just}). As detailed below, the two
classifications agree for 70\% of the sources.

Stellar parameters for Class\,II and III sources (according to
our own classification based on SED fits) were estimated from the
near-IR (2MASS) photometry. The procedure we have used follows closely
that adopted by \cite{bon01} and improved by \cite{nat06}. We assume
that the $J$-band emission from these sources is dominated by the
stellar photosphere and that it is only marginally contaminated by the
emission from circumstellar material and that the IR colors of
Class~II sources can be approximated by emission from a passive
circumstellar disk as described by \cite{mey97}. These assumptions
obviously do not apply to Class~I sources and for this reason we do
not derive photospheric parameters for these sources. 

Our procedure starts with the dereddening of each object in the $J-H$
vs. $H-K$ diagram to the locus of cTTS. As opposed to the procedure
used by \cite{nat06} we have used the \cite{car89} extinction law. 
Two sources have colors slightly bluer than those of reddened
main-sequence stars, presumably due to photometric uncertainties.
Dereddening these sources extrapolating the colors of Class\,II and
III sources would produce an overestimation of the extinction.  To
minimize this effect, we have dereddened these sources to $J-H=0.578$.

The values we derive for the $J$-band extinction $A_{\rm J}$ (col.~4
of Table~\ref{tab:target_litdata}) are very similar to the numbers in
\cite{nat06}. The one exception is WL\,16 for which our procedure
produces a significantly higher extinction.

Bolometric luminosities (col.~5 of Table~\ref{tab:target_litdata})
were estimated from the dereddened $J$ band magnitudes and the
bolometric correction used by \citet{nat06}: $\log L_{\rm bol} = 1.24
+ 1.1 \log L_{\rm J}$. Stellar masses and effective temperatures
(col.~6 and 7) were obtained from $L_{\rm bol}$ assuming that stars
lie on the $0.5$\,Myr isochrone of the \cite{dan97} evolutionary
tracks.  

As part of the DROXO program, we have obtained complementary IR
spectroscopy at the VLT using the ISAAC instrument and the same
observing modes described in \cite{nat06}. Low-resolution spectra
($\lambda/\Delta\lambda\sim900$) in the $J$ or $K$ band were obtained
for 12 of the 13 YSOs in our sample that had not been observed by
\citet{nat06}, the exception being WL\,19/GY\,277. These spectra
comprise the Pa$\beta$ and Br$\gamma$ lines that we use to measure
accretion luminosities and mass accretion rates. For the reduction of
the ISAAC data and the measurements  of the Pa$\beta$ or Br$\gamma$
line we followed the procedures described by \cite{nat06}.  Accretion
luminosities (or upper limits), both from the new near-IR spectra and
from those of \citet{nat06}, were then estimated from empirical
relations with the Pa$\beta$ or Br$\gamma$ luminosities
\citep{nat04b,cal04}. They are listed in col.~8 of
Table~\ref{tab:target_litdata}. Mass accretion rates (or upper
limits), listed in col.~9 of the same table, were calculated from
$L_{\rm acc}$ and the photospheric parameters derived from the near-IR
photometry. They were therefore computed only for YSOs for which these
latter parameters are available, i.e. Class~II and Class~III objects
with complete 2MASS photometry. As a result the new near-IR
spectra add only one accretion rate (for IRS\,54/GY\,378) and two
upper limits (for GY\,253 and IRS\,51/GY\,315) to the values in
\citet{nat06}.

\begin{table*}\begin{center}
\caption{Stellar/circumstellar data for the
objects in our sample (see \S \ref{sect:anc_SED}).}
\label{tab:target_litdata}
\begin{tabular}{lrr@{ - }lrrrrrr}\hline
Object    & ISO   & \multicolumn{2}{c}{IR - SED$^\dag$}& $A_{\rm J}$ & $\log{\frac{L}{L_\odot}}$ & $\log{T_{\rm eff}}$ & ${\rm M_*}$      & $\log{L_{\rm acc}}$ & $\log{\dot{M}_{\rm acc}}$  \\
          & Src.  & \multicolumn{2}{c}{Class}     & [mag]	    &				& [K]		     & [${\rm M_\odot}$] & [${\rm L_\odot}$]   & [${\rm M_\odot/yr}$]	    \\  \hline
DoAr25/GY17   &  38 &	II&   II &   0.7 &   0.15 &   3.63 &  -0.27 & $<$-2.34 & $<$-9.23\\
 IRS14/GY54   &  47 & III?&  III &   5.2 &  -0.23 &   3.55 &  -0.57 & $<$-1.54 & $<$-8.15\\
 WL12/GY111   &  65 &	 I&    I &    -- &     -- &	-- &	 -- & $ $   -- & $ $   --\\
 WL22/GY174   &  90 &	II&   II &    -- &     -- &	-- &	 -- & $ $   -- & $ $   --\\
 WL16/GY182   &  92 &	II&   II &  10.0 &   2.26 &   4.04 &   0.56 & $ $ 2.07 & $ $-5.43\\
 WL17/GY205   & 103 &	II&   II &  11.3 &   0.69 &   3.68 &   0.09 & $ $   -- & $ $   --\\
 WL10/GY211   & 105 &	II&   II &   4.5 &   0.51 &   3.67 &  -0.04 & $ $-0.89 & $ $-7.92\\
 EL29/GY214   & 108 &	 I&    I &    -- &     -- &	-- &	 -- & $ $   -- & $ $   --\\
      GY224   & 112 &	II&   II &   8.6 &   0.54 &   3.67 &  -0.01 & $ $   -- & $ $   --\\
 WL19/GY227   & 114 &	II&  III &  16.3 &   1.88 &   3.90 &   0.51 & $ $   -- & $ $   --\\
 WL11/GY229   & 115 &	II&   II &   4.3 &  -0.94 &   3.47 &  -0.92 & $ $-2.39 & $ $-8.84\\
 WL20/GY240   & 121 &	II&    I &    -- &     -- &	-- &	 -- & $ $   -- & $ $   --\\
IRS37/GY244   & 124 &	II&    I &    -- &     -- &	-- &	 -- & $ $   -- & $ $   --\\
  WL5/GY246   & 125 &  III&  III &  16.8 &   2.22 &   4.02 &   0.55 & $ $   -- & $ $   --\\
IRS42/GY252   & 132 &	II&   II &   7.5 &   0.69 &   3.68 &   0.09 & $<$-1.14 & $<$-8.23\\
      GY253   & 133 &  III&  III &   8.8 &   0.36 &   3.66 &  -0.14 & $<$-1.53 & $<$-8.52\\
  WL6/GY254   & 134 &	 I&   II &  18.6 &   2.43 &   4.12 &   0.59 & $ $   -- & $ $   --\\
     CRBR85   & 137 &	 I&   II &    -- &     -- &	-- &	 -- & $ $   -- & $ $   --\\
IRS43/GY265   & 141 &	 I&    I &    -- &     -- &	-- &	 -- & $ $   -- & $ $   --\\
IRS44/GY269   & 143 &	 I&    I &    -- &     -- &	-- &	 -- & $ $   -- & $ $   --\\
IRS45/GY273   & 144 &	II&   II &   6.6 &   0.07 &   3.62 &  -0.32 & $<$-1.59 & $<$-8.45\\
IRS46/GY274   & 145 &	 I&    I &    -- &     -- &	-- &	 -- & $ $   -- & $ $   --\\
IRS47/GY279   & 147 &	II&   II &   7.4 &   0.63 &   3.68 &   0.05 & $<$-1.55 & $<$-8.61\\
      GY289   & 152 &  III&   II &   7.3 &   0.19 &   3.64 &  -0.24 & $<$-1.23 & $<$-8.14\\
      GY291   & 154 &	II&   II &   7.4 &   0.21 &   3.64 &  -0.23 & $<$-1.65 & $<$-8.58\\
IRS48/GY304   & 159 &	 I&    I &    -- &     -- &	-- &	 -- & $ $   -- & $ $   --\\
IRS51/GY315   & 167 &	 I&   II &  12.9 &   2.29 &   4.06 &   0.57 & $<$ 1.56 & $<$-5.95\\
IRS54/GY378   & 182 &	 I&   II &   6.2 &   0.35 &   3.66 &  -0.15 & $ $ 0.02 & $ $-6.95\\
\hline
\end{tabular}\end{center}
Notes: $^\dag$ IR class from the spectral slope between $2\,\mu$m and
$14\,\mu$m \citep{bon01} and SED Class from fitting of the SED
(\S\,\ref{sect:anc_SED}, Table\, \ref{tab:sed_results});
\end{table*}

Given the fragmentary nature of the above described system parameters
we have striven to obtain a more complete and homogeneous set of
estimates by fitting the SEDs of our objects with star/disk/envelope
models. The details of the procedure, as well as the tests we have
performed to ascertain its usefulness, are described in the Appendix.
Table \ref{tab:sed_results} lists, for each source, the quality of the
fit (the $\chi^2$ of the ``best-fit'' model) and the adopted values,
with uncertainties\footnote{As noted in the Appendix, \S
\ref{ap:sed_method}, the statistical significance of uncertainties is
not easily assessed; the plausibility of the uncertainties on disk
mass accretion rates is, however, indicated by a comparison with
independent estimates from the literaure for a control sample of stars
in the Taurus-Aurigae region (cf. Fig.\,\ref{fig:rob06_mdot}).}, for
the following stellar and circumstellar parameters: extinction (the
sum of interstellar and envelope extinction), stellar effective
temperature and mass, disk mass, disk and envelope accretion rates.
The last column indicates the evolutionary stage assigned following
the criteria given by \citet{rob07}, and reported in the Appendix 
(\S \ref{ap:sed_rhoOph}), based on the disk and envelope accretion
rates and on the disk mass. These definitions are meant to reproduce
in most cases the {\em classical} classification based on the SED
slope (i.e. Class\,I, II, and III), which is often used to describe
the evolutionary status of YSOs. The stages, being based on physical
quantities, have however the advantage of not depending on the
inclination of the system with respect to the line of sight or on the
effective temperature of the central object.

Comparing the evolutionary stages from the SED fits with the IR
classes derived from the ISOCAM photometry
(Tab.\,\ref{tab:target_litdata} and \ref{tab:sed_results}) we find
agreement in 19 out of 27 cases\footnote{As described in the Appendix
we rejected the result of the SED fits for one of our 28 YSOs,
WL5/GY246. We consider it as a Class\,III object, i.e. the same Class
given by the ISOCAM photometry, and derive its parameters from the
spectral type and NIR photometry.} (6, 11 and 2 Class/Stage I, II, and
III objects, respectively): 2 Class\,II objects according to the
ISOCAM classification are reclassified as Stage\,I and one as
Stage\,III; 4 Class\,I and 1 Class\,III are reclassified as Stage\,II.
In the following we will base our discussion on the evolutionary
stages defined according to the results of the SED fits. However, in
order to use a more familiar terminology, when referring to the
`stages', we will improperly adopt the term `class'.

\begin{table*}
\begin{center}
\caption{Parameters from SED fits(cf. \S\,\ref{ap:sed_rhoOph}).}
\label{tab:sed_results}
\begin{tabular}{lrr@{}lr@{}lr@{}lr@{}lr@{}lr@{}lr}
\hline
                Name&     $\chi^2_{best}$&\multicolumn{2}{c}{                         $A_V$}&\multicolumn{2}{c}{            $\log T_{\rm eff}$}&\multicolumn{2}{c}{                $M_{\rm star}$}&\multicolumn{2}{c}{           $\log M_{\rm disk}$}&\multicolumn{2}{c}{     $\log \dot{M}_{\rm disk}$}&\multicolumn{2}{c}{      $\log \dot{M}_{\rm env}$}&    Stage/Class\\
                    &                    &\multicolumn{2}{c}{                         [mag]}&\multicolumn{2}{c}{                           [K]}&\multicolumn{2}{c}{                   [$M_\odot]$}&\multicolumn{2}{c}{                   [$M_\odot]$}&\multicolumn{2}{c}{           [$M_\odot yr^{-1}]$}&\multicolumn{2}{c}{           [$M_\odot yr^{-1}]$}&               \\
\hline
         DoAr25/GY17&              $1.01$&     $1.9$&  $_{      0.76}^{       3.0}$&     $3.5$&  $_{       3.5}^{       3.6}$&    $0.33$&  $_{      0.18}^{      0.95}$&    $-2.0$&  $_{      -2.3}^{      -1.6}$&    $-7.6$&  $_{      -8.2}^{      -7.3}$&    $-7.4$&               $^{      -5.2}$&   II\\
          IRS14/GY54&              $0.11$&     $18.$&  $_{       17.}^{       18.}$&     $3.5$&  $_{       3.5}^{       3.6}$&    $0.35$&  $_{      0.26}^{      0.49}$&    $-6.5$&  $_{      -8.2}^{      -3.6}$&   $<-9.0$&                            $$&      $--$&                            $$&  III\\
          WL12/GY111&              $5.46$&     $56.$&  $_{       45.}^{       63.}$&     $3.4$&  $_{       3.4}^{       3.5}$&    $0.11$&  $_{      0.11}^{      0.15}$&    $-2.2$&  $_{      -3.3}^{      -1.8}$&    $-6.3$&  $_{      -6.4}^{      -5.9}$&    $-6.1$&  $_{      -6.1}^{      -6.0}$&    I\\
          WL22/GY174&             $31.65$&     $63.$&  $_{       62.}^{   1.6e+03}$&     $3.7$&  $_{       3.6}^{       3.7}$&     $1.5$&  $_{      0.99}^{       2.7}$&    $-1.3$&  $_{      -1.5}^{      -1.2}$&    $-6.2$&  $_{      -7.0}^{      -6.1}$&    $-8.3$&  $_{      -8.3}^{      -3.9}$&   II\\
          WL16/GY182&              $9.53$&     $31.$&  $_{       28.}^{       33.}$&     $3.7$&  $_{       3.7}^{       4.0}$&     $3.8$&  $_{       2.9}^{       4.2}$&    $-3.1$&  $_{      -5.1}^{      -1.2}$&    $-8.0$&  $_{      -11.}^{      -7.6}$&    $-7.8$&               $^{      -7.1}$&   II\\
          WL17/GY205&              $2.03$&     $42.$&  $_{       40.}^{       46.}$&     $3.7$&  $_{       3.6}^{       3.7}$&     $1.7$&  $_{      0.80}^{       2.9}$&    $-1.5$&  $_{      -1.7}^{      -1.4}$&    $-6.3$&  $_{      -6.8}^{      -6.2}$&    $-8.0$&  $_{      -9.3}^{      -7.2}$&   II\\
          WL10/GY211&              $0.51$&     $16.$&  $_{       15.}^{       17.}$&     $3.6$&  $_{       3.6}^{       3.7}$&     $1.0$&  $_{      0.42}^{       1.7}$&    $-3.2$&  $_{      -4.2}^{      -2.2}$&   $<-8.3$&                            $$&    $-8.2$&               $^{      -7.1}$&   II\\
          EL29/GY214&              $6.33$&     $42.$&  $_{       37.}^{       45.}$&     $3.8$&  $_{       3.7}^{       4.0}$&     $4.9$&  $_{       3.6}^{       5.7}$&    $-1.7$&  $_{      -3.1}^{      -1.0}$&    $-7.0$&  $_{      -8.5}^{      -6.1}$&    $-5.2$&  $_{      -6.4}^{      -4.2}$&    I\\
               GY224&              $0.33$&     $36.$&  $_{       35.}^{       38.}$&     $3.7$&  $_{       3.7}^{       3.7}$&     $2.2$&  $_{       1.5}^{       2.5}$&    $-3.7$&  $_{      -5.1}^{      -2.2}$&   $<-8.2$&                            $$&   $<-7.9$&                            $$&   II\\
          WL19/GY227&              $2.70$&     $53.$&  $_{       53.}^{       55.}$&     $3.8$&  $_{       3.8}^{       3.8}$&     $3.2$&  $_{       2.8}^{       3.5}$&    $-6.8$&  $_{      -7.5}^{      -6.6}$&   $<-9.0$&                            $$&      $--$&                            $$&  III\\
          WL11/GY229&              $0.04$&     $18.$&  $_{       18.}^{       19.}$&     $3.5$&  $_{       3.5}^{       3.5}$&    $0.15$&  $_{      0.12}^{      0.18}$&    $-4.2$&  $_{      -5.3}^{      -3.0}$&   $<-8.9$&                            $$&    $-8.7$&               $^{      -6.8}$&   II\\
          WL20/GY240&              $0.81$&     $24.$&  $_{       21.}^{       35.}$&     $3.5$&  $_{       3.5}^{       3.6}$&    $0.30$&  $_{      0.17}^{      0.60}$&    $-2.7$&  $_{      -3.8}^{      -1.9}$&    $-7.9$&  $_{      -8.8}^{      -6.9}$&    $-5.5$&  $_{      -6.0}^{      -4.8}$&    I\\
         IRS37/GY244&              $2.02$&     $42.$&  $_{       38.}^{       45.}$&     $3.6$&  $_{       3.5}^{       3.7}$&    $0.63$&  $_{      0.34}^{       1.7}$&    $-2.2$&  $_{      -3.7}^{      -1.6}$&    $-8.1$&  $_{      -9.8}^{      -7.0}$&    $-5.9$&  $_{      -8.1}^{      -5.1}$&    I\\
    WL5/GY246$^\dag$&                $--$&     $52.$&  $_{       44.}^{       65.}$&     $3.8$&  $_{       3.8}^{       3.8}$&     $3.0$&  $_{       1.5}^{       5.0}$&      $--$&                            $$&      $--$&                            $$&      $--$&                            $$&  III\\
         IRS42/GY252&              $1.87$&     $30.$&  $_{       29.}^{       42.}$&     $4.1$&  $_{       4.0}^{       4.1}$&     $3.0$&  $_{       2.6}^{       3.4}$&    $-3.6$&  $_{      -4.4}^{      -2.5}$&   $<-7.5$&                            $$&      $--$&                            $$&   II\\
               GY253&              $5.04$&     $29.$&  $_{       28.}^{       29.}$&     $3.6$&  $_{       3.6}^{       3.6}$&    $0.71$&  $_{      0.39}^{      0.96}$&    $-8.0$&  $_{      -8.5}^{      -7.3}$&   $<-9.0$&                            $$&      $--$&                            $$&  III\\
           WL6/GY254&              $2.10$&     $53.$&  $_{       50.}^{       65.}$&     $4.1$&  $_{       4.0}^{       4.1}$&     $3.3$&  $_{       2.8}^{       3.5}$&    $-3.7$&  $_{      -4.7}^{      -2.9}$&   $<-7.9$&                            $$&      $--$&                            $$&   II\\
              CRBR85&              $0.25$&     $67.$&  $_{       66.}^{       69.}$&     $3.7$&  $_{       3.6}^{       3.7}$&     $2.0$&  $_{      0.79}^{       3.0}$&    $-3.2$&  $_{      -4.7}^{      -2.0}$&   $<-7.9$&                            $$&    $-6.5$&  $_{      -8.2}^{      -5.6}$&   II\\
         IRS43/GY265&              $1.55$&     $47.$&  $_{       44.}^{       51.}$&     $3.6$&  $_{       3.6}^{       3.7}$&     $2.0$&  $_{       1.4}^{       3.4}$&    $-1.5$&  $_{      -3.1}^{      -1.1}$&    $-7.1$&  $_{      -8.5}^{      -6.3}$&    $-5.0$&  $_{      -5.9}^{      -4.4}$&    I\\
         IRS44/GY269&              $2.22$&     $57.$&  $_{       51.}^{       63.}$&     $3.5$&  $_{       3.5}^{       3.6}$&    $0.31$&  $_{      0.26}^{      0.41}$&    $-2.2$&  $_{      -2.5}^{      -2.1}$&    $-6.1$&  $_{      -7.4}^{      -5.8}$&    $-5.7$&  $_{      -5.8}^{      -5.6}$&    I\\
         IRS45/GY273&              $0.81$&     $29.$&  $_{       27.}^{       31.}$&     $3.7$&  $_{       3.6}^{       4.0}$&     $1.9$&  $_{      0.97}^{       2.5}$&    $-4.9$&  $_{      -6.4}^{      -2.7}$&   $<-9.0$&                            $$&   $<-7.2$&                            $$&   II\\
         IRS46/GY274&              $1.23$&     $32.$&  $_{       30.}^{       34.}$&     $3.6$&  $_{       3.5}^{       3.7}$&    $0.61$&  $_{      0.36}^{       1.5}$&    $-2.1$&  $_{      -3.2}^{      -1.6}$&    $-7.8$&  $_{      -9.2}^{      -6.9}$&    $-5.8$&  $_{      -7.2}^{      -5.0}$&    I\\
         IRS47/GY279&              $4.86$&     $31.$&  $_{       29.}^{       33.}$&     $3.7$&  $_{       3.7}^{       3.7}$&     $2.7$&  $_{       2.1}^{       3.6}$&    $-1.9$&  $_{      -3.5}^{      -1.2}$&    $-8.4$&  $_{      -9.9}^{      -8.3}$&    $-6.4$&  $_{      -7.1}^{      -6.0}$&   II\\
               GY289&              $2.32$&     $23.$&  $_{       22.}^{       25.}$&     $3.6$&  $_{       3.5}^{       3.6}$&    $0.47$&  $_{      0.20}^{       1.0}$&    $-5.8$&  $_{      -6.3}^{      -4.9}$&   $<-9.0$&                            $$&   $<-7.0$&                            $$&   II\\
               GY291&              $0.19$&     $24.$&  $_{       23.}^{       26.}$&     $3.6$&  $_{       3.5}^{       3.7}$&    $0.50$&  $_{      0.23}^{       1.4}$&    $-4.5$&  $_{      -5.9}^{      -2.8}$&   $<-8.6$&                            $$&   $<-7.5$&                            $$&   II\\
         IRS48/GY304&              $4.73$&     $26.$&  $_{       18.}^{       35.}$&     $3.6$&  $_{       3.6}^{       3.7}$&    $0.94$&  $_{      0.46}^{       2.1}$&    $-2.0$&  $_{      -3.0}^{      -1.5}$&    $-6.8$&  $_{      -8.3}^{      -5.9}$&    $-5.1$&  $_{      -5.6}^{      -4.5}$&    I\\
         IRS51/GY315&             $12.91$&     $30.$&  $_{       27.}^{       33.}$&     $3.7$&  $_{       3.6}^{       3.7}$&     $2.1$&  $_{      0.80}^{       2.5}$&    $-1.5$&  $_{      -2.0}^{      -1.3}$&    $-6.7$&  $_{      -7.9}^{      -6.2}$&    $-8.1$&               $^{      -6.5}$&   II\\
         IRS54/GY378&              $1.17$&     $26.$&  $_{       25.}^{       28.}$&     $3.7$&  $_{       3.6}^{       3.7}$&     $2.0$&  $_{      0.63}^{       2.6}$&    $-1.9$&  $_{      -2.2}^{      -1.5}$&    $-7.1$&  $_{      -7.7}^{      -6.8}$&    $-7.2$&  $_{      -8.4}^{      -6.5}$&   II\\
\hline
\end{tabular}
\end{center}
$\dag$: Parameters not derived from SED fits, but from the spectral type
and NIR photometric (see Appendix).
\end{table*}

\section{Analysis}
\label{sect:analysis}

We here describe the steps taken to derive X-ray and
[Ne\,II]/[Ne\,III] luminosities from the {\em XMM-Newton} and the {\em
Spitzer} IRS data, respectively.

\subsection{X-ray luminosities}
\label{sect:ana_X}

We discuss separately the X-ray luminosities of the 21 YSOs for which
an analysis of the X-ray spectra from the DROXO observation was
possible (cf. \S\,\ref{sect:data_xmm}), and those of the remaining 7
objects for which we either make use of previous {\em Chandra} ACIS
observations or we compute upper limits from the DROXO data. We will
then discuss possible biases and uncertainties on the X-ray
luminosities due to the high source absorption and to their intrinsic
variability.

\subsubsection{Spectral analysis of DROXO sources}
\label{sect_ana_X_DROXO}

For the 21 YSOs with usable DROXO data the observed low-resolution
X-ray spectra were fitted with simple emission models convolved with
the detector response using XSPEC v.12.3.1 \citep{arn96} as described
by Pillitteri et al. (2009). We analyzed the time-averaged spectra
accumulated during times of low background, i.e. excluding the intense
background flares due to solar soft protons. This is the same time
filter used by Pillitteri et al. (2009) for source detection, as it
maximized the sensitivity to faint sources. It is not, however, the
same filter used by Pillitteri et al. (2009) for spectral analysis.
This latter differs from source to source and was devised to maximize
the S/N by including times of high background when the source is
bright enough to contribute positively to the S/N. Although the
resulting spectra have higher S/N with respect to those based on the
universal time filter we use here, the ensuing luminosities are not
suitable for our purpose as they do not scale linearly with the
time-averaged luminosities. 

As done by Pillitteri et al. (2009) in many cases we fitted
simultaneously data from all three EPIC instruments. In other cases
the combined fits were statistically unsatisfactory because of
cross-calibration issues\footnote{One or both of the following: $i$)
the source falls on a gap between the CCDs in one of the detectors,
and we are unable to properly account for the missing part of the PSF;
$ii$) the source is intense and the statistical uncertainties per
spectral bin are lower than the precision of the cross-calibration.}
and we excluded one or two of the detectors. The choice of detectors
is the same as that of Pillitteri et al. (2009). In all cases, a model
of isothermal plasma emission (the APEC model in XSPEC) subject to
photoelectric absorption (WABS) from material in the line of sight is
found to be adequate. We adopted the plasma elemental abundances
derived by \citet{mag07} for YSOs in the Orion Nebula Cluster (ONC).
In one case (EL29/GY214) the abundances of metals (all elements other
than H and He) had to be increased by a factor of 3.5, with respect to
the \citet{mag07} values, in order to obtain a reasonable fit.  The
spectra were fitted with a variety of initial parameters to avoid
ending up in a local minimum of the $\chi^2$ space. The resulting fits
are all statistically reasonable, with a mean $\chi^2_{red}$=1.1 and a
maximum of 1.7. Results for the 21 usable DROXO sources (along with
X-ray luminosities or upper limits for the other 7 YSOs; see below),
are presented in Table \ref{tab:droxo_irs_results}, cols.\,6-9. For
each source we indicate the detector(s) used for the spectral
analysis, the $\rm N_H$ and kT values from the spectral fits, and the
absorption corrected X-ray luminosity in the $0.3-10$\,keV band.
Statistical 90\% confidence intervals for these quantities are also
given. For $\rm N_H$ and kT these were obtained within XSPEC with the
{\sc error} command, while for $L_{\rm X}$ they were propagated from
those on the plasma emission measures.

\subsubsection{Other objects} 
\label{sect_ana_X_noDROXO}

We then estimated fluxes, or upper limits, for the 7 YSOs without a
usable DROXO detection, marked in Table \ref{tab:droxo_irs_results} by
footnotes in the `Instr.' column indicating the source of the quoted
$N_{\rm H}$, $kT$, and $L_{\rm X}$ values. IRS42/GY252 was detected in
DROXO but, being contaminated by the wings of another bright source,
we prefer to use the luminosity  obtained by \citet{ima01} from the
analysis of a {\em Chandra} ACIS spectrum, corrected for the different
distance assumptions and choice of energy bands. IRS37/GY244 was
also detected by \citet{ima01} but not in the DROXO data, possibly
because it lies close to the edge of the EPIC field and in the
PSF-wings of a brighter X-ray source. The X-ray luminosity given by
\citet{ima01} for IRS37/GY244 is, however, based on a poorly
constrained spectral fit and we, therefore, decided to estimate
$L_{\rm X}$ from the ACIS count rate and a suitable conversion factor
(see below). The same method was adopted to estimate the $L_{\rm X}$ 
of IRS14/GY54 and WL16/GY182. These were detected as faint X-ray
sources by \cite{fla03d} in a re-analysis of the {\em Chandra} data
but no spectral analysis is possible due to the low photon statistics.
The remaining three objects, IRS48/GY304, WL11/GY229, and CRBR85, are
not detected in any X-ray dataset and we estimated upper limits to
their $L_{\rm X}$ from the upper limit to their {\em XMM-Newton}
(DROXO) count-rate.

Count-rate to luminosity conversion factors were thus employied for
six objects for which no reliable spectral anlaysis was possible, i.e.
three {\em Chandra} ACIS detections and three {\rm XMM-Newton} upper
limits. The conversion factors were computed with the {\em Portable,
Interactive Multi-Mission Simulator}\footnote{
http://heasarc.gsfc.nasa.gov/Tools/w3pimms.html}(PIMMS) assuming an
isothermal plasma emission. This requires the assumption of a plasma
temperature, $kT$, and, more crucially, of an absorption column
density, $N_{\rm H}$. For the temperature we took $kT = 3.4$\,keV, the
median of the $kT$ values obtained from the X-ray spectral fits of the
DROXO detections in our sample. Absorption estimates were derived from
the $A_{\rm J}$ values in Table \ref{tab:target_litdata}, when
available, and from the $A_{\rm V}$ values in
Table\,\ref{tab:sed_results} in the remaining cases. $A_{\rm J}$ and
$A_{\rm V}$ were converted to $N_{\rm H}$  following \citet{vuo03}:
$\rm N_{\rm H}= 5.6\times10^{21} A_J = 1.6\times10^{21}
A_V$\,cm$^{-2}$. To the three $L_{\rm X}$ estimates from the {\em
Chandra} ACIS count-rates we assign a 90\% uncertainty of 0.5\,dex. 

\begin{table*}
\begin{center}
\caption{Results from the analysis of the {\em Spitzer}/IRS and DROXO
datasets (cf. \S\,\ref{sect:ana_X} and \S\,\ref{sect:ana_Ne}).}
\label{tab:droxo_irs_results}
\begin{tabular}{lr@{}lrr@{}lr|cr@{}lr@{}lr@{}l}
\hline
           Name&       \multicolumn{2}{c}{$\rm L_{\rm [NeII]}$}&        $\rm F_{\rm [NeII]}^{cont}$&      \multicolumn{2}{c}{$\rm L_{\rm [NeIII]}$}&       $\rm F_{\rm [NeIII]}^{cont}$&         Instr.&\multicolumn{2}{c}{$\rm N_H$}&   \multicolumn{2}{c}{kT}&                  \multicolumn{2}{c}{$\rm L_X$}\\
               & \multicolumn{2}{c}{[$10^{28}$\,erg\,s$^{-1}$]}&                     [Jy]& \multicolumn{2}{c}{[$10^{28}$\,erg\,s$^{-1}$]}&                     [Jy]&               &\multicolumn{2}{c}{[$10^{22}$cm$^{-2}$]}&\multicolumn{2}{c}{[keV]}& \multicolumn{2}{c}{[$10^{28}$\,erg\,s$^{-1}$]}\\
\hline
    DoAr25/GY17&              $1.00$&$_{  0.80}^{  1.19}$&      0.26&             $<0.49$&                  $$&      0.28&            pn      &                   $0.97$&     $_{  0.92}^{  1.03}$&                   $2.74$&     $_{  2.54}^{  2.95}$&  $180.64$&        $_{ 170.89}^{ 193.12}$\\
    IRS14/GY54 &             $<5.43$&                  $$&      0.28&             $<2.18$&                  $$&      0.12&            ACIS$^1$&                       $$&                       $$&                       $$&                       $$&    $2.14$&        $_{   1.07}^{   3.21}$\\
    WL12/GY111 &             $43.29$&$_{ 21.48}^{ 59.14}$&     39.20&            $<20.44$&                  $$&     28.55&            all     &                  $14.62$&     $_{  7.12}^{ 39.18}$&                  $64.00$&                       $$&   $12.14$&        $_{   6.88}^{  49.82}$\\
    WL22/GY174 &           $<397.59$&                  $$&     73.63&            $<26.61$&                  $$&     11.97&            m1+pn   &                  $14.11$&     $_{ 10.74}^{ 18.30}$&                   $2.35$&     $_{  1.72}^{  3.49}$&   $98.80$&        $_{  49.56}^{ 216.32}$\\
    WL16/GY182 &            $<80.07$&                  $$&     29.03&            $<36.24$&                  $$&      8.29&            ACIS$^1$&                       $$&                       $$&                       $$&                       $$&    $2.19$&        $_{   0.69}^{   6.92}$\\
    WL17/GY205 &            $<23.40$&                  $$&      2.01&            $<17.84$&                  $$&      2.09&            all     &                   $3.73$&     $_{  2.20}^{  5.91}$&                   $3.43$&     $_{  1.97}^{ 10.71}$&    $7.88$&        $_{   3.97}^{  17.53}$\\
    WL10/GY211 &              $5.44$&$_{  4.43}^{  6.45}$&      0.41&             $<2.79$&                  $$&      0.40&            pn      &                   $2.80$&     $_{  2.29}^{  3.30}$&                   $4.85$&     $_{  3.73}^{  7.56}$&   $39.66$&        $_{  30.92}^{  49.58}$\\
    EL29/GY214 &            $<78.98$&                  $$&     67.46&            $<40.82$&                  $$&     47.74&            m2      &                   $6.37$&     $_{  5.81}^{  7.15}$&                   $4.23$&     $_{  3.57}^{  4.90}$&  $159.79$&        $_{ 131.08}^{ 200.10}$\\
    GY224      &             $<5.51$&                  $$&      1.36&             $<4.33$&                  $$&      1.38&            m1+m2   &                   $3.42$&     $_{  2.66}^{  4.50}$&                   $8.62$&     $_{  4.35}^{ 32.06}$&   $20.58$&        $_{  17.21}^{  30.18}$\\
    WL19/GY227 &            $<15.17$&                  $$&      1.63&             $<9.32$&                  $$&      1.06&            all     &                   $9.92$&     $_{  8.47}^{ 11.55}$&                   $3.49$&     $_{  2.84}^{  4.53}$&   $75.45$&        $_{  54.90}^{ 104.74}$\\
    WL11/GY229 &             $<0.31$&                  $$&      0.03&             $<0.37$&                  $$&      0.04&            all$^2$ &                       $$&                       $$&                       $$&                       $$&   $<2.69$&                            $$\\
    WL20/GY240 &             $19.80$&$_{ 19.00}^{ 20.60}$&      1.74&             $<4.87$&                  $$&      3.09&            pn      &                   $2.28$&     $_{  2.16}^{  2.39}$&                   $2.48$&     $_{  2.41}^{  2.64}$&  $115.98$&        $_{ 107.26}^{ 124.97}$\\
    IRS37/GY244&             $20.49$&$_{ 14.12}^{ 30.52}$&      1.59&             $<4.72$&                  $$&      1.50&            ACIS$^3$&                   $5.80$&     $_{  4.00}^{  8.80}$&                  $>1.70$&                       $$&   $18.20$&        $_{   5.75}^{  57.54}$\\
    WL5/GY246  &              $5.25$&$_{  4.05}^{  6.45}$&      0.81&              $1.40$&$_{  1.09}^{  1.71}$&      0.38&            all     &                   $6.62$&     $_{  5.93}^{  7.30}$&                   $4.05$&     $_{  3.37}^{  5.20}$&  $621.86$&        $_{ 496.86}^{ 762.79}$\\
    IRS42/GY252&             $<7.73$&                  $$&      6.32&             $<6.89$&                  $$&      5.48&            ACIS$^4$&                   $3.90$&     $_{  2.70}^{  5.20}$&                   $1.30$&     $_{  0.90}^{  2.00}$&   $28.18$&        $_{  12.02}^{  72.44}$\\
    GY253      &             $<0.63$&                  $$&      0.01&             $<0.43$&                  $$&      0.01&            pn      &                   $3.42$&     $_{  3.21}^{  3.62}$&                   $2.62$&     $_{  2.44}^{  2.83}$&  $122.82$&        $_{ 110.46}^{ 135.85}$\\
    WL6/GY254  &            $<41.10$&                  $$&     28.77&            $<22.26$&                  $$&     19.70&            all     &                   $5.73$&     $_{  4.22}^{  7.36}$&                   $8.00$&     $_{  4.53}^{ 43.45}$&   $40.21$&        $_{  31.00}^{  59.38}$\\
    CRBR85     &            $<10.72$&                  $$&      4.98&             $<7.31$&                  $$&      4.21&            all$^2$ &                       $$&                       $$&                       $$&                       $$&   $<7.41$&                            $$\\
    IRS43/GY265&            $261.59$&$_{251.96}^{271.32}$&     12.30&            $<35.94$&                  $$&     12.44&            pn      &                   $4.57$&     $_{  4.40}^{  4.75}$&                   $3.00$&     $_{  2.85}^{  3.16}$&  $277.03$&        $_{ 258.23}^{ 296.42}$\\
    IRS44/GY269&             $90.62$&$_{ 63.85}^{117.43}$&     62.53&            $<87.12$&                  $$&     69.45&            pn      &                   $7.17$&     $_{  6.81}^{  7.55}$&                   $3.71$&     $_{  3.42}^{  4.04}$&  $242.14$&        $_{ 218.22}^{ 267.08}$\\
    IRS45/GY273&              $9.99$&$_{  8.12}^{ 11.86}$&      1.22&             $<2.82$&                  $$&      1.15&            all     &                   $1.00$&     $_{  0.33}^{  1.86}$&                   $4.77$&     $_{  1.59}^{  0.00}$&    $6.58$&        $_{   3.66}^{  16.27}$\\
    IRS46/GY274&            $<13.05$&                  $$&      6.63&            $<11.69$&                  $$&      5.44&            pn      &                   $8.43$&     $_{  5.63}^{ 13.20}$&                   $4.78$&     $_{  2.99}^{  8.92}$&   $14.72$&        $_{   8.75}^{  29.62}$\\
    IRS47/GY279&             $10.29$&$_{  7.68}^{ 12.89}$&      4.74&             $<5.73$&                  $$&      3.81&            all     &                   $1.95$&     $_{  1.30}^{  2.90}$&                   $2.12$&     $_{  1.23}^{  3.83}$&   $18.83$&        $_{  10.90}^{  39.40}$\\
    GY289      &             $<0.49$&                  $$&      0.05&             $<0.42$&                  $$&      0.06&            all     &                   $1.93$&     $_{  1.68}^{  2.20}$&                   $3.09$&     $_{  2.55}^{  3.85}$&   $27.79$&        $_{  23.16}^{  33.36}$\\
    GY291      &             $<1.01$&                  $$&      0.19&             $<0.70$&                  $$&      0.16&            all     &                   $2.33$&     $_{  2.04}^{  2.60}$&                   $2.50$&     $_{  2.14}^{  3.07}$&   $34.87$&        $_{  28.39}^{  41.50}$\\
    IRS48/GY304&            $<37.55$&                  $$&     15.80&            $<18.56$&                  $$&     23.48&            all$^2$ &                       $$&                       $$&                       $$&                       $$&  $<25.70$&                            $$\\
    IRS51/GY315&            $<19.38$&                  $$&      7.23&            $<13.53$&                  $$&      6.31&            m1      &                   $3.46$&     $_{  3.09}^{  3.83}$&                   $2.67$&     $_{  2.33}^{  3.20}$&  $111.89$&        $_{  91.87}^{ 133.90}$\\
    IRS54/GY378&            $<19.34$&                  $$&      2.76&             $<8.87$&                  $$&      3.21&            all     &                  $20.05$&     $_{  6.48}^{ 42.87}$&                   $2.32$&                       $$&   $45.21$&        $_{   0.00}^{1368.17}$\\
\hline
\end{tabular}
\end{center}
Notes: IR and X-ray fluxes and luminosities are corrected for
extinction. Col. 6 indicates the {\rm XMM-Newton}/EPIC or {\em
Chandra} (ACIS) detector(s) whose data was used for the fitting of the
X-ray spectra: m1=MOS1, m2=MOS2, pn=pn, all=MOS1+MOS2+pn. Notes
indicate the origin of $N_{\rm H}$, $kT$, and $L_{\rm X}$ for objects
with no usable DROXO detection  (\S \ref{sect_ana_X_noDROXO}): $^1$
$L_{\rm X}$ from ACIS count-rate; $^2$  $L_{\rm X}$ upper limit from
DROXO data; $^3$ $N_{\rm H}$ and $kT$ from the spectral analysis of
{\em Chandra} ACIS data by \citet{ima01}, $L_{\rm X}$ from ACIS
count-rate; $^4$ $N_{\rm H}$, $kT$, and $L_{\rm X}$ from
\citet[][$L_{\rm X}$ corrected for the differences in the assumed
distance and energy band]{ima01}.
\end{table*}

\subsubsection{Biases and uncertainties}

Given the high absorption to which $\rho$ Ophiuchi members are
subject, we may wonder whether some or all of our X-ray luminosities
are biased by the fact that low temperature emission components may be
fully absorbed and therefore unaccounted for in the spectral fits. A
very soft X-ray emission like that of the evolved CTTS TW Hya
\citep[e.g.][]{ste04}, $kT$=0.2-0.3\,keV, would indeed have remained
undetected in $\rho$ Ophiuchi, as for a typical $N_{\rm
H}=4\times10^{22}$\,cm$^{-2}$ the observed flux in the {\em
XMM-Newton} band is reduced by a factor $\sim5\times10^4$, with
respect to the unabsorbed case. The X-ray spectrum of TW\,Hya is,
however, quite peculiar among YSOs. In the $\sim$1\,Myr old ONC, for
example, the {\em Chandra Orion Ultradeep Project} (COUP) observation
\citep{get05a} indicates that, based on a sample of $\sim$100 members
subject to little absorption ($N_{\rm H}<10^{21}$\,cm$^{-2}$) and
whose X-ray spectra are well fit by 2-T models, the high-temperature
component dominates the emission in most cases (80\%) and, indeed, the
mean $kT$ (weighted by the emission measures of the two components) is
$>$1.07\,keV in 95\% of the cases. We therefore argue that, if low
temperature components similar to those observed in the ONC should
remain indeed unobserved with our data, the resulting underestimation
of the X-ray luminosities would typically be less than a factor of
two.

Another source of uncertainty on the X-ray luminosities is their
intrinsic time variability. While a full study of YSO variability in
$\rho$ Ophiuchi is beyond the scope of the present work, assessing its
effect is  important when we correlate $L_{\rm X}$ with other
quantities observed non-simultaneously with the X-ray observation.
Pillitteri et al. (2009) compare the average X-ray emission during the
DROXO observation with that detected during previous {\em Chandra} and
{\em XMM-Newton} observations of $\rho$ Ophiuchi. The comparison
indicates that the activity levels, averaged over $\sim$1 day, the
typical length of the previous observations, usually vary by less than
a factor of 2 (1$\sigma$) over the timescale of years. The variability
within each X-ray observation, i.e. on the timescale of hours, can
however be much larger due to flares that can reach up to $\sim$100
times the quiescent X-ray luminosity. These large flares are however
not frequent. For the YSOs in the ONC, for example, an analysis of the
lightcurves in the COUP dataset along the lines of \citet{wol05} and
\citet{car07} indicates that the X-ray flux is above the quiescent
level\footnote{More precisely the ``characteristic level'' as defined
by \citet{wol05} and \citet{car07}} by a factor of 2 or more for
10-15\% of the time, and by a factor of 5 or more for 2-3\% of the
time. Making the simplifying assumption that the {\em Spitzer} IRS
observations are much shorter than the timescale of the X-ray
variability, we can take these fractions as the fractions of objects
for which the {\em Spitzer} observations coincided with an X-ray
emission level above the characteristic level by more than the
specified factor. For the 28 objects in our sample this implies 3-4
objects with a difference in $L_{\rm X}$ of a factor of $>$2 and $\le
1$ with a difference of a factor of $>$5.

\subsection{[Ne\,II] and [Ne\,III] line luminosities}
\label{sect:ana_Ne}

The estimation of neon line luminosities is performed in two steps:
the direct measurement of fluxes from the reduced IRS spectra and the
correction for extinction.

\subsubsection{Fluxes}

We measured the [Ne\,II] and [Ne\,III] line fluxes, $F_{\rm II}$ and
$F_{\rm III}$, by integrating the spectra in the
$\lambda$=12.79-12.83$\mu$m and $\lambda$=15.53-15.57$\mu$m intervals,
respectively. The underlying continuum was subtracted by fitting a
polynomial to two intervals on the left and right of the lines:
12.71-12.78$\mu$m and 12.84-12.91$\mu$m for [Ne\,II] and  
15.45-15.53$\mu$m and 15.58-15.65$\mu$m for [Ne\,III]. The degree of
the polynomial ranged between 1 and 3 and the fit was repeated after
excluding datapoints that deviated more than 2$\sigma$ from a first
fit. The 1$\sigma$ uncertainties on the fluxes, $\delta F_{\rm II}$
and $\delta F_{\rm III}$, were then estimated by propagating the
uncertainties on the individual spectral bins, taken as the maximum
between the formal uncertainties given by the reduction process and
the 1$\sigma$ dispersion of the continuum fit. The lines were
considered detected if the signal to noise ratio, $\delta F/F$, was
$>$3. In the opposite case upper limits were computed as
$\max(F,0.0)$+3\,$\delta F$. As indicated in
\S\,\ref{sect:data_spitzer}, four YSOs have spectra from two separate
observations: since the Ne lines are not detected in any of the
spectra of these four objects we report the most stringent of the
upper limits and the continuum fluxes of the relative spectra. 

The [Ne\,II] line was detected in 10 YSOs, i.e. $\sim 36$\% of our
sample, while the [Ne\,III] line is detected only in one star,
WL5/GY246, interestingly the brightest in X-rays in our sample (\S
\ref{sect:results_neIII}) and likely a Class III object (see below).
Figure~\ref{fig:neII_lines} shows the 10 detected [Ne\,II] lines and
the single detected [Ne\,III] line. Gaussian profiles centered at the
nominal line wavelengths and with normalizations from the measured
fluxes are superimposed on the observed spectra.

One of our YSOs, IRS\,51/GY\,315, was also included in the study of
\citet{lah07} using c2d data and we here use the same spectrum. Our
3$\sigma$ upper limits on [Ne\,II] and [Ne\,III] fluxes are $\sim$20\%
and $\sim$50\% higher than implied by the 1$\sigma$ upper limits of
\citet{lah07}. We attribute the discrepancy to the differences between
the two measuring procedures.

\begin{figure}
\begin{center}
\epsfig{file=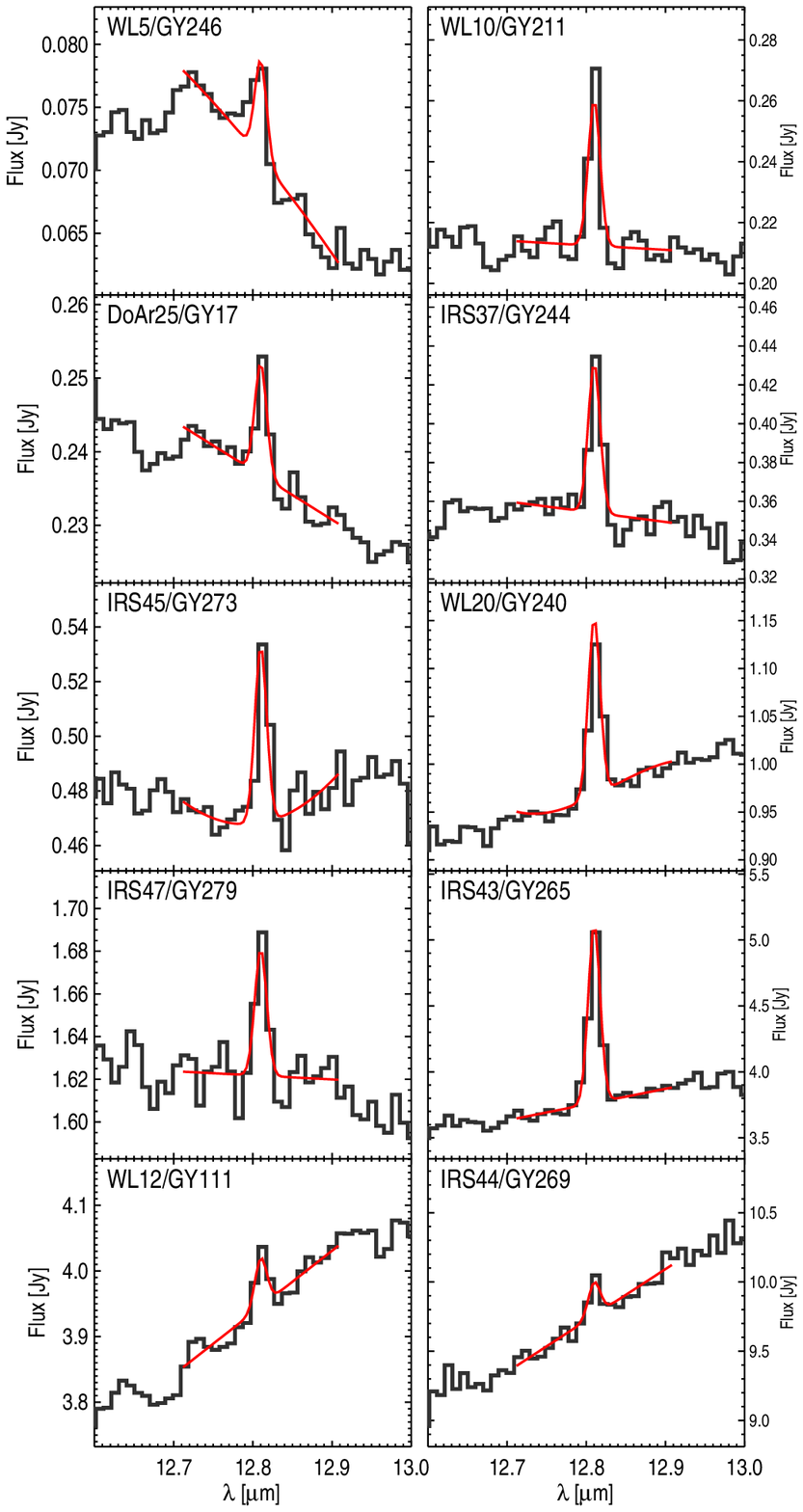, width=8.8cm}
\epsfig{file=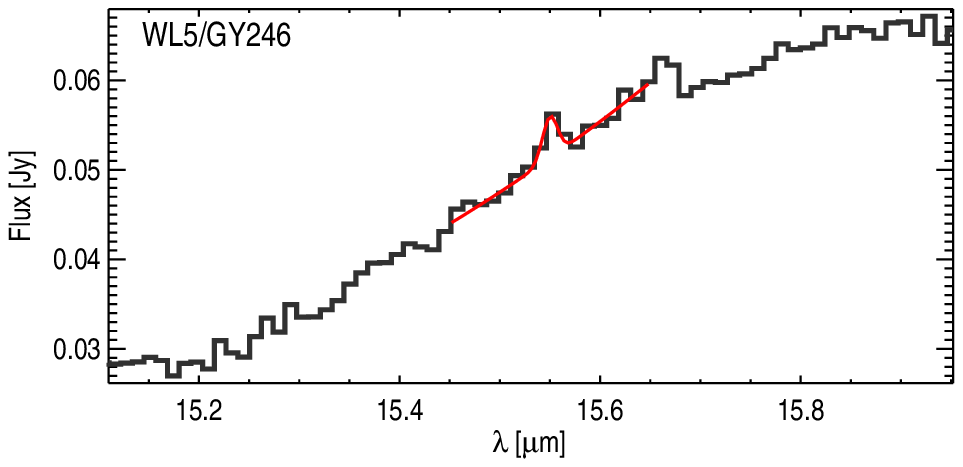, width=8.8cm, bb=14 515 325 650} 
\end{center}
\caption{{\em Spitzer}/IRS spectra in the [Ne\,II] 12.81$\mu$m region
for the ten YSOs for which the line was detected (upper panels), and
in the [Ne\,III] 15.55$\mu$m region for the single detection of this
line (lower panel). The smooth (red) lines are the sum of the
continuum fits used to measure the lines and Gaussians (of fixed width
and cetenred at the nominal line wavelengths) with integral equal to
the line fluxes measured by summing over the line spectral bins (see
text). Object names are given in each panel.}
\label{fig:neII_lines}
\end{figure}

\subsubsection{Extinction correction and luminosities}
\label{sect:extinction}

In order to correct the line (and continuum) fluxes for extinction we
have chosen, for each of our YSO, a best-guess extinction ($A_{\rm
J}$) from the up-to three estimates at our disposal. The $A_{\rm J}$
values from Tab.\,\ref{tab:target_litdata}, estimated from the 2MASS
photometry, were adopted when available. We otherwise estimated
$A_{\rm J}$ from the $N_{\rm H}$ values given by the X-ray spectral
fits (Tab.\,\ref{tab:droxo_irs_results}), converted according to
\citet{vuo03} ($\rm A_J=1.8\cdot 10^{-22}$\,$N_{\rm H}$). Finally, in
the absence of the previous two estimates, we computed $A_{\rm J}$ 
from the $A_{\rm V}$ values given by the SED model fits and listed in
Tab.\,\ref{tab:sed_results}: $A_{\rm J}=0.282$\,$A_{\rm V}$
\citep{rie85}. We make an exception to this rule for WL12/GY111, for
which the $N_{\rm H}$ value is very uncertain and we prefer to use the
extinctions from the SED fit. Table \ref{tab:extinction} lists these
three estimates, along with uncertainties for the latter two, and the
adopted $A_{\rm J}$ value.  For extinctions taken from
Tab.\,\ref{tab:target_litdata} we adopted a 1\,$\sigma$ uncertainty of
1 magnitude.

Critical for the derivation of unabsorbed fluxes is the choice of
extinction-law, i.e., in the cases of the [Ne\,II] and [Ne\,III]
lines, the two ratios $A_{12.81}/A_J$ and $A_{15.55}/A_J$. The
extinction law at these wavelengths, in between two strong silicates
absorption features at 9.7$\mu$m and 18$\mu$m, is not well established
and seems to depend significantly on the grain characteristics
\citep{wei01,dra03}. \citet{cha09} have recently established that, for
stars in the $\rho$ Ophiuchi region with low absorption ($A_K<0.5$),
the $R_V=3.1$ extinction law computed by \citet{wei01} for a
``standard'' grain size distribution fits the measurements between
1.25$\mu$m to 24$\mu$m.  The extinction law of highly absorbed stars
($A_K>2$), however, is better reproduced by the $R_V$=5.5
\citet{wei01} extinction law, implying grain growth in the dense parts
of the cloud. Since, with one exception, the stars in our sample are
highly extincted we adopt the $R_V$=5.5 extinction law and
specifically $A_{12.81}/A_J=0.16$ and $A_{15.55}/A_J=0.13$.

With $A_{\rm J}$  ranging from 0.7 to 25 mag, the resulting correction
factors for the [Ne\,II] 12.81$\mu$m line fluxes range from $\sim$1.1
to $\sim$39 (mean: 5.9). Note that the difference with the $R_V=3.1$
extinction law is significant: had we adopted it 
($A_{12.81}/A_J=0.097$) the 12.81$\mu$m correction factor would have
ranged from 1.1 to 9.5 (mean 2.7)\footnote{The correction factors
computed from the two extinction laws differ significantly only for
high extinctions: for DoAr\,25, $A_{\rm J}=0.7$, the difference is
only $\sim$4\%.}.

\begin{table}
\begin{center}
\caption{Different estimates of the $J$-band extinction for the objects 
in our sample, and adopted value (cf. \S\,\ref{sect:extinction}).}
\label{tab:extinction}
\begin{tabular}{lrr@{}lr@{}lr}
\hline
\scriptsize
&\multicolumn{6}{c}{$A_J$ [mag]}\\
\cline{2-7}
        Name &                          Lit. &     \multicolumn{2}{c}{$N_H$} &       \multicolumn{2}{c}{SED} &   Adopted \\
\hline
 DoAr25/GY17 &       0.7 &            1.7 & $_{       1.6}^{       1.8}$ &            0.5 & $_{       0.2}^{       0.8}$ &       0.7 \\
 IRS14/GY54  &       5.2 &                                             & &            5.0 & $_{       4.9}^{       5.0}$ &       5.2 \\
 WL12/GY111  &           &           26.1 & $_{      12.7}^{      70.0}$ &           15.8 & $_{      12.8}^{      17.7}$ &      15.8 \\
 WL22/GY174  &           &           25.2 & $_{      19.2}^{      32.7}$ &           17.9 & $_{      17.4}^{     456.7}$ &      25.2 \\
 WL16/GY182  &      10.0 &                                             & &            8.6 & $_{       7.8}^{       9.2}$ &      10.0 \\
 WL17/GY205  &      11.3 &            6.7 & $_{       3.9}^{      10.5}$ &           11.9 & $_{      11.1}^{      13.0}$ &      11.3 \\
 WL10/GY211  &       4.5 &            5.0 & $_{       4.1}^{       5.9}$ &            4.5 & $_{       4.2}^{       4.7}$ &       4.5 \\
 EL29/GY214  &           &           11.4 & $_{      10.4}^{      12.8}$ &           11.9 & $_{      10.5}^{      12.6}$ &      11.4 \\
 GY224       &       8.6 &            6.1 & $_{       4.7}^{       8.0}$ &           10.2 & $_{       9.8}^{      10.6}$ &       8.6 \\
 WL19/GY227  &      16.3 &           17.7 & $_{      15.1}^{      20.6}$ &           14.9 & $_{      14.8}^{      15.4}$ &      16.3 \\
 WL11/GY229  &       4.3 &                                             & &            5.2 & $_{       5.0}^{       5.3}$ &       4.3 \\
 WL20/GY240  &           &            4.1 & $_{       3.9}^{       4.3}$ &            6.7 & $_{       6.0}^{       9.9}$ &       4.1 \\
 IRS37/GY244 &           &           10.4 & $_{       7.1}^{      15.7}$ &           11.8 & $_{      10.8}^{      12.7}$ &      10.4 \\
 WL5/GY246   &      16.8 &           11.8 & $_{      10.6}^{      13.0}$ &           14.5 & $_{      12.3}^{      18.3}$ &      16.8 \\
 IRS42/GY252 &       7.5 &            7.0 & $_{       4.8}^{       9.3}$ &            8.3 & $_{       8.2}^{      11.9}$ &       7.5 \\
 GY253       &       8.8 &            6.1 & $_{       5.7}^{       6.5}$ &            8.1 & $_{       7.9}^{       8.3}$ &       8.8 \\
 WL6/GY254   &      18.6 &           10.2 & $_{       7.5}^{      13.1}$ &           14.9 & $_{      14.2}^{      18.4}$ &      18.6 \\
 CRBR85      &           &                                             & &           19.0 & $_{      18.6}^{      19.6}$ &      19.0 \\
 IRS43/GY265 &           &            8.2 & $_{       7.9}^{       8.5}$ &           13.4 & $_{      12.5}^{      14.4}$ &       8.2 \\
 IRS44/GY269 &           &           12.8 & $_{      12.2}^{      13.5}$ &           16.1 & $_{      14.5}^{      17.8}$ &      12.8 \\
 IRS45/GY273 &       6.6 &            1.8 & $_{       0.6}^{       3.3}$ &            8.3 & $_{       7.7}^{       8.9}$ &       6.6 \\
 IRS46/GY274 &           &           15.1 & $_{      10.1}^{      23.6}$ &            9.1 & $_{       8.4}^{       9.6}$ &      15.1 \\
 IRS47/GY279 &       7.4 &            3.5 & $_{       2.3}^{       5.2}$ &            8.7 & $_{       8.3}^{       9.4}$ &       7.4 \\
 GY289       &       7.3 &            3.5 & $_{       3.0}^{       3.9}$ &            6.6 & $_{       6.1}^{       7.0}$ &       7.3 \\
 GY291       &       7.4 &            4.2 & $_{       3.6}^{       4.7}$ &            6.9 & $_{       6.6}^{       7.2}$ &       7.4 \\
 IRS48/GY304 &           &                                             & &            7.4 & $_{       5.0}^{       9.9}$ &       7.4 \\
 IRS51/GY315 &      12.9 &            6.2 & $_{       5.5}^{       6.8}$ &            8.6 & $_{       7.6}^{       9.2}$ &      12.9 \\
 IRS54/GY378 &       6.2 &           35.8 & $_{      11.6}^{      76.5}$ &            7.4 & $_{       7.0}^{       7.9}$ &       6.2 \\
\hline
\end{tabular}
\end{center}
\end{table}

[Ne\,II] and [Ne\,III] line luminosities were finally derived from the
measured fluxes, assuming a distance of 120\,pc. Resulting line
luminosities and upper limits for the whole sample are listed in Table
\ref{tab:droxo_irs_results}, along with absorption-corrected continuum
flux densities at the nominal line wavelengths. The reported
uncertainties reflect measurement errors as well as uncertainties in
$A_{\rm J}$, but neglect possible systematic uncertainties related to
the extinction law.

We conclude this section with a cautionary note: the adopted
extinction values refer to the central objects. The luminosity
corrections are thus valid only if the bulk of the [Ne\,II] and
[Ne\,III] lines originates in the proximity of the YSOs. This
assumptions may be false for emission from shocks associated with jets
and outflows.

\section{Results}
\label{sect:results}

\subsection{The [Ne\,II] line}
\label{sect:results_neII}

We detect [Ne\,II]\,$12.8\,\mu$m line emission in 10 out of the 28
YSOs observed with {\em Spitzer}/IRS within the DROXO field of view
(cf. Fig. \ref{fig:neII_lines}). In one case, WL5/GY246, we also
detect the [Ne\,III]\,$15.5\,\mu$m line. All the [Ne\,II] detections
in $\rho$\,Oph are X-ray sources: 9 are DROXO sources and the tenth,
IRS37/GY244, was detected in an earlier  {\em Chandra} observation
\citep[][see \S\,\ref{sect_ana_X_noDROXO}]{ima01}. Conversely the line
is not detected in any of the three X-ray undetected objects.

\begin{figure}
\begin{center}
\epsfig{file=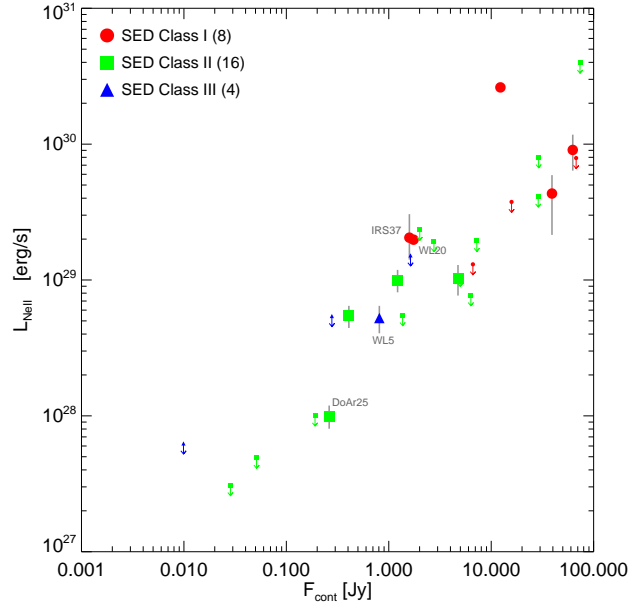, height=8.0cm} 
\caption{[Ne\,II] 12.81$\mu$m line luminosity vs. the continuum flux
density at the same wavelength. Both quantities are corrected for
interstellar extinction. Class\,I, Class\,II, and Class\,III objects,
according to the classification derived from the SED fits, are
indicated with different symbols (see legend).}
\label{fig:neII_vs_c}
\end{center}
\end{figure}

We investigated possible relations between the [Ne\,II] line emission
and other physical parameters of the systems. First, however, we
discuss an important observational bias, namely the dependence of our
line detection sensitivity on the continuum intensity. Figure
\ref{fig:neII_vs_c} shows the relation between the [Ne\,II] line
luminosity and the continuum flux density at the same wavelength. Both
quantities are corrected for interstellar extinction and stars of
different evolutionary classes are plotted with different symbols. The
lower boundary of detections and upper limits clearly shows a positive
correlation, most likely due to the expected anti-correlation between
the detection sensitivity and the counting-statistic uncertainties
(that increase with the continuum). The upper envelope however also
shows a correlation, which is independent of this detection bias and
whose physical origin is to be understood. Figure \ref{fig:neII_vs_c}
also indicates that Class\,I objects have higher continuum flux
densities and [Ne\,II] line luminosities than Class\,II and Class\,III
objects. This is not immediately interpretable in terms of the X-ray
excitation mechanism discussed in the introduction.

\begin{figure}
\begin{center}
\epsfig{file=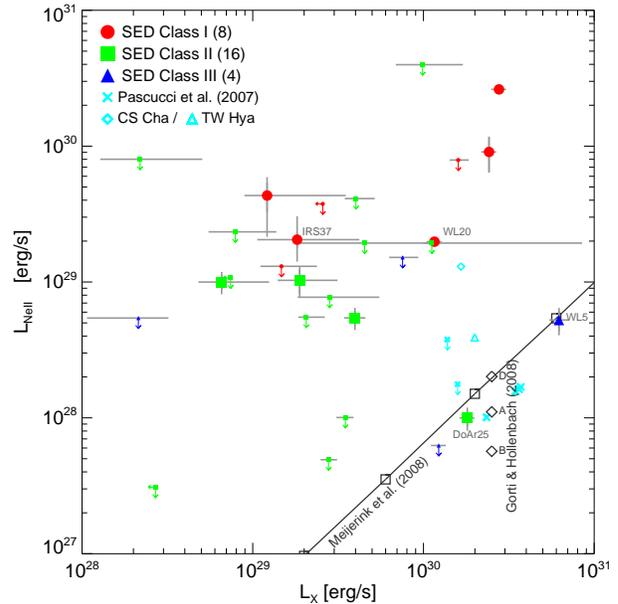, height=8.0cm} 

\caption{[Ne\,II] luminosity vs. $L_{\rm X}$. Symbols as in Fig.
\ref{fig:neII_vs_c}. The 1\,$\sigma$ error bars for the [Ne\,II]
luminosity account for uncertainties in the measurement errors and in
the extinction correction. For the X-ray luminosities estimated from
DROXO data errors come from the statistical uncertainty of the
normalization of the best-fit emission models, while for the {\em
Chandra} ACIS detections we assumed a 50\% uncertainty. Also plotted
here are the six T-Tauri stars in the \citet{pas07} sample and two
stars from \citet{esp07} (CS\,Cha and TW\,Hya). The model predictions
of MGN\,08 are indicated by the diagonal line; those of GH\,08, for
their models `A', `B', and `D' by the three diamond symbols.}

\label{fig:x_vs_neII}
\end{center}
\end{figure}

Figure~\ref{fig:x_vs_neII} shows the scatter plot between the [Ne\,II]
line luminosity and $L_{\rm X}$ in the $0.3-10$\,keV band\footnote{The
choice of energy band is not particularly important: had we 
restricted $L_{\rm X}$ to E$>0.87$\,keV, i.e. to photons relevant for
the K-shell ionization of Ne, the points in Fig.~\ref{fig:x_vs_neII}
would have shifted downward by 0.05-0.15\,dex depending on the plasma
temperature.}. We also plot the six T-Tauri stars in the \citet{pas07}
sample (four [Ne\,II] detections and two upper limits) and two stars
(CS\,Cha and TW\,Hya) from \citet{esp07}. Also shown are the
theoretical predictions by MGN\,08, calculated, as a function of
$L_{\rm X}$, assuming the \citet{dal99} disk model ($M_* = 0.5
M_\odot$, $R_* = 2.0 R_\odot$ , $T_* = 4000 K$, $\dot{M} = 10^{-8}
M_\odot$ yr$^{-1}$), and the predictions of GH\,08 for their fiducial
model (model ``A'': $M_* = 1.0 M_\odot$, $R_* = 2.61 R_\odot$ , $T_* =
4278 K$, $\dot{M} = 3\cdot 10^{-8} M_\odot$ yr$^{-1}$, $L_{\rm
FUV}=10^{31.7}$ erg/s) and two variations: model ``B'' (with 100 times
lower dust opacity, taken to represent an evolved disk)  and  model
``D'' (with 10 times higher FUV flux).

Three conclusions are apparent: i) no overall trend of increasing
[Ne\,II] line luminosity with $L_{\rm X}$ is apparent; ii) two of
the 10 measured [Ne\,II] luminosities, those of DoAr\,25 and
WL\,5/GY\,246, as well as the 18 upper limits, are  consistent with
predictions by current models for X-ray ionization of the disk; iii)
the remaining  8 measured [Ne\,II] luminosities are 1 to 3
orders of magnitude brighter than predicted. With respect to this
latter point it is important to note, however, that the authors of
both calculations stress that their models refer to objects with
specific star and disk parameters and are moreover affected by several
uncertainties, e.g. in the atomic physics, in the simplified disk
models that do not include holes, gaps, and rims, and in the current
lack of EUV photoexcitation. Our findings indeed confirm these
suspicions, and indicate that physical parameters other than $L_{\rm
X}$ are likely to be important in determining the [Ne\,II] line
luminosity. 

Rather than a connection with $L_{\rm X}$, Fig.~\ref{fig:x_vs_neII}
indeed suggests that the [Ne\,II] flux might be related to the
evolutionary state of the YSOs, Class\,I being the strongest and
Class\,III the faintest emitters. The position of TW\,Hya and of the
six \citet{pas07} `transition disk'-systems,  seems consistent with
this interpretation. CS\,Cha, also believed to host a transition disk
has hower a strong [Ne\,II] emission in line with that of most of the
[Ne\,II] detections in our sample.  Within our sample, the higher
[Ne\,II] luminosity of Class\,I objects with respect to Class\,II ones
is confirmed with significances ranging from 99.8\% to 99.99\% by the
five two-population tests for censored data implemented in the {\sc
ASURV} package \citep{fei85,iso90}.

\begin{figure*}
\centerline{
\epsfig{file=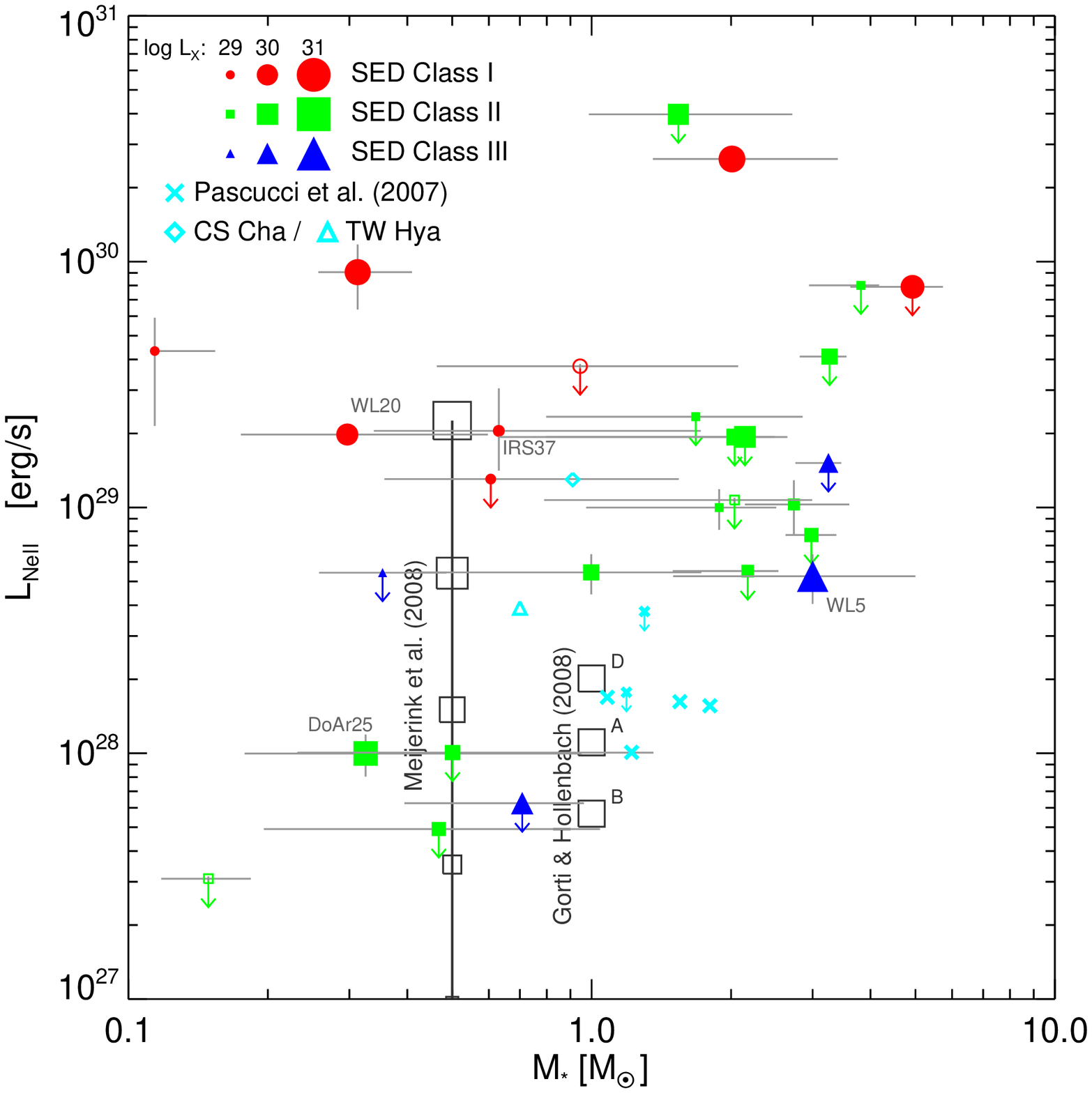, height=8.0cm} 
\epsfig{file=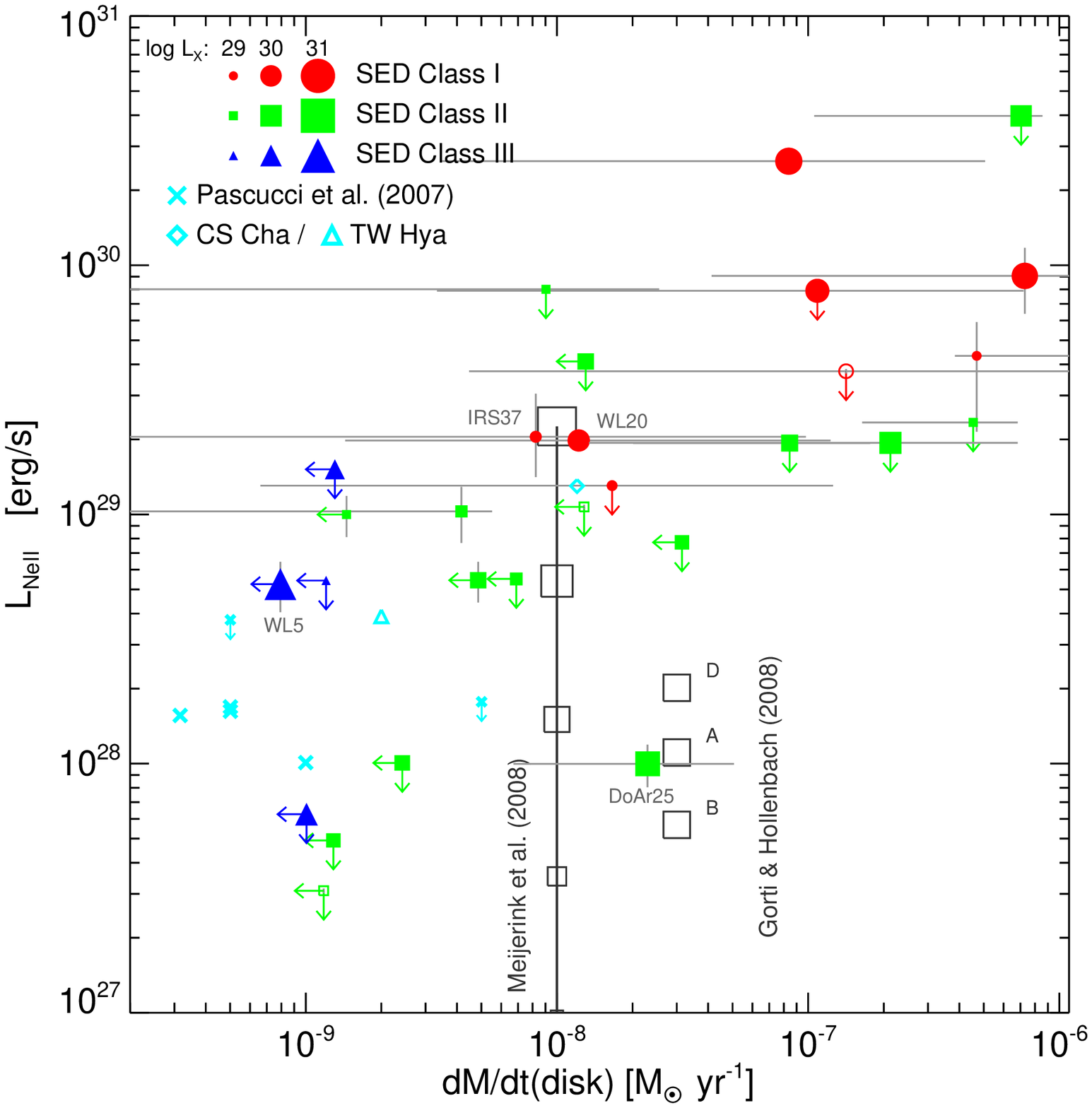, height=8.0cm} 
}
\caption{[Ne\,II] luminosity vs. $M_{*}$ (left panel) and
$\dot{M}_{\rm disk}$ (right panel). Symbols as in Fig.
\ref{fig:neII_vs_c} and \ref{fig:x_vs_neII}, but with size
proportional to $\log L_{\rm X}$, as exemplified in the legend. The
three X-ray undetected objects are indicated by empty symbols. Model
results are plotted as in Fig. \ref{fig:x_vs_neII}.}
\label{fig:neII_vs_mdm}
\end{figure*}

The overall lack of correlation with X-ray luminosity leads us to
investigate possible correlations of the [Ne\,II] emission with other
stellar and circumstellar parameters. Figure \ref{fig:neII_vs_mdm}
shows the relations with disk mass accretion rate and with stellar
mass, both estimated from the SED fits. Also shown are the MGN\,08 and
the GH\,08 model predictions. The most fundamental of stellar
parameters, mass, does not seem to influence the [Ne\,II] luminosity.
At any given mass, Class\,I objects have significantly higher line
luminosities with respect to Class\,II and Class\,III ones; this
points toward a role of parameters related to the YSO evolution, such
as matter inflows and outflows. The [Ne\,II] luminosity seems indeed
to to correlate with $\dot{M}$. Statistical tests for censored data,
the Generalized Kendall's $\tau$ and the Spearman's $\rho$ as
implemented in the {\sc ASURV} package, confirm the existence of a
correlation with $\sim$99.5\% confidence. The stars from \citet{pas07}
and \citet{esp07}, shown in Fig. \ref{fig:neII_vs_mdm} but not used
for the correlation tests, appear compatible with our sample. The
correlation with $\dot{M}$ may also explain the correlation of the
[Ne\,II] luminosity with the continuum flux at 12.81\,$\mu$m (Fig.
\ref{fig:neII_vs_c}) as this latter correlates strongly with $\dot{M}$
(not shown, confidence $\sim$99.99\%). The  [Ne\,II] - $\dot{M}$
correlation might also explain, at least in part, the difference in
the [Ne\,II] luminosity among the different evolutionary classes as
Class\,I objects have statistically higher accretion rates. 

As for the discrepancy with the X-ray excitation models, we note that
three out of five Class\,I [Ne\,II] detections have nominal accretion
rates that are higher than those used as inputs for both of the models
considered here. This might be the reason for their higher than
predicted line luminosities. The other Class\,I [Ne\,II] detections,
WL\,20 and IRS\,37, however, have $\dot{M}$ estimates that, although
with large uncertainties, are  similar to those assumed by the X-ray
ionization models and still their line luminosity is much larger than
predicted. It is also possible that the [Ne\,II]-$\dot{M}$ correlation
simply results from Class\,I objects having brighter line emission due
to a mechanism unrelated to disk mass accretion. One good candidate
might be the defining characteristics of Class\,I objects, i.e. high
envelope accretion rates and/or their associated outflows. Indeed the
[Ne\,II] luminosity also shows a significant correlation, at the
$\sim$99.9\% level, with the $\dot{M}_{\rm env}$ values derived from
the SED fits.

\subsection{The [Ne\,III] line}
\label{sect:results_neIII}

A spectral feature at $\sim15.55$\,$\mu$m, likely associated with
a [Ne\,III] transition, is detected in only one star, WL\,5. An
alternative identification for the observed feature might be a water
rotational transition at 15.57\,$\mu$m, as detected e.g. by
\citet{car08} and \citet{sal08} in three CTTs. The line observed on
WL\,5 is however well centered at 15.55\,$\mu$m (cf.
Fig.~\ref{fig:neII_lines}) and the wavelength difference with the
water line is significant: $\sim$2 spectral bins or about the FWHM
spectral resolution. Moreover the many other water lines that are seen
in the spectra published by \citet{car08} and \citet{sal08} are not
visible in the part of the WL\,5 spectrum shown in
Fig.~\ref{fig:neII_lines}, with the exception of a likely H$_2$O line at
15.67\,$\mu$m. We are therefore confident in the identification of the
line with [Ne\,III].

WL\,5 is a Class\,III object and, according to its F7 spectral type
\citep{gre95}, one of the most massive/hottest objects of the sample
of [Ne\,II] detections (see the Appendix for a detailed discussion of
its properties). For both Neon lines, WL\,5 has the lowest {\em
observed} (i.e. absorbed) continuum flux among the stars in which
[Ne\,II] was detected, thus facilitating the detections of the lines.
The luminosities of both Neon lines compare reasonably well with the
prediction of MGN\,08 for the  $L_{\rm X}$ of the object
(6.2$\times$10$^{30}$\,erg\,$s^{-1}$): the [Ne\,II] line is only 7\%
fainter than predicted (well within 1$\sigma$) and the [Ne\,III] line
is 70\% brighter than predicted (within 2$\sigma$).

All the other stars with [Ne\,II] detection in our sample have
[Ne\,III] upper limits that are significantly larger and therefore
compatible with the predictions of MGN\,08. If we assume that the line
ratios predicted by MGN\,08, rather than the luminosities, are correct
and use the measured [Ne\,II] line luminosities to predict [Ne\,III]
luminosities, we conclude that, for 8 out of 9 stars, our detection
sensitivity for [Ne\,III] is too low by a factor 2.4-8. For the
remaining case, IRS\,43/GY\,265, the star in our sample with the
brightest [Ne\,II] line, the measured upper limit is only 10\% higher
than the predicted [Ne\,III] flux.

\section{Summary and discussion}
\label{sect:conclusions}

We investigated the origin of the [Ne\,II] and [Ne\,III] fine
structure lines by studying a sample of 28 $\rho$ Ophiuchi members in
the field of view of the DROXO deep X-ray observation and with 
available {\em Spitzer} IRS data. The [Ne\,II] 12.81\,$\mu$m and the
[Ne\,III] 15.55\,$\mu$m lines were detected in ten and one YSOs,
respectively; absorption corrected line luminosities and upper limits
for non-detections were computed and compared with predictions of
X-ray disk ionization models. Finally, we explored empirical relations
between [Ne\,II] line luminosity and stellar and circumstellar
parameters estimated by fitting the SEDs of the objects with
star/disk/envelope models.

The luminosities of the 10 detected [Ne\,II] lines are, for the most
part, 1-3 dex higher than predicted by models of X-ray irradiated (and
ionized) circumstellar disks. Moreover, the [Ne\,II] luminosities do
not correlate with the X-ray luminosities. We conclude that, if these
lines are indeed produced by X-ray ionization, factors other than
$L_{\rm X}$ are also important for the line production. Published
models might still be valid: since they assume given star and disk
characteristics (or few variations) it is possible that some of these
assumptions are critical and that they do not correspond to the
characteristics of most of our stars. Other excitation mechanisms
might, however, turn out to be more important than X-rays, such as
strong shocks resulting from the interaction of the stellar wind and
jets with circumstellar material \citep{har89,hol89,van09}.

Interestingly, the [Ne\,II] luminosities of two of the objects in our
sample, DoAr\,25 and WL\,5/GY\,246, match the theoretical prediction
for X-ray irradiated disks remarkably well. DoAr\,25 is a Class\,II
object with mass accretion rate similar to that assumed by the models
we have used for comparison. WL\,5 is, based on its SED between 1 and
8\,$\mu$m, a Class III object, but we cannot exclude the presence of a
gas disk or a dust disk with a large inner hole. It might be similar
to the four stars with transitional disks studied by
\citet{bro07}, for one of which, T\,Cha, the [Ne\,II] line was
detected by \citet{lah07}. WL\,5 is, moreover, together with Sz\,102
\citep{lah07}, the second YSO for which a detection of the  [Ne\,III]
line has been reported. As for the [Ne\,II] line, the luminosity of
the [Ne\,III] line  of WL\,5 is roughly consistent with theoretical
predictions for the X-ray irradiation mechanism. Also largely
consistent with this mechanism are the upper limits to the
luminosities of the undetected [Ne\,II] and [Ne\,III] lines. 

The [Ne\,II] 12.81\,$\mu$m line luminosity correlates with the
continuum flux at the same wavelength. While the lower envelope of the
relation may be explained with a sensitivity bias, the upper envelope
likely reflects a correlation with some physical characteristics to
which the continuum flux is related. Disk accretion rate, which we
have found to correlate with the continuum flux at 12.81\,$\mu$m, is
one candidate, and indeed the [Ne\,II] luminosity correlates with it.
A tentative physical explanation of the correlation might involve the
increased EUV flux produced in the accretion shock, provided that this
latter is able to reach and significantly ionize neon atoms above the
accretion disk. Alternatively, given the correlation generally found
between accretion and outflow rates, the correlation might result from
the [Ne\,II] emission being produced in outflow-related shocks as
mentioned above. Finally, the statistical correlation might be
unphysical, and simply driven by the higher accretion rates of
Class\,I objects combined with their higher [Ne\,II] luminosities.

Indeed, Class\,I objects, i.e. those with significant envelope
accretion, have significantly higher [Ne\,II] luminosities than
Class\,II objects. We propose that the presence of a circumstellar
envelope and/or envelope accretion and/or the strong associated
outflows, i.e. the defining characteristics of Class\,I objects, plays
a role in determining the line emission. Larger YSO samples and more
sophisticated theoretical models are needed to pinpoint the production
mechanism of these gas-tracing lines and to derive a consistent
picture of the environment around YSOs at different evolutionary
stages.

\begin{acknowledgements}

We acknowledge financial support by the {\em Agenzia Spaziale
Italiana}.  This work is based in part on observations made with the
{\em Spitzer} Space Telescope and with {\em XMM-Newton}. {\em Spitzer}
is operated by the Jet Propulsion Laboratory, California Institute of
Technology under a contract with NASA. {\em XMM-Newton} is an ESA
science mission with instruments and contributions directly funded by
ESA Member States and NASA. We thank the anonymous referee for his
numerous and insightful comments and suggestions.

\end{acknowledgements}

\bibliographystyle{aa} %aa.bst\
\bibliography{bibtex.bib}

\appendix

\section{(Circum)Stellar parameters from SED fits}
\label{ap:sed_just}

In this appendix we describe how we constrained some stellar and
circumstellar parameters of the objects in our sample by comparing
their SEDs with the theoretical models of \citet{rob06}. These consist
of a grid of 200,000 model SEDs that include contributions from the
central star, the circumstellar disk, and the envelope, parametrized
with 14 parameters. The models that best approximate the observed SEDs
were found with the aid of the Web based tool presented by
\citet{rob07}. As stated by \citet{rob07}, and in accord with basic
principles, this method does not allow the simultaneous determination
of all the 14 physical parameters, since the SEDs are often defined by
less than 14 independent fluxes. However, depending on the available
fluxes, {\em some} of the parameters can be constrained more narrowly
than others. We are here interested, in particular, in obtaining the
range of values compatible with the observed SEDs for: $i$) the
extinction toward our objects, $ii$) their disk accretion rates.

\subsection{The method and its validation}
\label{ap:sed_method}

Our procedure follows closely that of \citet{rob07}: from the Web
interface we obtain, for each object, a list of the 1000 models that
best approximate the observed SEDs, i.e. those with the smallest
$\chi^2$. Our ``best guess'' parameter values and associated
confidence intervals are then derived by selecting a set of {\em
statistically reasonable} models and computing the median and the
$\pm1\sigma$ quantiles of the parameter values for these models. The
statistically reasonable models were defined as those with reduced
$\chi^2 < (\chi^2_{\rm best}+3)$, where $\chi^2_{\rm best}$ refers to
the best fit model, or in cases this condition results in less than 10
models, the 10 models with smallest $\chi^2$. Note that, because
the uncertainties on the observed SEDs are not well defined (see
below), and the parameter space is sampled only discretely by the
adopted grid of models, the statistical significance of the thus
derived confidence intervals cannot be easily assessed.

\begin{figure*}[]
\centerline{
 \epsfig{file=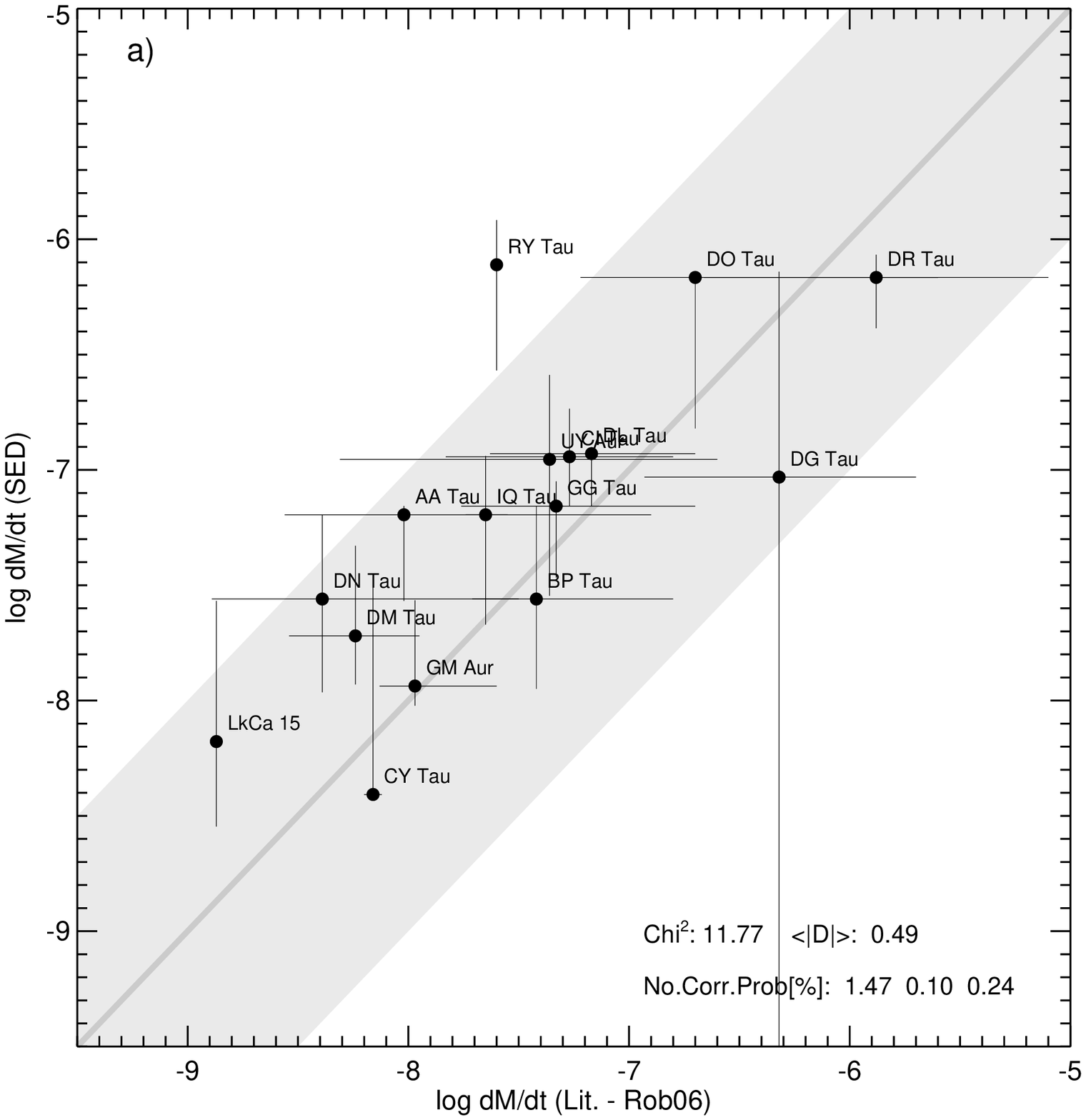   , height=6.0cm} 
 \epsfig{file=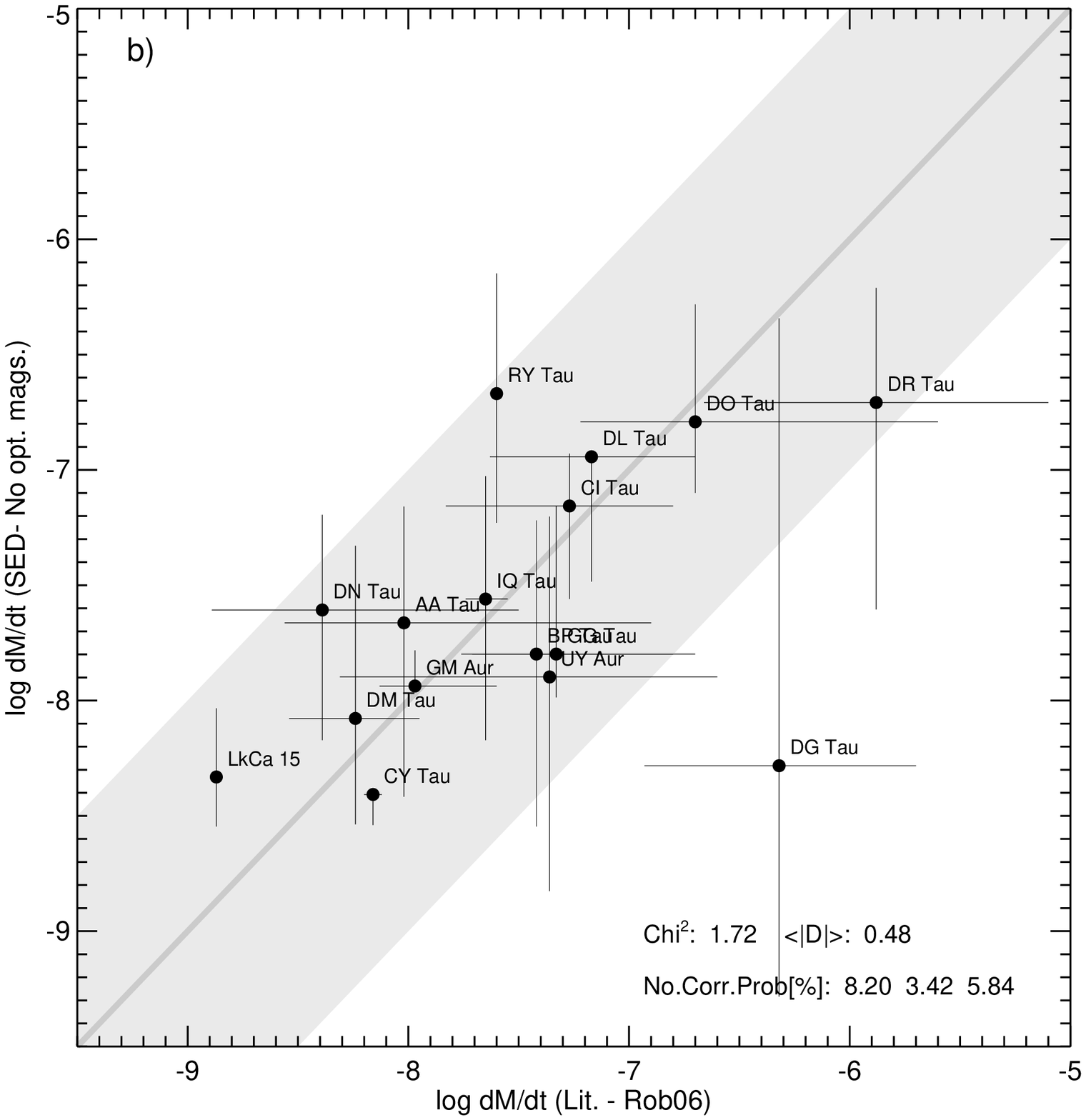 , height=6.0cm}
 \epsfig{file=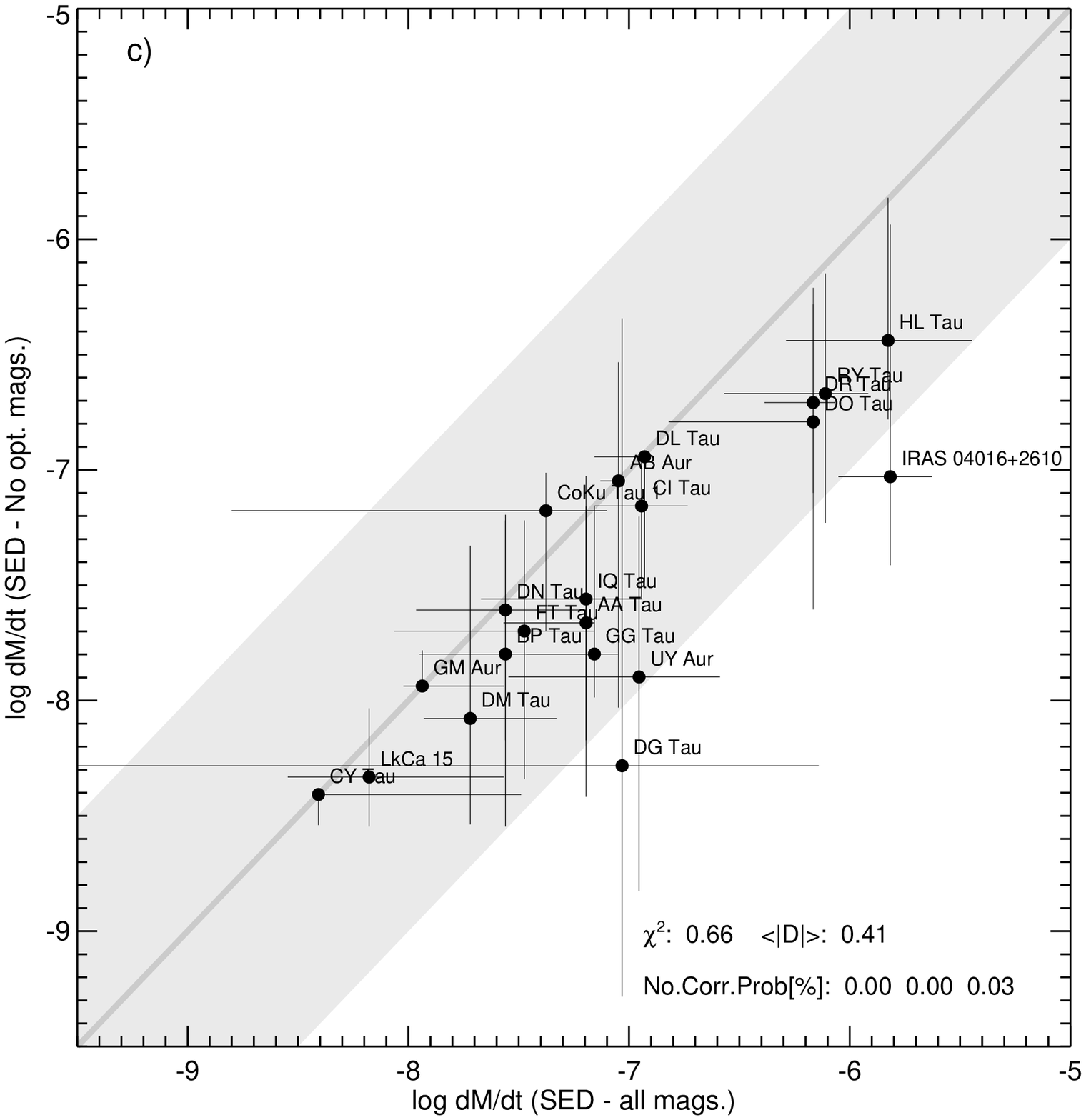, height=6.0cm} 
}
\caption{Comparison between mass accretion rates from the literature
and those derived from SED fits for the sample of T-Tauri stars
considered in \citet{rob06}. SED fits and determination of parameter
ranges were performed as for the $\rho$ Ophiuchi objects discussed in this
paper. Panel $a)$ compares the literature data with results of
SED fits using all the available photometry, including optical bands.
Panel $b)$ is analogous, but only photometry longward of 1$\mu$m
was used for the SED fits. Panel $c)$ finally compares the
results of SED fits with and without optical photometry. Reduced
$\chi^2$ values and mean absolute distances from the bisector, both
computed considering uncertainties on the abscissa only, are reported
within each panel.}
\label{fig:rob06_mdot}
\end{figure*}

A similar method\footnote{\citet{rob07} took the confidence interval
for each parameter as the full range of values in the selected models,
and the ``best guess'' values as those of the model with minimum
$\chi^2$.} was tested by \citet{rob07} by considering a sample of
Taurus-Auriga objects for which stellar and circumstellar parameters
had been derived independently in the literature and comparing these
parameters with those obtained from fitting the SEDs, defined from the
optical to millimeter wavelengths. In the case of our heavily absorbed
$\rho$ Ophiuchi YSOs the SEDs lack, with the exception of one star,
data in the optical bands, i.e. those more directly affected by the
accretion-shock emission. In order to test our ability to constrain
the accretion rates in the absence of optical information, we repeated
the SED fits of the Taurus-Auriga stars of \citet{rob07}, using the
same datapoints to define the SEDs, and both including and excluding
the optical magnitudes. The results are shown in
Fig.\,\ref{fig:rob06_mdot}. Panel $a)$, analogous to Fig\,2b in
\citet{rob07}, compares the accretion rates derived from the SED fits,
including optical data, with independent values from the literature.
Panel $b)$ compares the results of the SED fits without the optical
magnitudes with the literature data. The agreement between the two
quantities is acceptable and may actually be considered better than in
the former panel: the reduced $\chi^2$, computed from the identity
relation considering only uncertainties on $\dot{M}_{SED}$, is indeed
reduced from $\sim$12 to 1.7. This can in part be attributed to the
increased error bars; note, however, that the average of the unsigned
differences, abs($\dot{M}_{SED}$-$\dot{M}_{Lit.}$), is almost
unchanged, 0.49\,dex for panel $a)$ and 0.48\,dex for panel $b)$.
Panel $c)$ compares the $\dot{M}$ from the SED fits with and without
optical magnitudes, showing that the two sets of values agree within
uncertainties.  We conclude that the SEDs defined from IR to
millimeter wavelengths are indeed sensitive to the accretion rate, at
least in the $\dot{M}$ range covered by the Taurus-Auriga sample:
log\,$\dot{M}$=[-8.5,-6]. 

\begin{figure}
\centerline{
 \epsfig{file=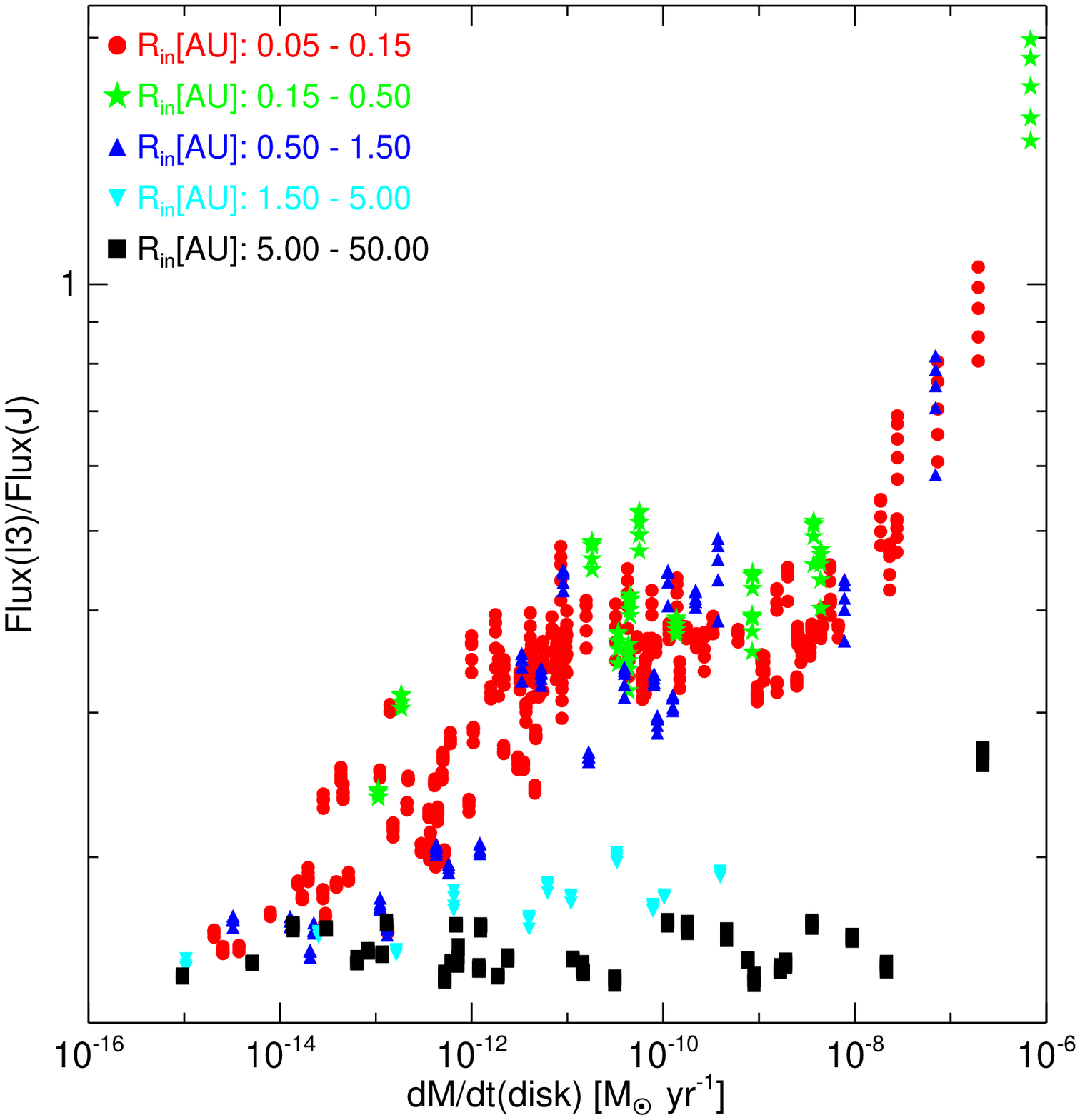, height=8.8cm}\\
}
\caption{Scatter plot of the ratio between the flux in the IRAC\,1
band over that in $J$, as a function of
disk accretion rate, according to the \citet{rob06} models for a solar
mass stars. Each point corresponds to one of the \citet{rob06} models
satisfying the following conditions: mass of the central object
between 0.7 and 1.3 M$_\odot$, age between 1 and 2\,Myr, and disk
inclination with respect to the line of sight $<$60$^\circ$. Different
symbols indicate models with inner disk radius in five ranges as
indicated in the legend.}
\label{fig:sed_colors}
\end{figure}

This is due to the effect of viscous heating affecting the disk
thermal structure. To exemplify this effect we plot in
Fig.\,\ref{fig:sed_colors}, as a function of accretion rate, the ratio
between the IRAC\,3 band and the $J$-band flux, for the \citet{rob06}
models for stars with mass between 0.7 and 1.3\,M$_\odot$, age between
1 and 2\,Myr (implying little or no circumstellar envelope), and low
disk inclination with respect to the line of sight ($i<$60$^\circ$).
We plot with different symbols models with disk inner radii in
different ranges, since the inner hole affects the flux at the IRAC\,3
wavelength (5.8\,$\mu$m). A relation between the two quantities is
seen for models with moderate inner disk holes, apparently
characterized by different regimes in three different $\dot{M}$
ranges: $\log(\dot{M}/M_{\odot})\lesssim-11$, 
$-11\lesssim\log(\dot{M}/M_{\odot})\lesssim-9$, and 
$\log(\dot{M}/M_{\odot})\gtrsim-9$. The factor of $\sim$2 scatter
around this relation may likely be attributed to model variations
within the specified parameter ranges and to the several other
unconstrained model parameters. Similar and even more pronounced
trends are apparent in analogous plots using fluxes in longer
wavelength IRAC and MIPS bands, with the expected difference that at
the longer wavelengths, emitted farther out in the disk, the size of
the inner hole has a much smaller effect. The three regimes in
Fig.\,\ref{fig:sed_colors} can be understood as follows: $i$) for
large accretion rates, $\log(\dot{M}/M_{\odot})\gtrsim-9$, the flux in
the IRAC band, originated in the inner disk (R$<$1\,AU), is
significantly affected by viscous accretion \citep{dal98,dal99}; $ii$)
for $-11\lesssim\log(\dot{M}/M_{\odot})\lesssim-9$ disk heating is
dominated by the stellar photospheric emission and, consequently, no
relation between the IRAC flux and $\dot{M}$ is observed; $iii$) for
$\log(\dot{M}/M_{\odot})\lesssim-11$ we again observe a direct
relation between the IRAC\,3 flux and $\dot{M}$, which we attribute to
the fact that these low accretion rates correspond, in the
\citet{rob06} model grid, to very low disk masses (M$_{\rm
disk}\lesssim10^{-6}$M\,$_\odot$ for the $\sim$1 solar mass stars
plotted in Fig.\,\ref{fig:sed_colors}). Since, in the model grid, disk
mass and accretion are directly correlated and such low mass disks are
optically thin \citep{rob06}, lower accretion rates imply lower disk
mass and lower emission in the IRAC band. The IRAC\,3 flux vs.
$\dot{M}$ correlation in this regime does not therefore imply that
that the mid-IR SED carries {\em direct} information on disk
accretion. 

As a result of this discussion, in the derivation of accretion rates
for our $\rho$ Ophiuchi sample from the SED fits, we decided not to
use values below  $10^{-9}$M\,$_{\odot}$\,yr$^{-1}$. In such cases we
instead conservatively assigned upper limits to $\dot{M}$ equal to the
maximum between $10^{-9}$M\,$_{\odot}$\,yr$^{-1}$ and the upper end of
the $\dot{M}$ confidence interval (see above).

\subsection{The $\rho$ Ophiuchi sample}
\label{ap:sed_rhoOph}

We collected photometric measurements and uncertainties (when
available) for our $\rho$ Ophiuchi sample from several sources: $J$,
$H$, and $K_s$ magnitudes (or upper limits) were taken for almost all
objects from 2\,MASS\footnote{The $J$-band flux of WL5/GY246 and the
$H$-band flux of CRBR85 were taken from \citet{all02} (converted from
the HST bands); the $J$-band upper limit for CRBR85 was taken from
\citet{bra00}.}; {\em Spitzer} IRAC (bands 1-4) and MIPS (bands 1 \&
2) photometry was collected from the c2d database\footnote{Photometry
extracted from the final (November 2007) c2d data delivery, selected
according to the following conditions on quality flags: `detection
quality flag' equal to `A', `B', or 'U'; `image type' equal to `0' for
MIPS2 and different from `-2' and `0' for IRAC and MIPS1; `flux
quality' flag equal to `A', `B', or empty.} \citep{eva03}; 1.2\,mm
fluxes were collected from \citet{stan06} and 1.3\,mm fluxes from
\citet{and94}\footnote{Seven total fluxes from spatially resolved maps
\citep[Tab. 2 in][]{and94}, and 15 peak fluxes or upper limits (from
Tab. 1 in the same work), converted to total flux with the factors
suggested by the authors.}. Optical $UBVR$ photometry for one object
with small absorption (DoAr\,25) was taken from \citet{yak92}. Table
\ref{tab:sed_input} list all the photometric flux densities collected
from the literature.

Finally we complement the photometric data with flux densities from
the IRS spectra (cf. \S\,\ref{sect:data_spitzer}). We computed flux
densities between 10 and 18$\mu$m, at regular wavelength intervals
spaced by 0.5$\mu$m. Each flux density was taken as the average of the
spectral bins in 0.2$\mu$m intervals centered at the nominal
wavelength. For the four stars with two IRS observations we have taken
the average of the two spectra. (In three cases the
wavelength-averaged fluxes differ by less than 0.1\,dex, while in one
case, EL29/GY214, the difference is 0.4\,dex. In all cases we verified
that the results of the model fits did not change appreciably choosing
either of the two spectra). Table \ref{tab:sed_input_IRS} lists the
flux densities from the IRS spectra. As stated in
\S\,\ref{sect:data_spitzer}  our sky subtraction procedure does not
take into account diffuse nebular emission. In order to assess the
significance of diffuse emission on the object flux densities, we have
considered the IRS spectra of the 13 YSOs in our sample observed in
the context of the {\em Spitzer} legacy program {\em From Molecular
Cores to Planet-Forming Disks} \citep[`c2d',][]{eva03}. As with the
entire c2d sample, the reduced/sky-subtracted IRS spectra have been
analyzed (and made publicly available) by the c2d team, using a
sophisticated extraction and sky subtraction method based on the
modelling of the cross dispersion profiles \citep{lah07}. We have
compared the flux densities derived from the c2d-reduced spectra with
those derived from the same spectra reduced by us. We find the spectra
to be similar, with both the maximum and the wavelength-averaged
discrepancy decreasing with object intensity. The maximum discrepancy
falls below 10\% for the 9 YSOs with c2d-reduced spectra that have
average flux $>$0.5\,Jy. Based on this comparison, and noting that the
c2d objects are representative of our sample as for their position
with respect to nebulosity seen in IRAC and MIPS maps, we decided to
use the IRS-derived fluxes for defining the SEDs of the 17 stars with
average IRS flux $>$0.5\,Jy. 

\begin{figure*}
\centerline{
\epsfig{file=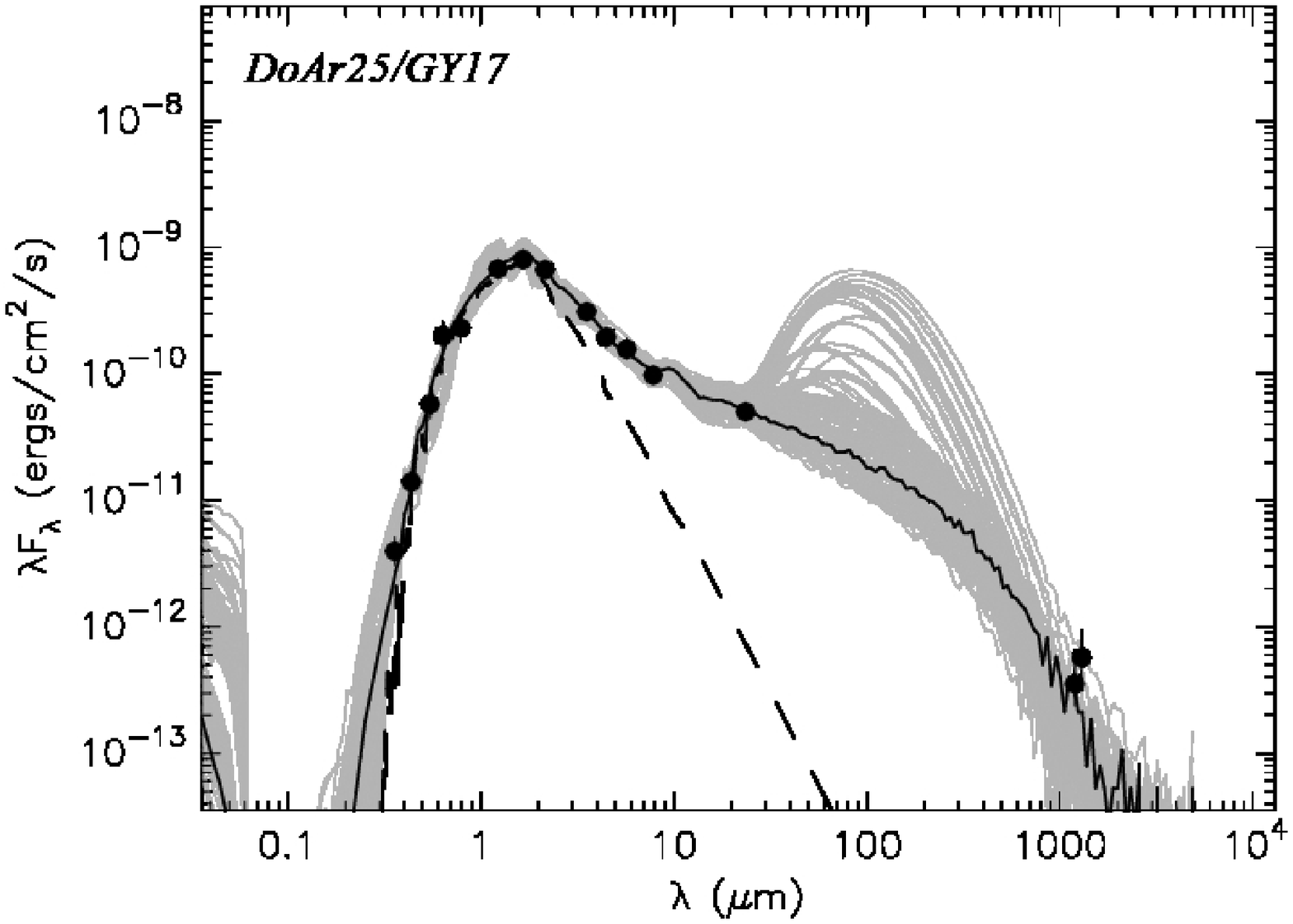 , width=5.80cm} 
\epsfig{file=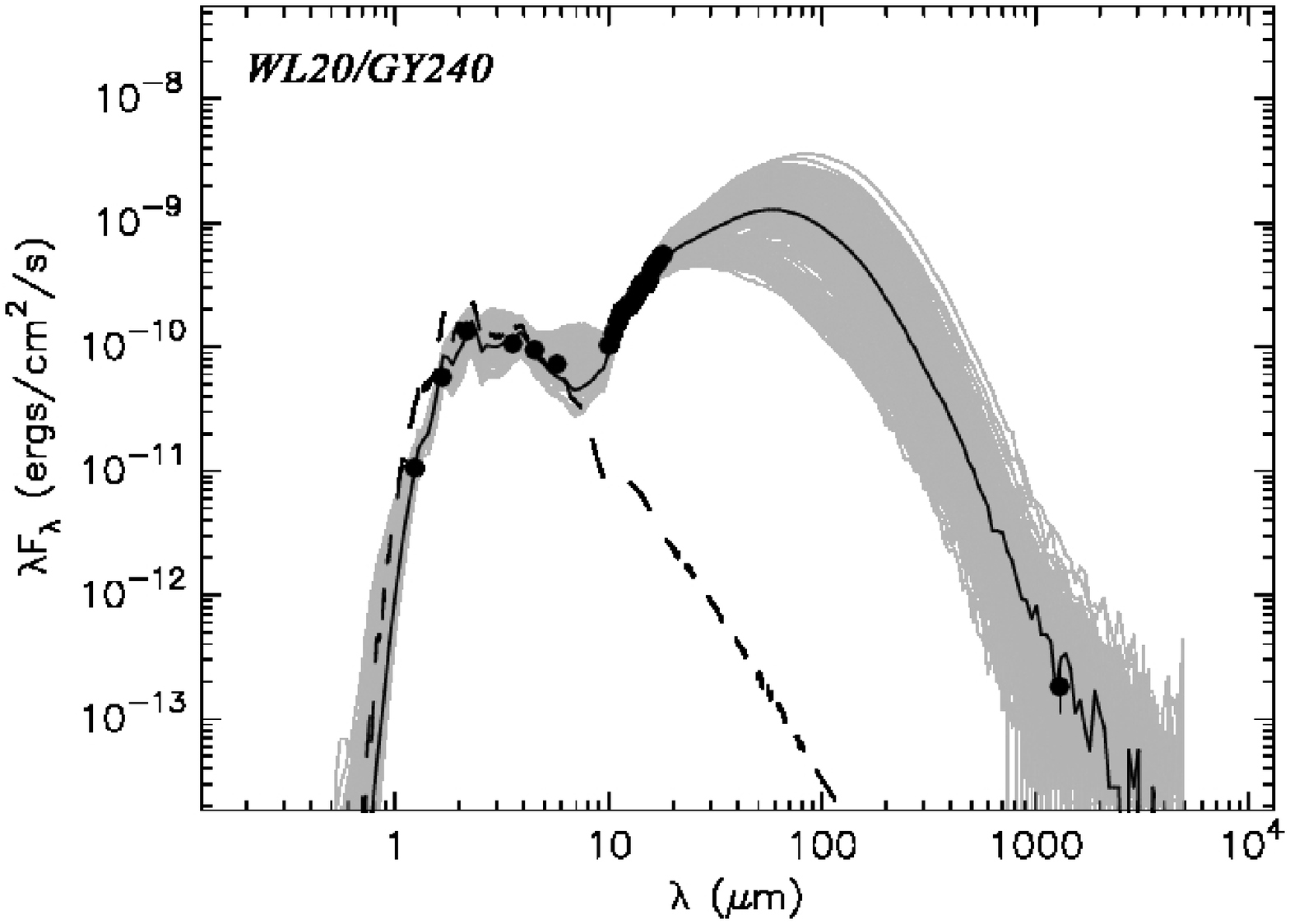, width=5.80cm} 
\epsfig{file=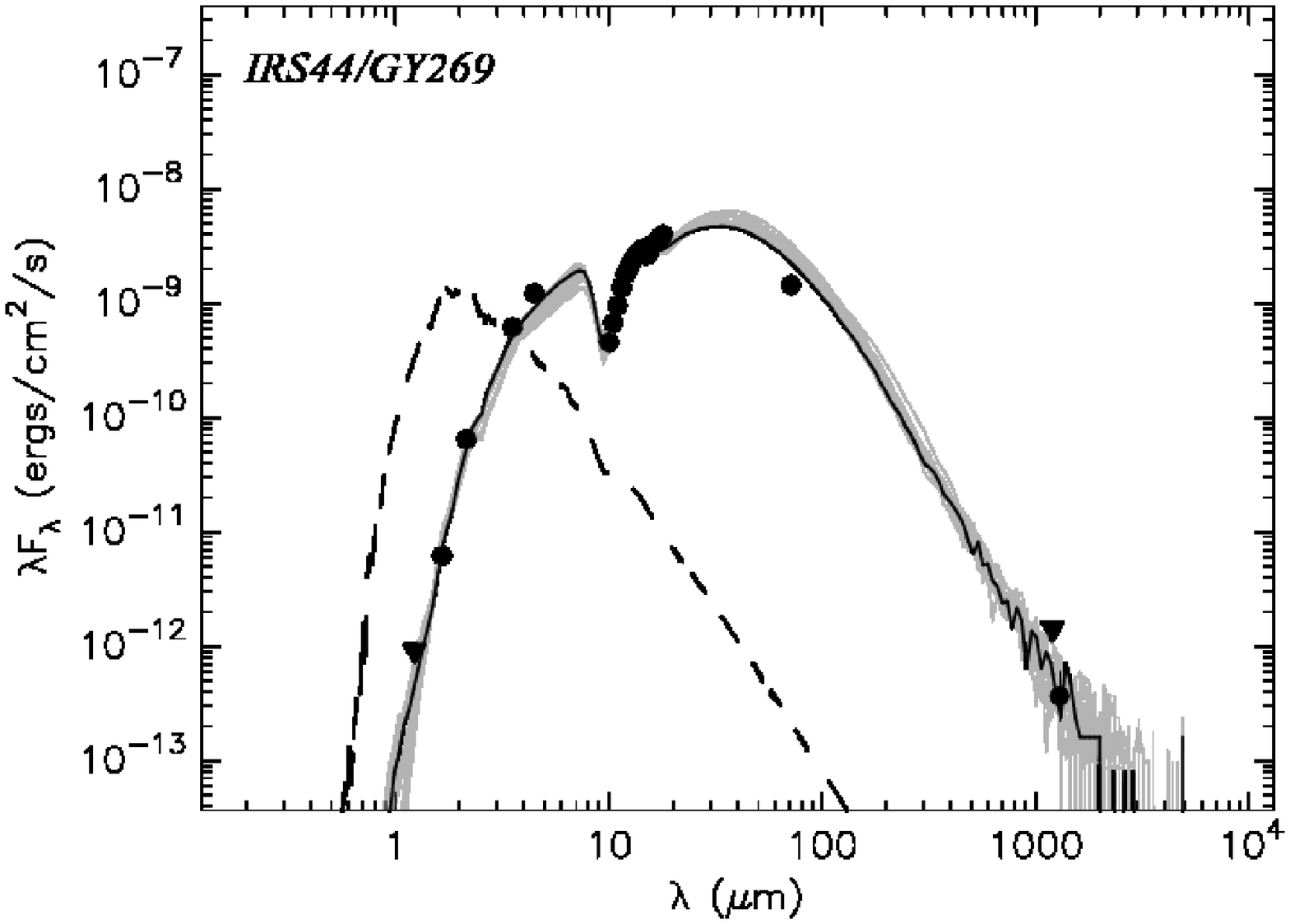, width=5.80cm}} 
\centerline{
\epsfig{file=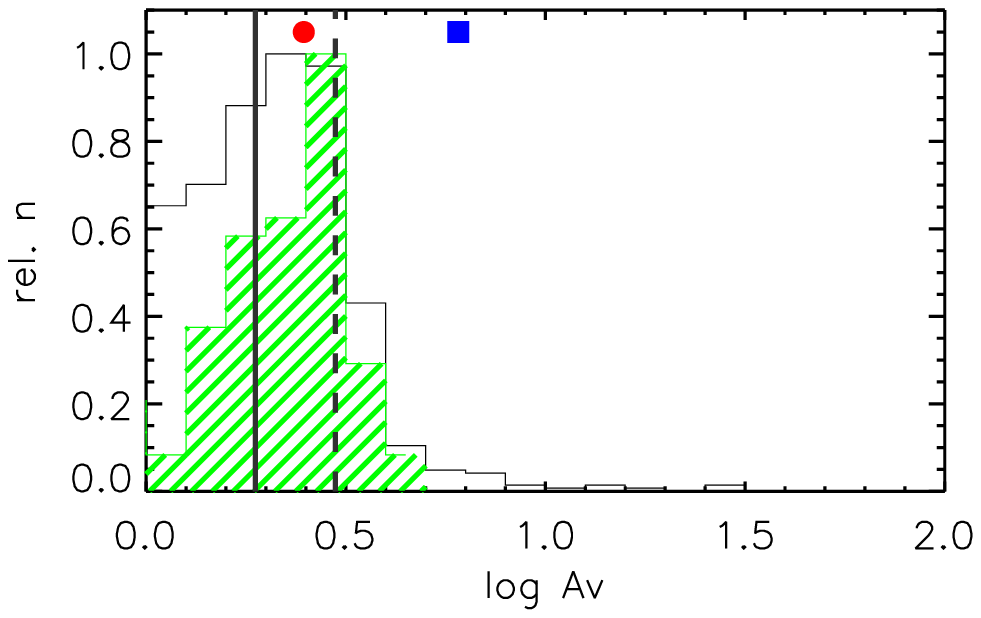 , width=5.95cm} 
\epsfig{file=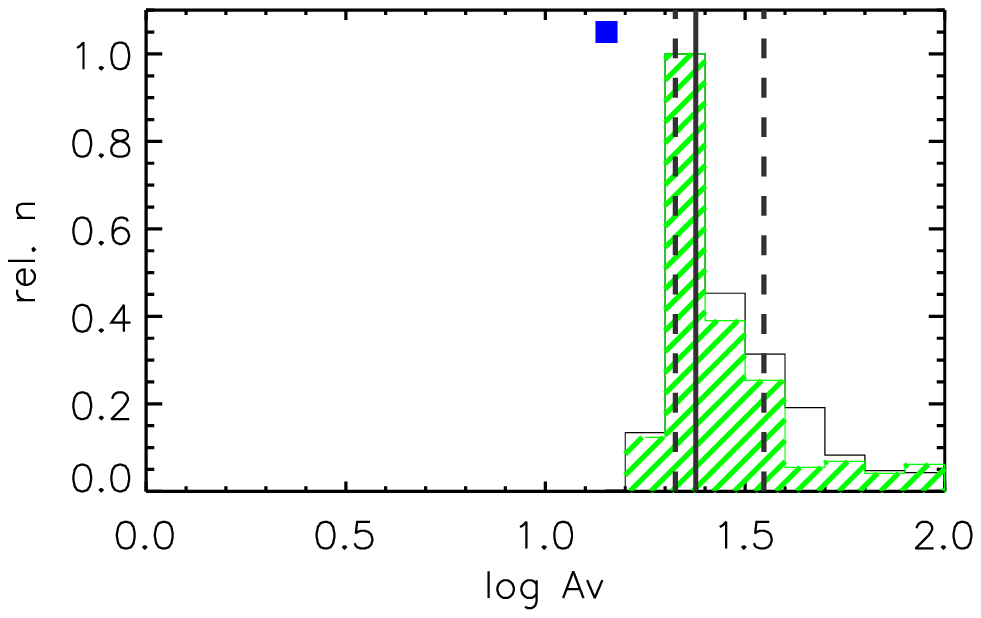, width=5.95cm} 
\epsfig{file=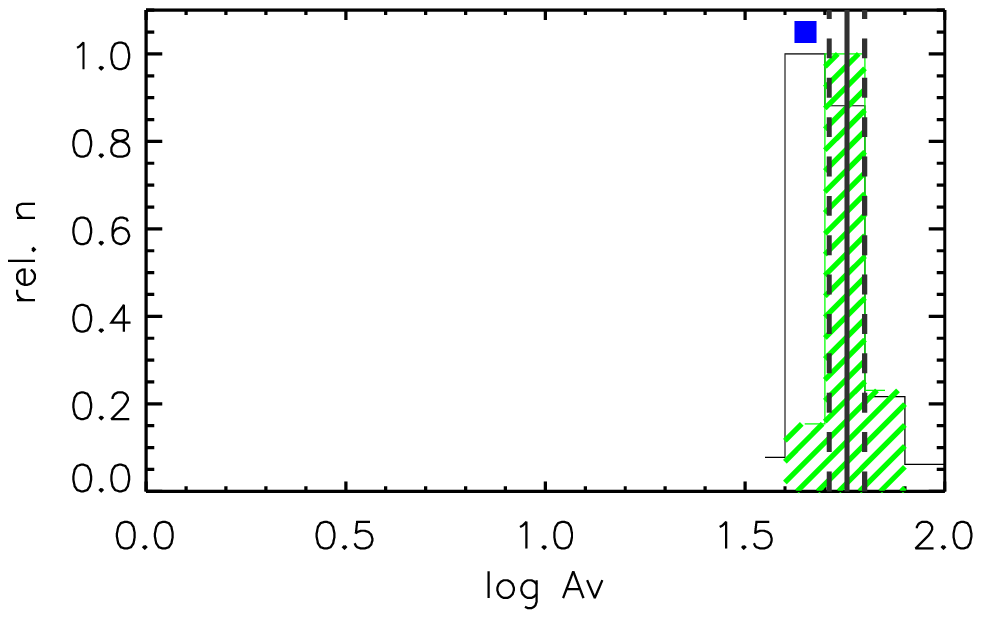, width=5.95cm}} 
\centerline{
\epsfig{file=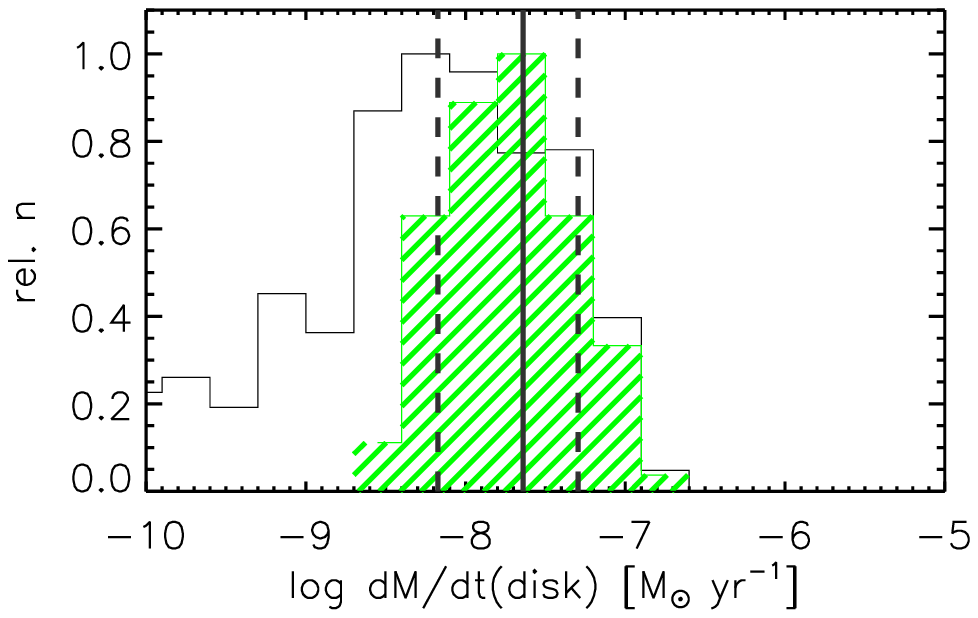 , width=5.95cm} 
\epsfig{file=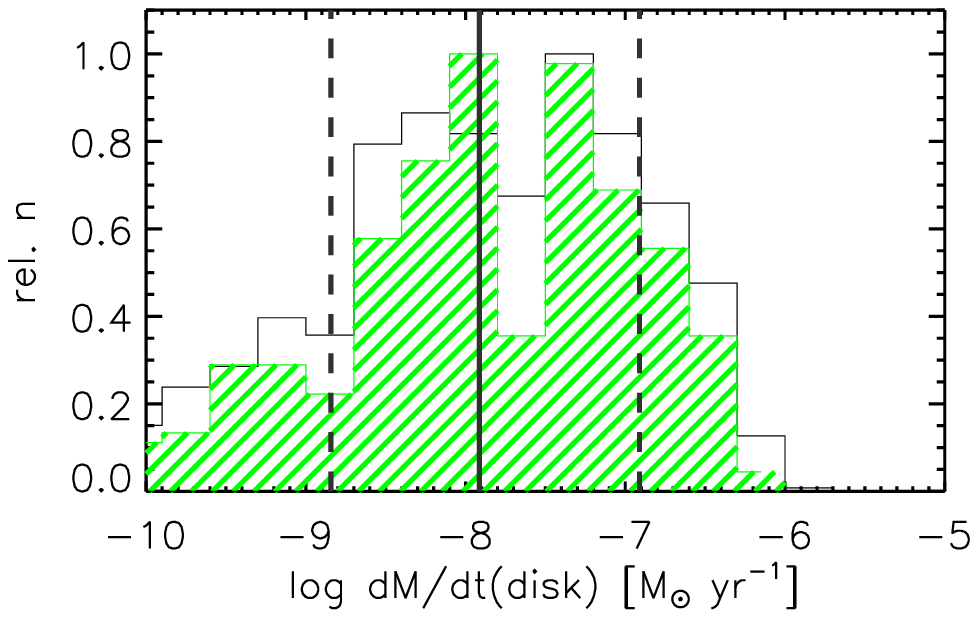, width=5.95cm} 
\epsfig{file=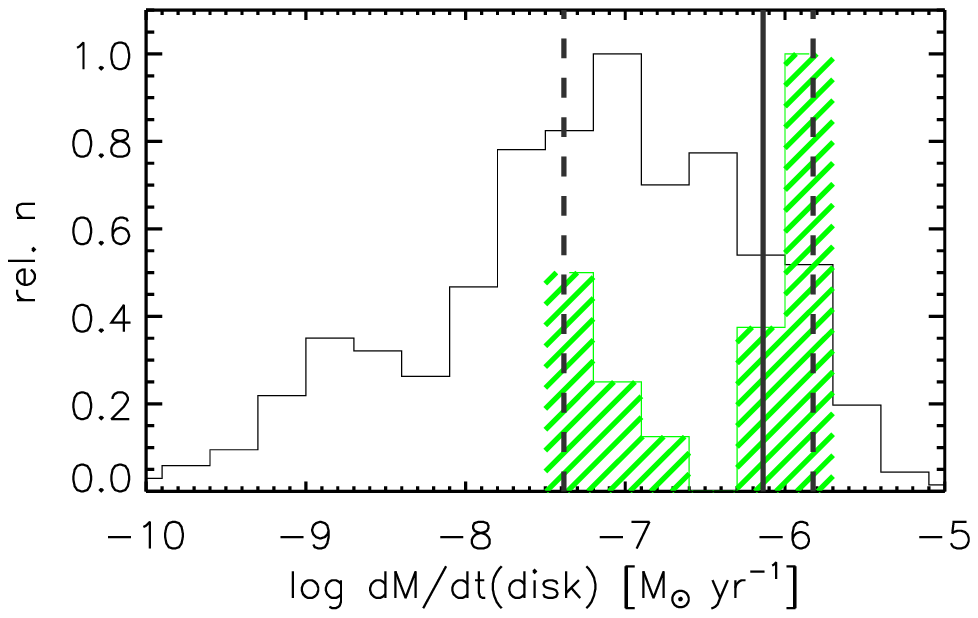, width=5.95cm}} 
\caption{Examples of SED fits for three objects in our sample with
[Ne\,II] detections. From left to right: DoAr25/GY17, WL20/GY240, and
IRS44/GY269. The first is classified as Stage/Class\,II, the other two as
Stage/Class\,I. The upper row shows the SEDs and the best fit models as
produced by the Web interface provided by \citet{rob06}. For the
datapoints, detections and upper limits are indicated by circles and
triangles, respectively. The lower two rows represent distributions of
two fit parameters, $A_{\rm V}$  and $\dot{M}_{\rm disk}$. The empty histograms
refer to the 1000 model fits with lowest $\chi^2$ and the green
histograms to the {\em statistically reasonable} samples of models defined in
\S\,\ref{ap:sed_method}. The solid and dashed vertical lines indicate
the median and the 1$\sigma$ dispersion for these latter samples. For the
panels in the second row the symbols close to the upper axis indicate
the $A_{\rm V}$ values inferred from the $A_{\rm J}$ in
Table\,\ref{tab:target_litdata} (circles) and from the X-ray derived
$N_{\rm H}$
in Table \ref{tab:droxo_irs_results} (squares). }
\label{fig:sed_examples}
\end{figure*}

As suggested by \citet{rob07}, in order to account for systematic
uncertainties, underestimation of the measurement errors, and
intrinsic object variability in time, a lower limit of 25\%, 10\%, and
40\% was imposed on the uncertainties of optical, NIR/MIR, and 
millimeter fluxes, respectively. 

\begin{figure*}
\centerline{
\epsfig{file=sed_disk_0.ps,  width=4.60cm} 
\epsfig{file=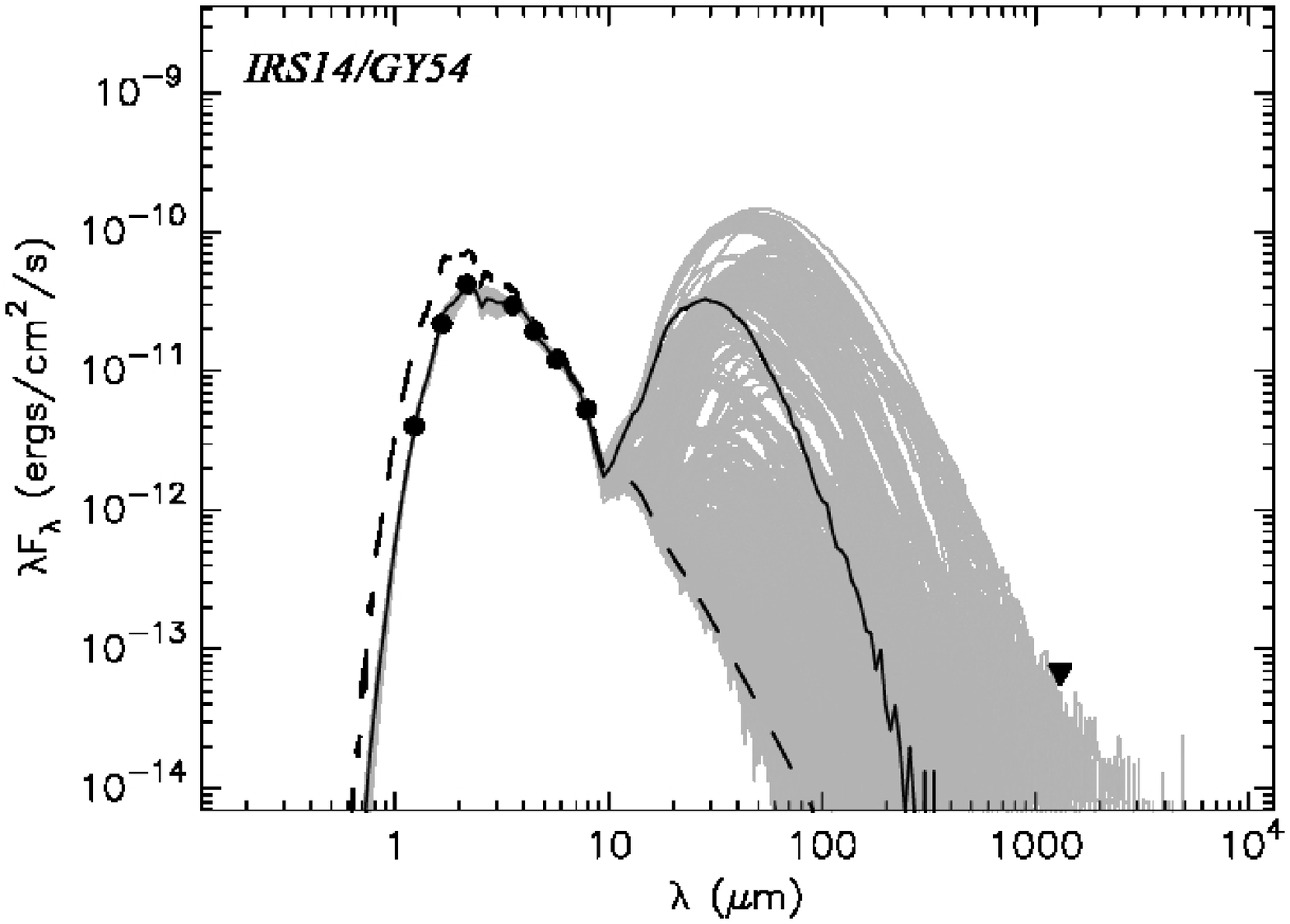,  width=4.60cm} 
\epsfig{file=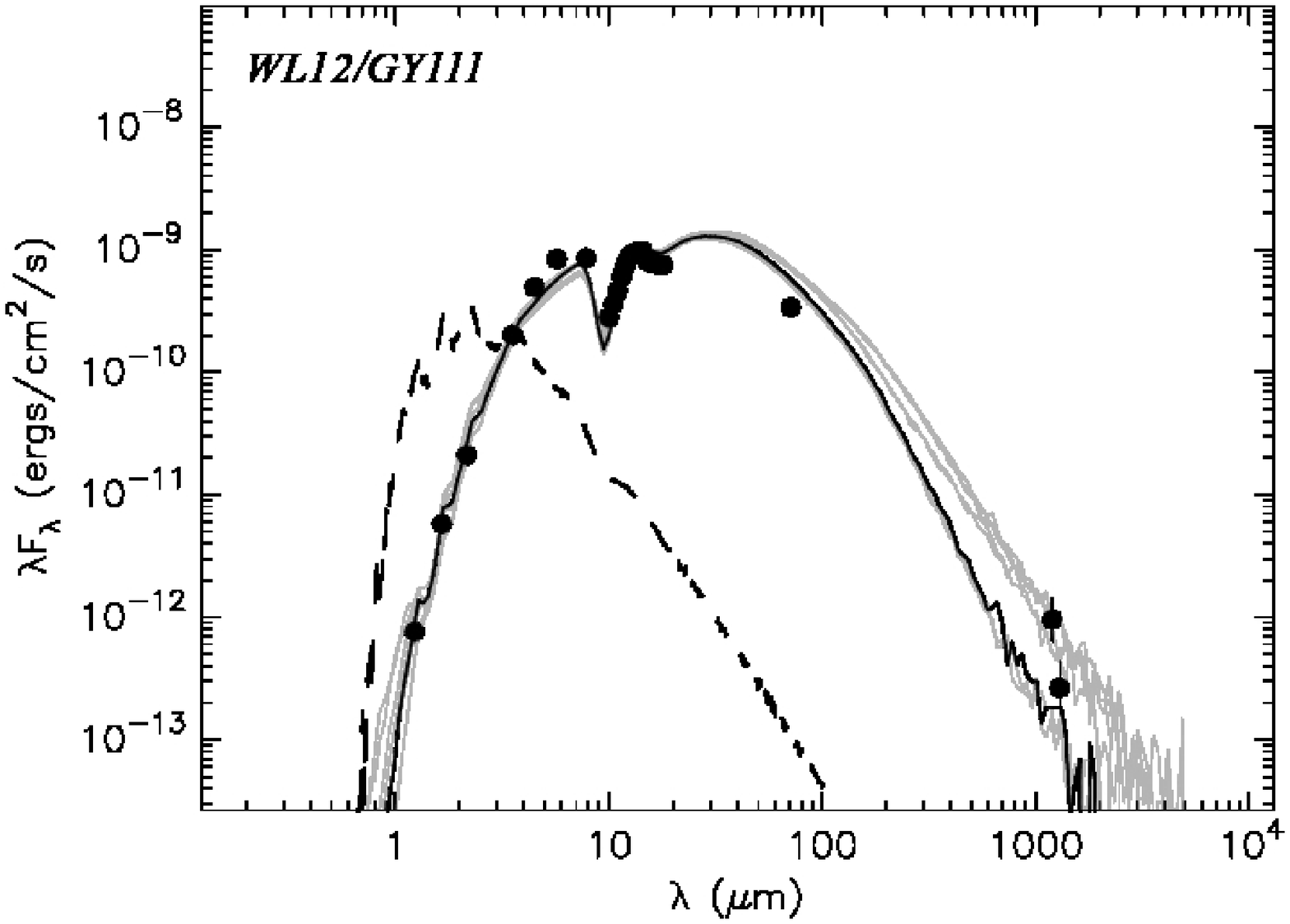,  width=4.60cm}
\epsfig{file=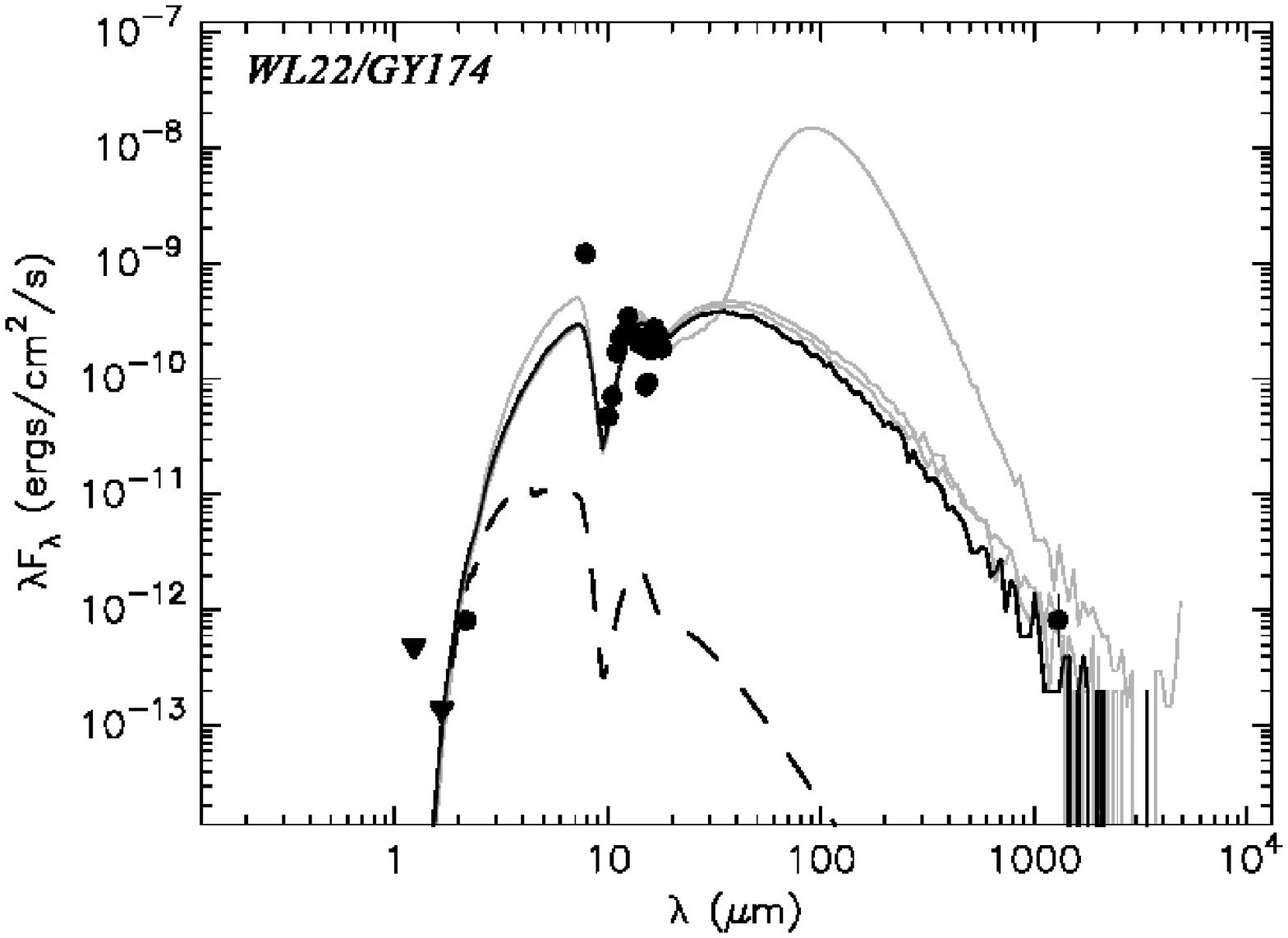,  width=4.60cm} 
}\centerline{
\epsfig{file=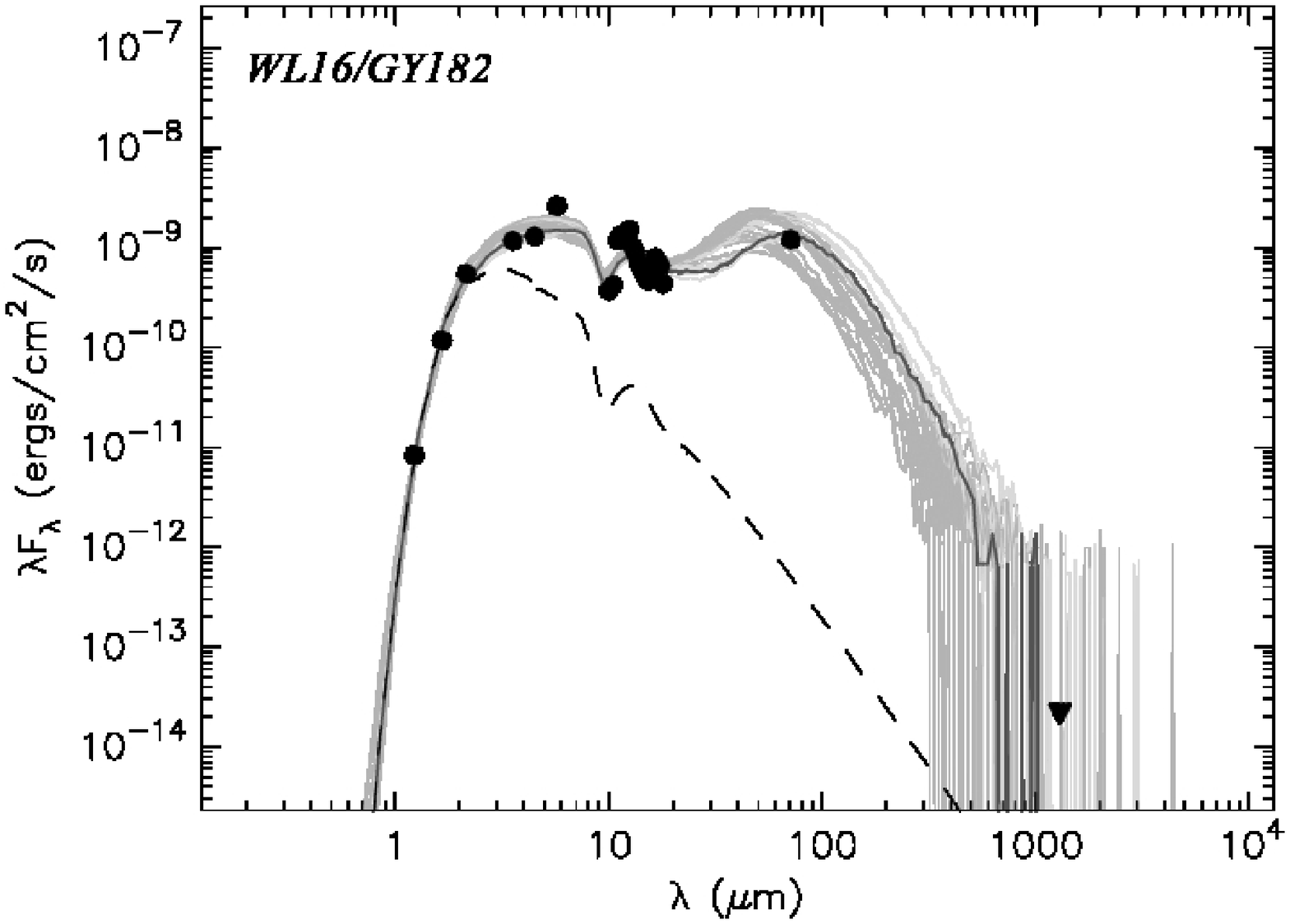,  width=4.60cm} 
\epsfig{file=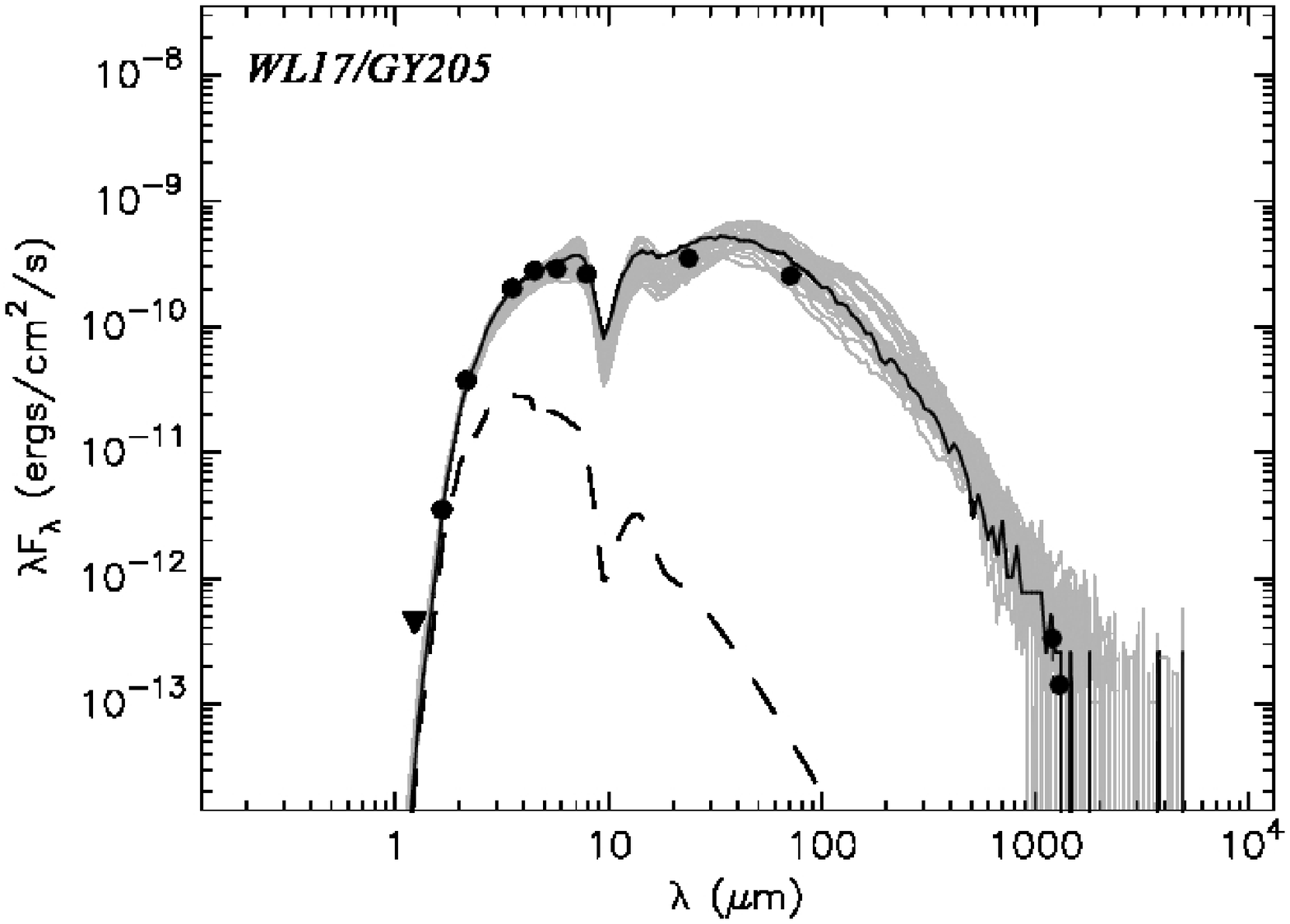,  width=4.60cm} 
\epsfig{file=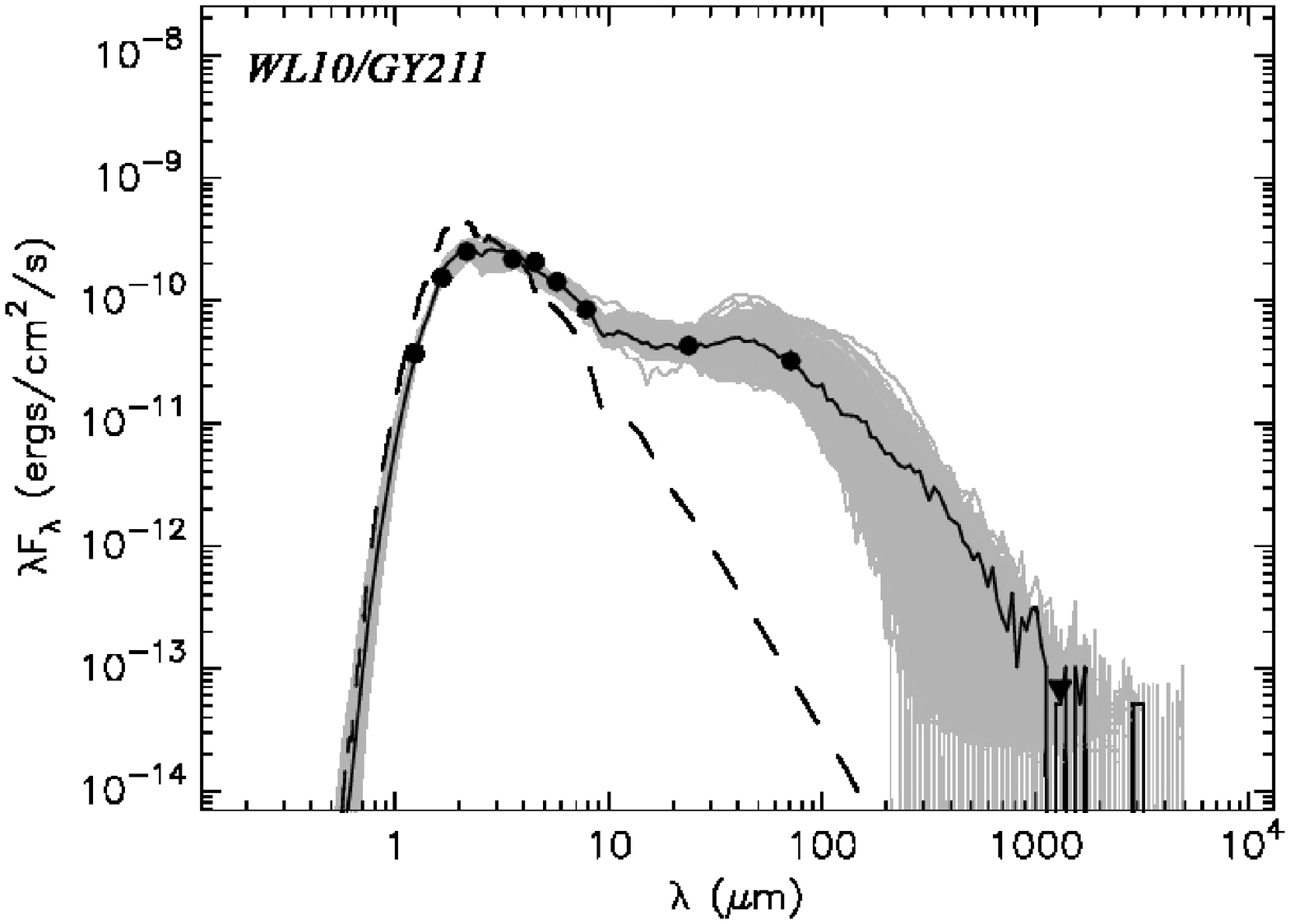,  width=4.60cm} 
\epsfig{file=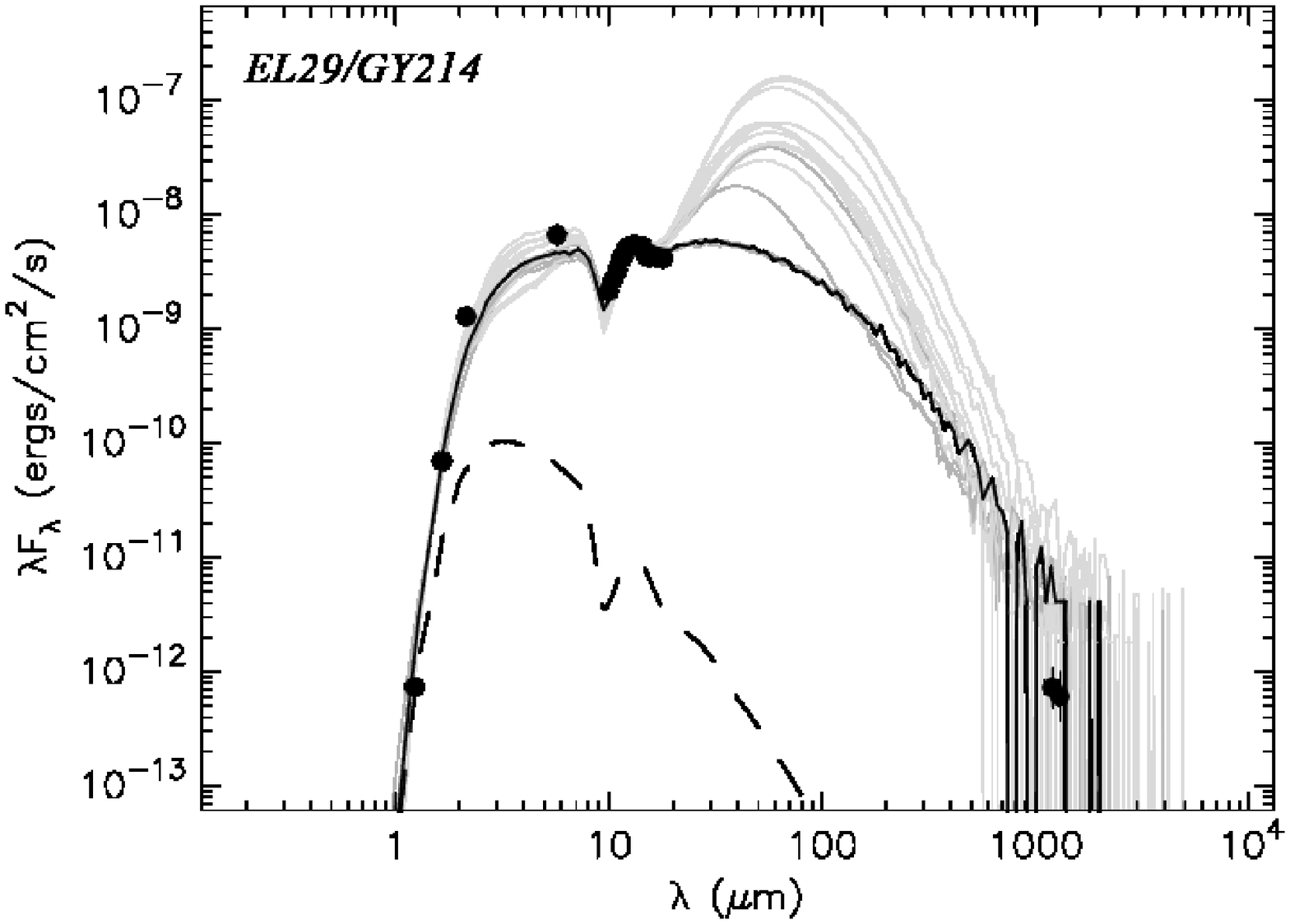,  width=4.60cm} 
}\centerline{
\epsfig{file=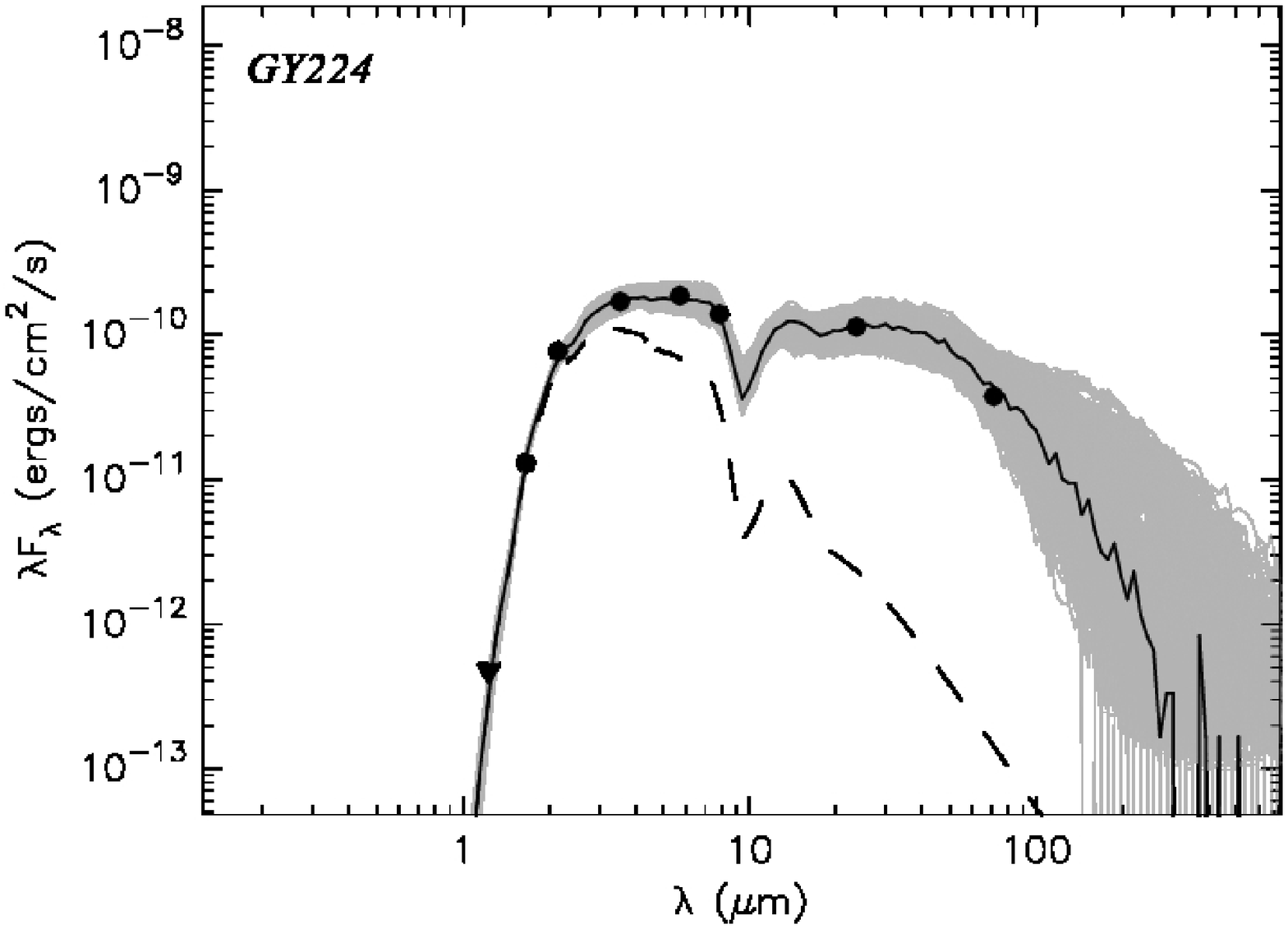,  width=4.60cm} 
\epsfig{file=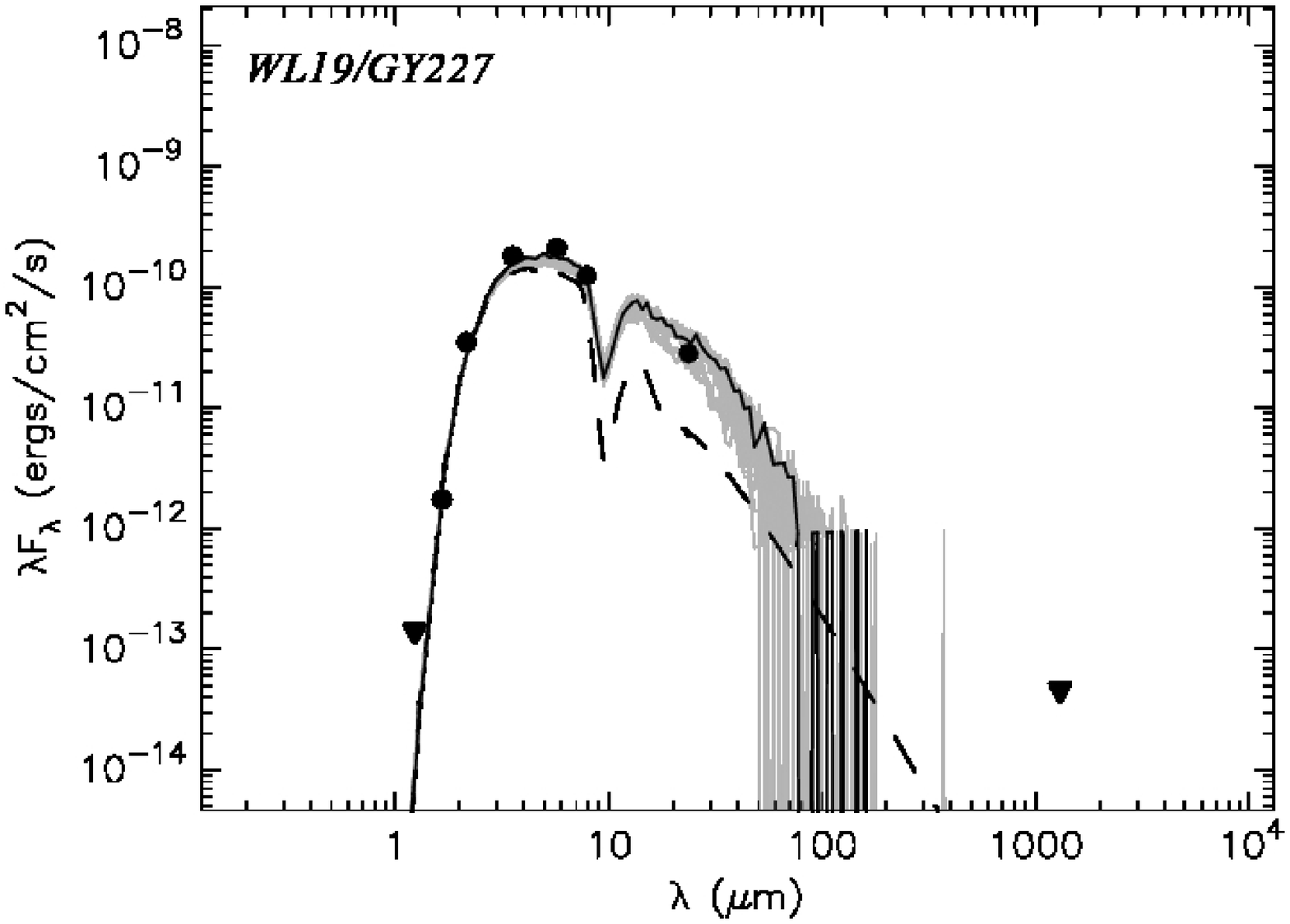 , width=4.60cm} 
\epsfig{file=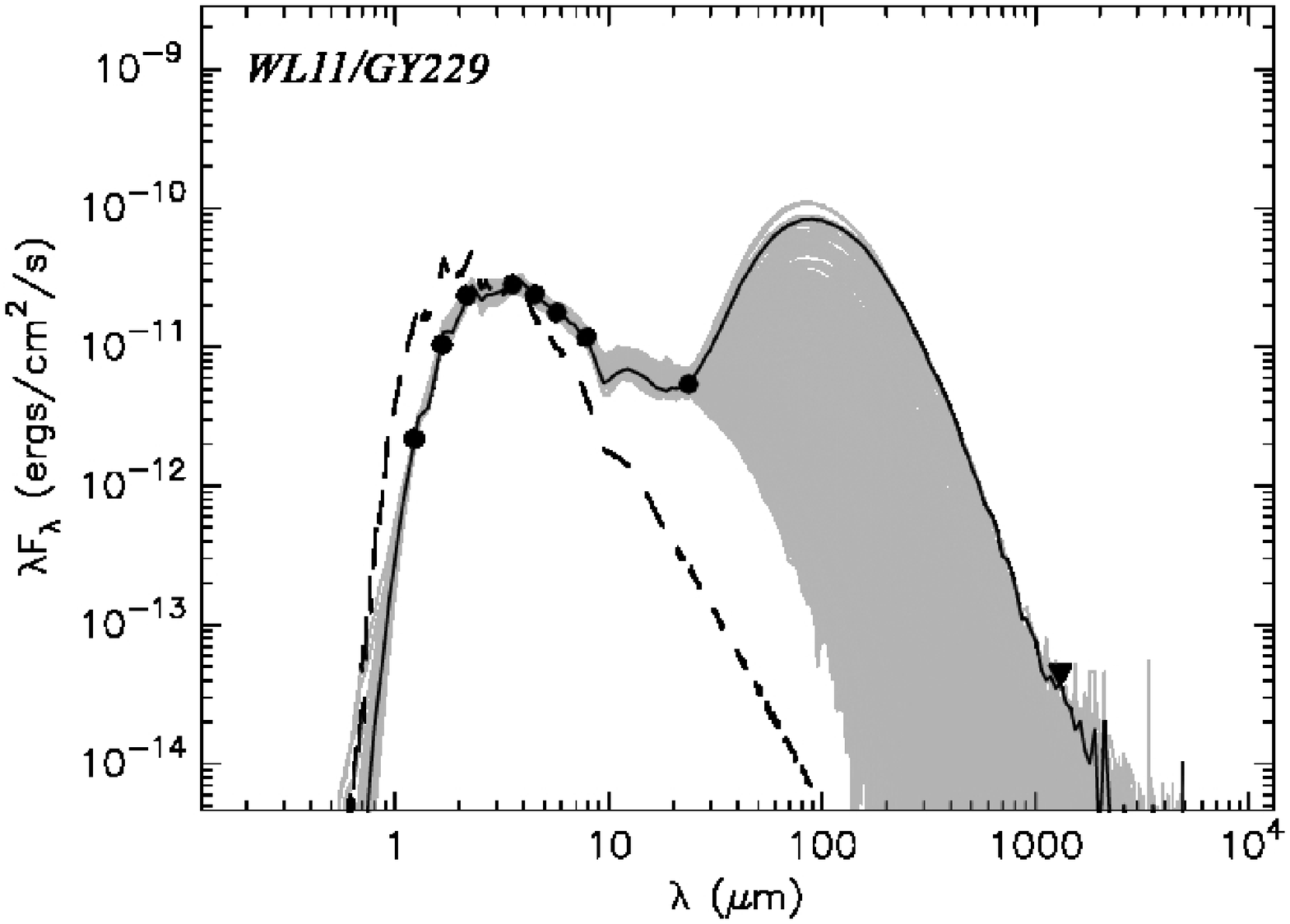, width=4.60cm} 
\epsfig{file=sed_disk_11.ps, width=4.60cm}
}\centerline{
\epsfig{file=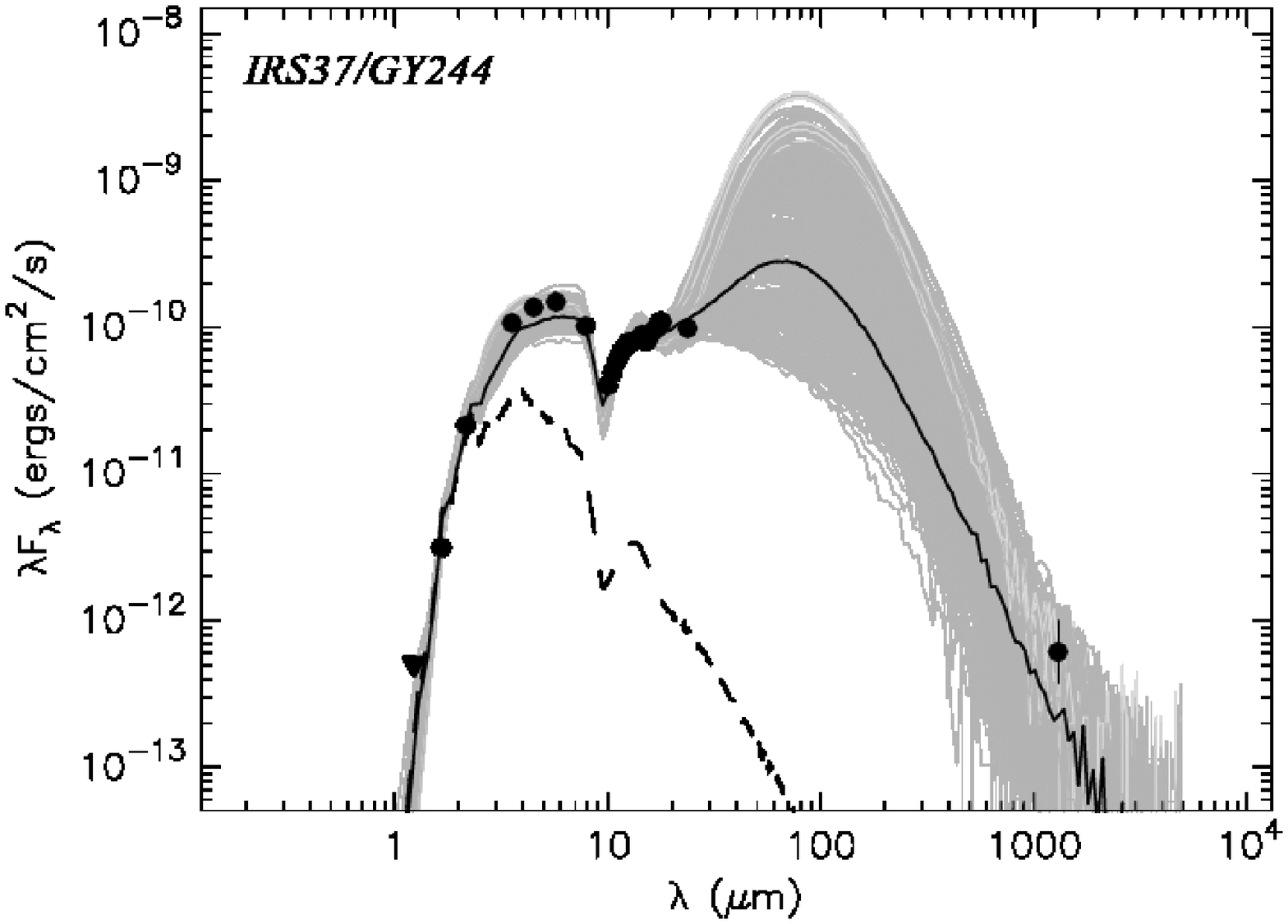, width=4.60cm} 
\epsfig{file=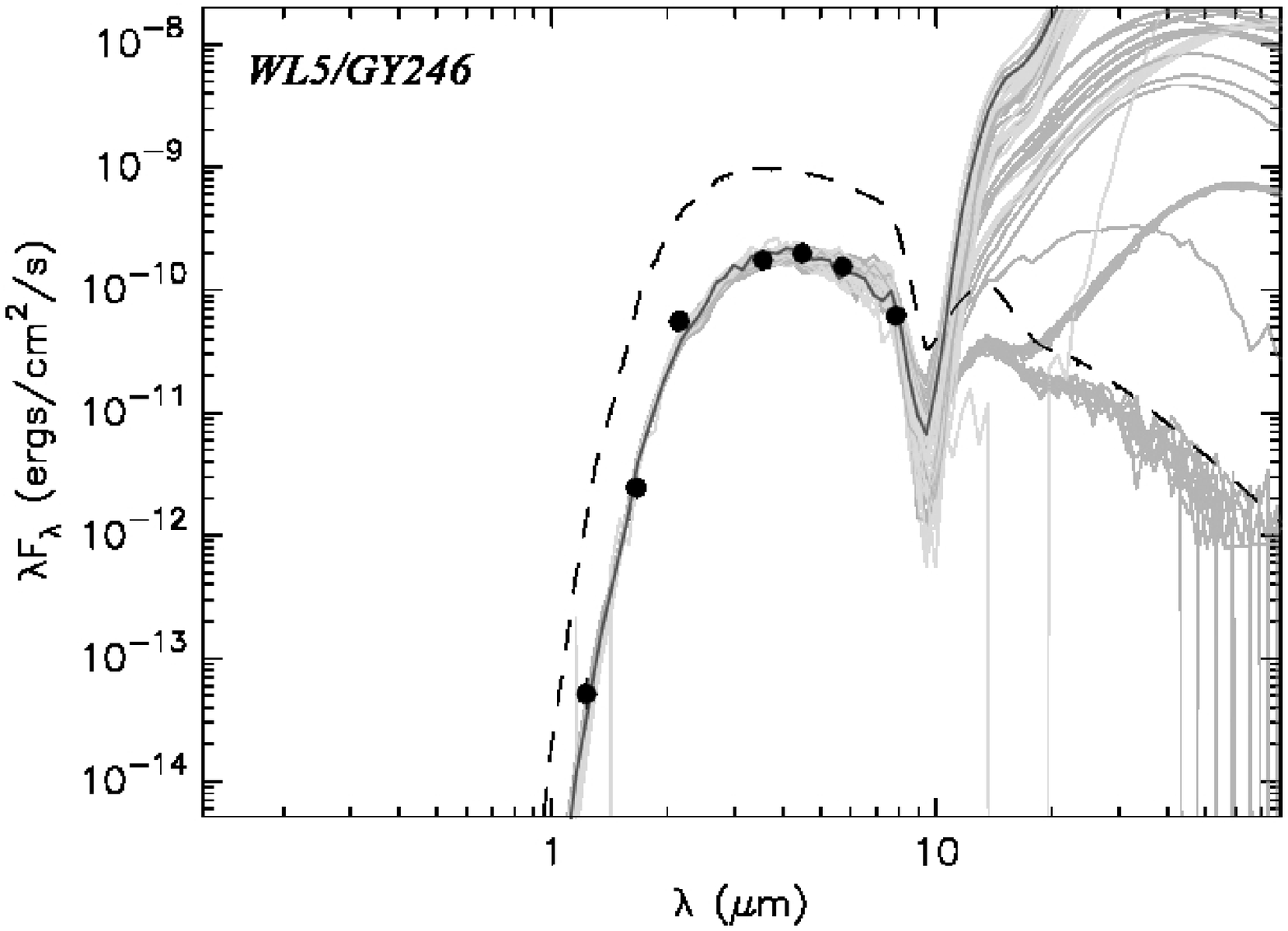, width=4.60cm} 
\epsfig{file=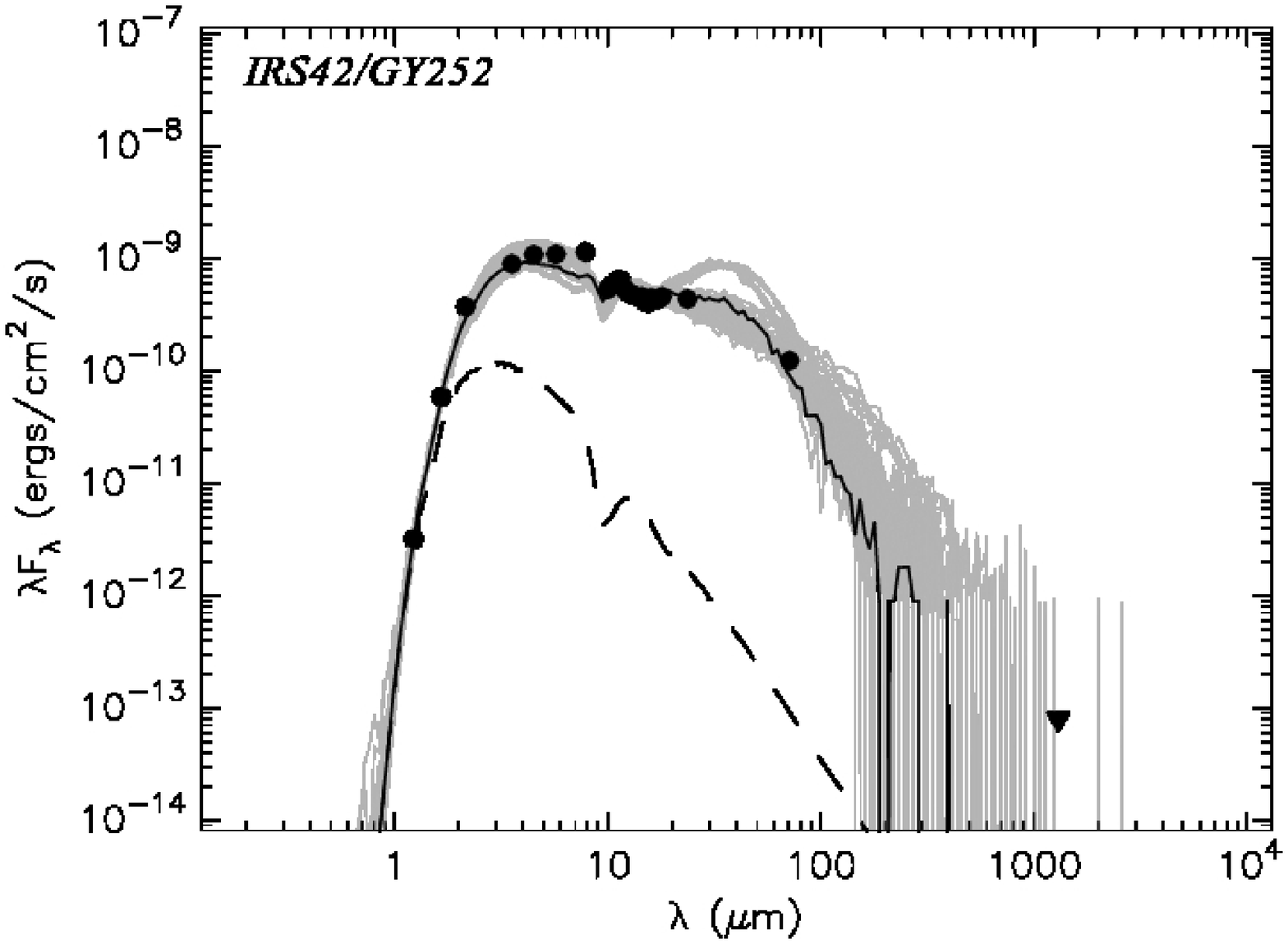, width=4.60cm} 
\epsfig{file=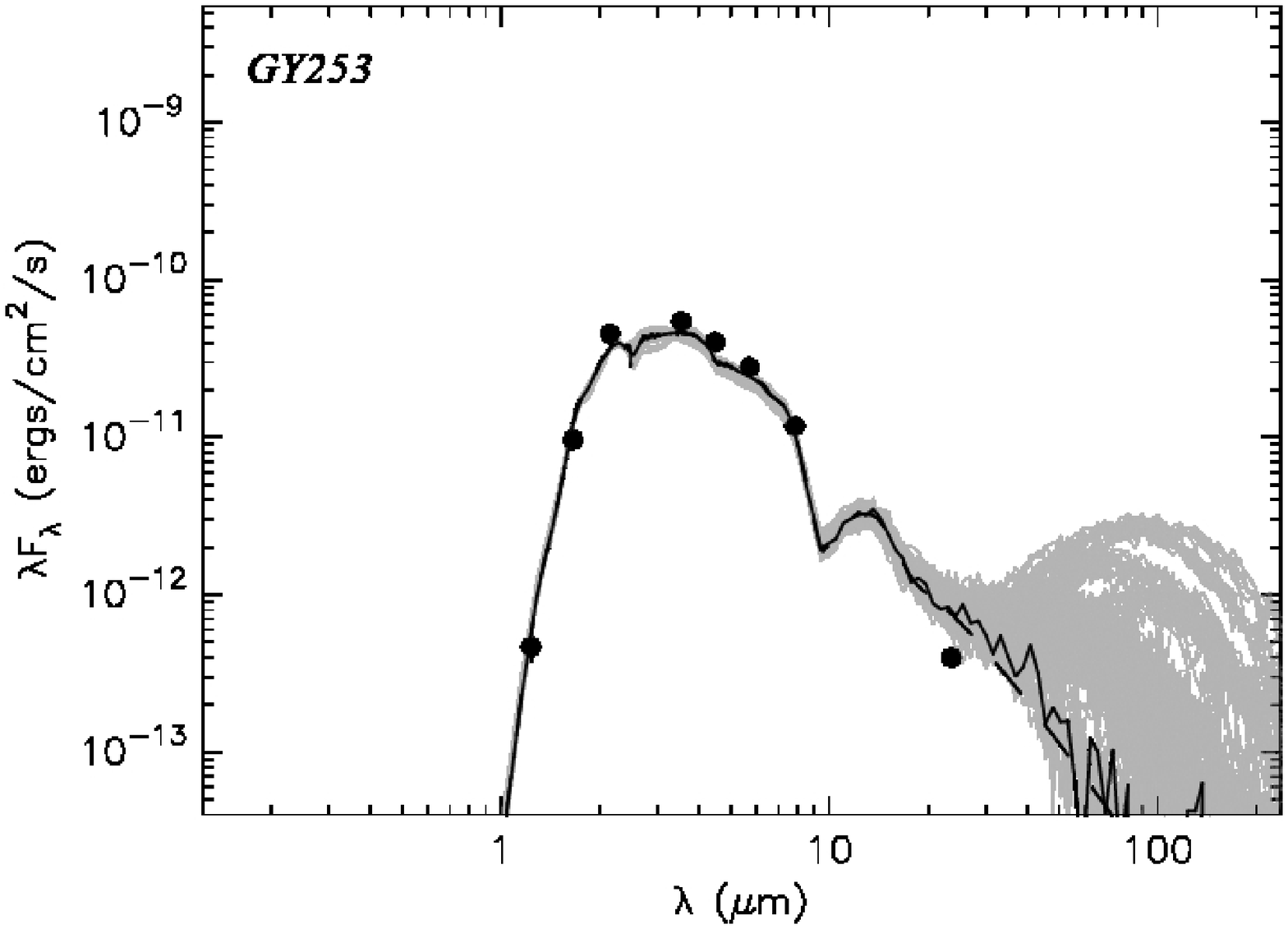, width=4.60cm} 
}\centerline{
\epsfig{file=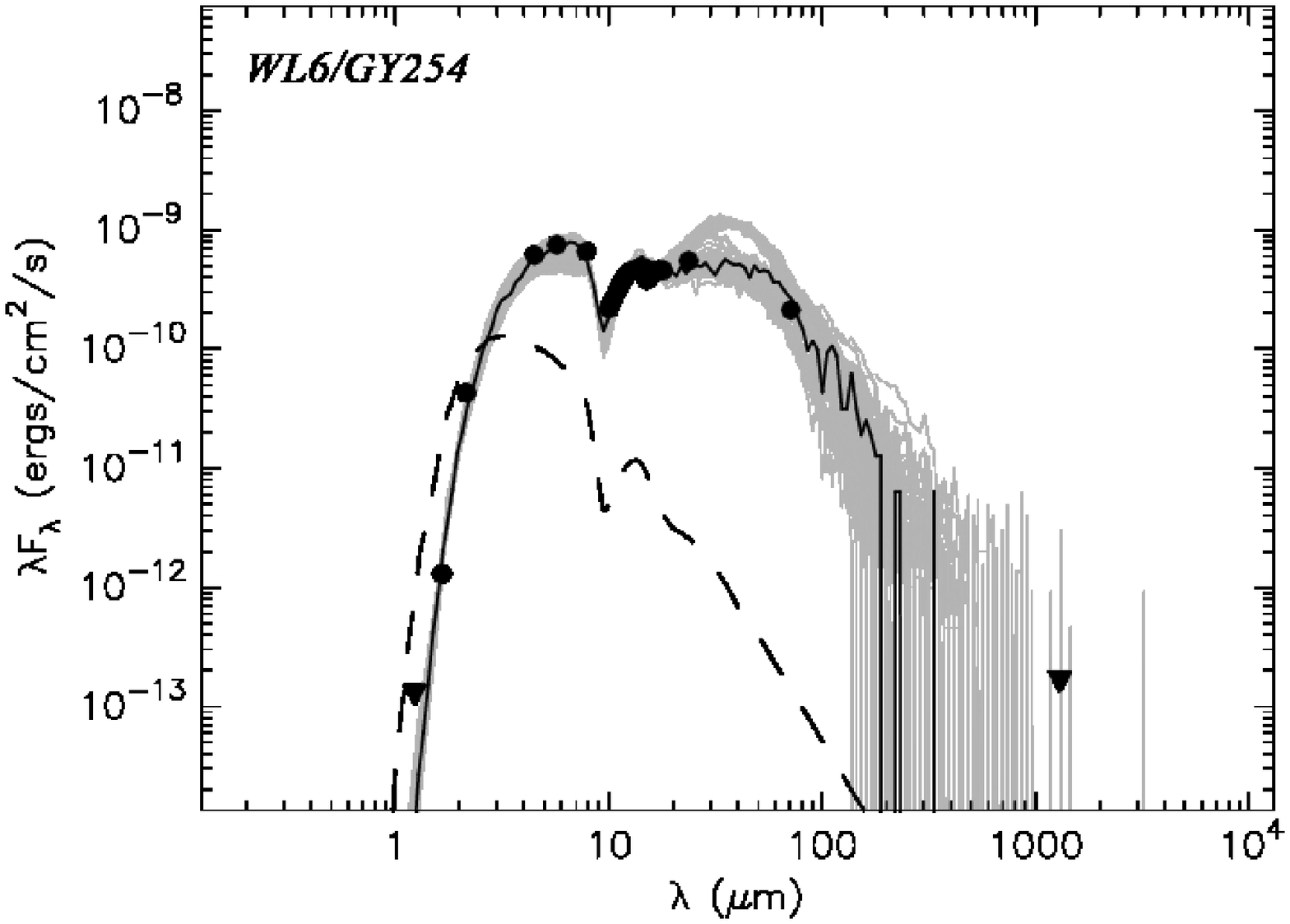, width=4.60cm} 
\epsfig{file=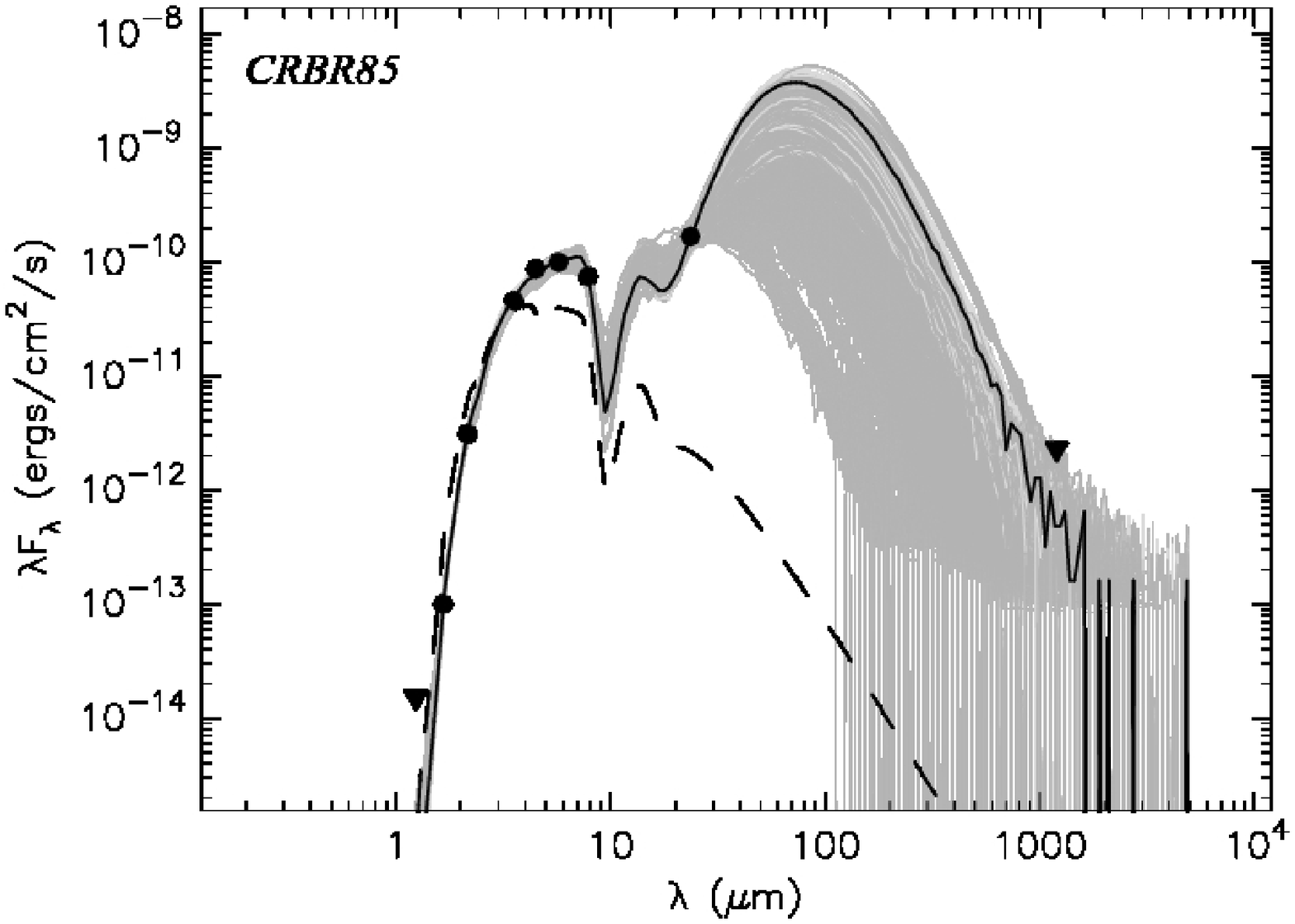, width=4.60cm}
\epsfig{file=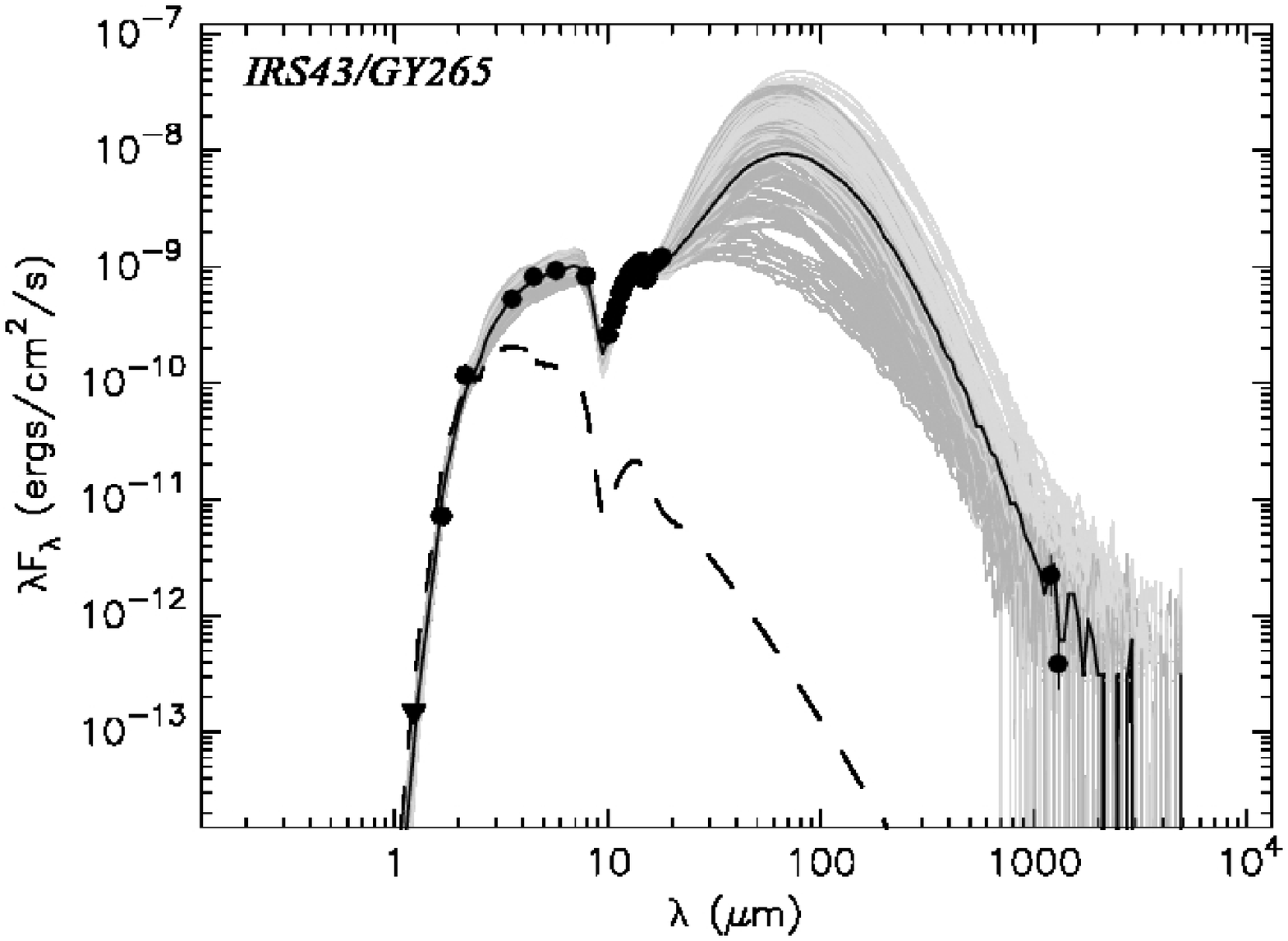, width=4.60cm} 
\epsfig{file=sed_disk_19.ps, width=4.60cm} 
}\centerline{
\epsfig{file=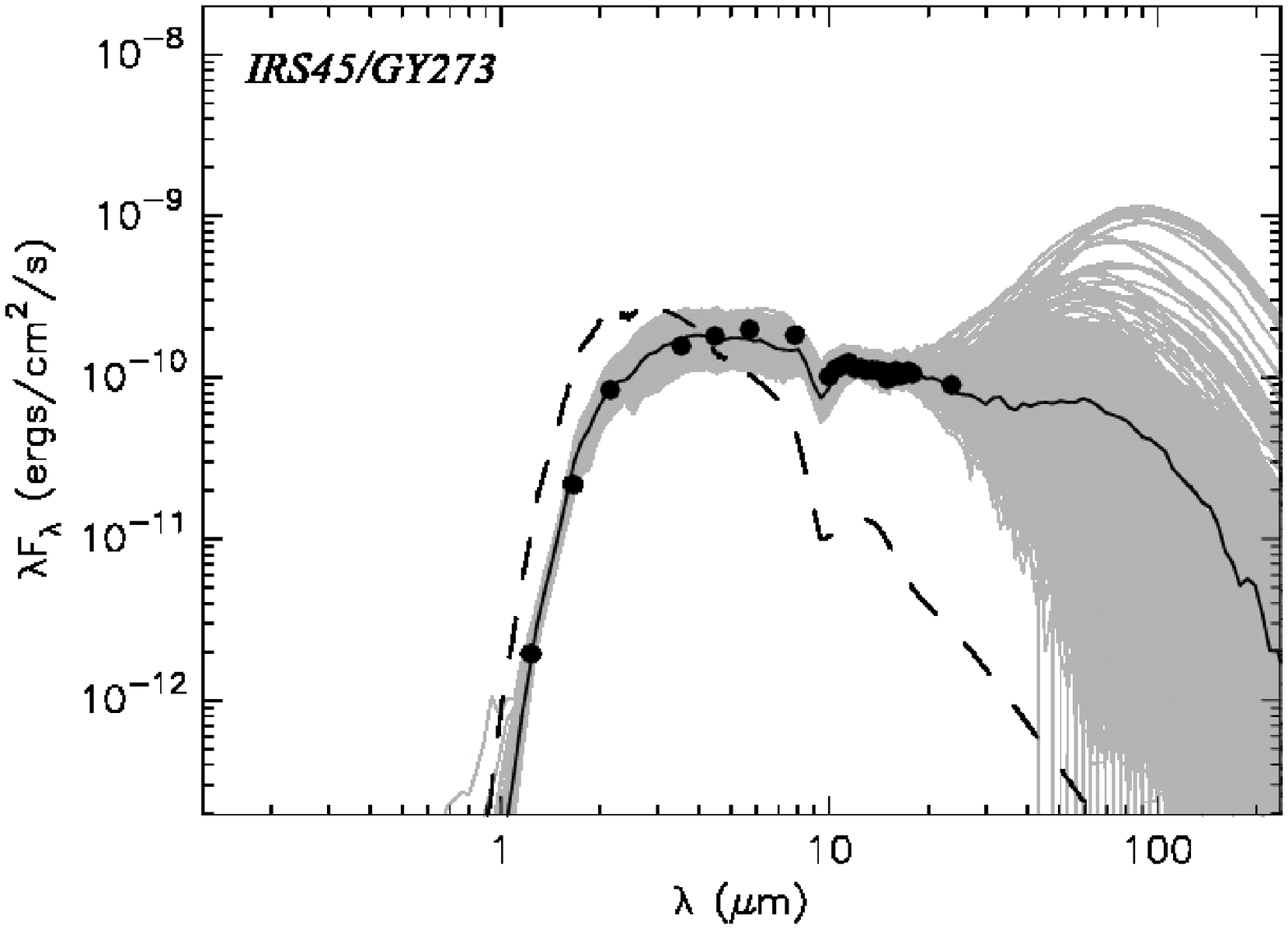, width=4.60cm} 
\epsfig{file=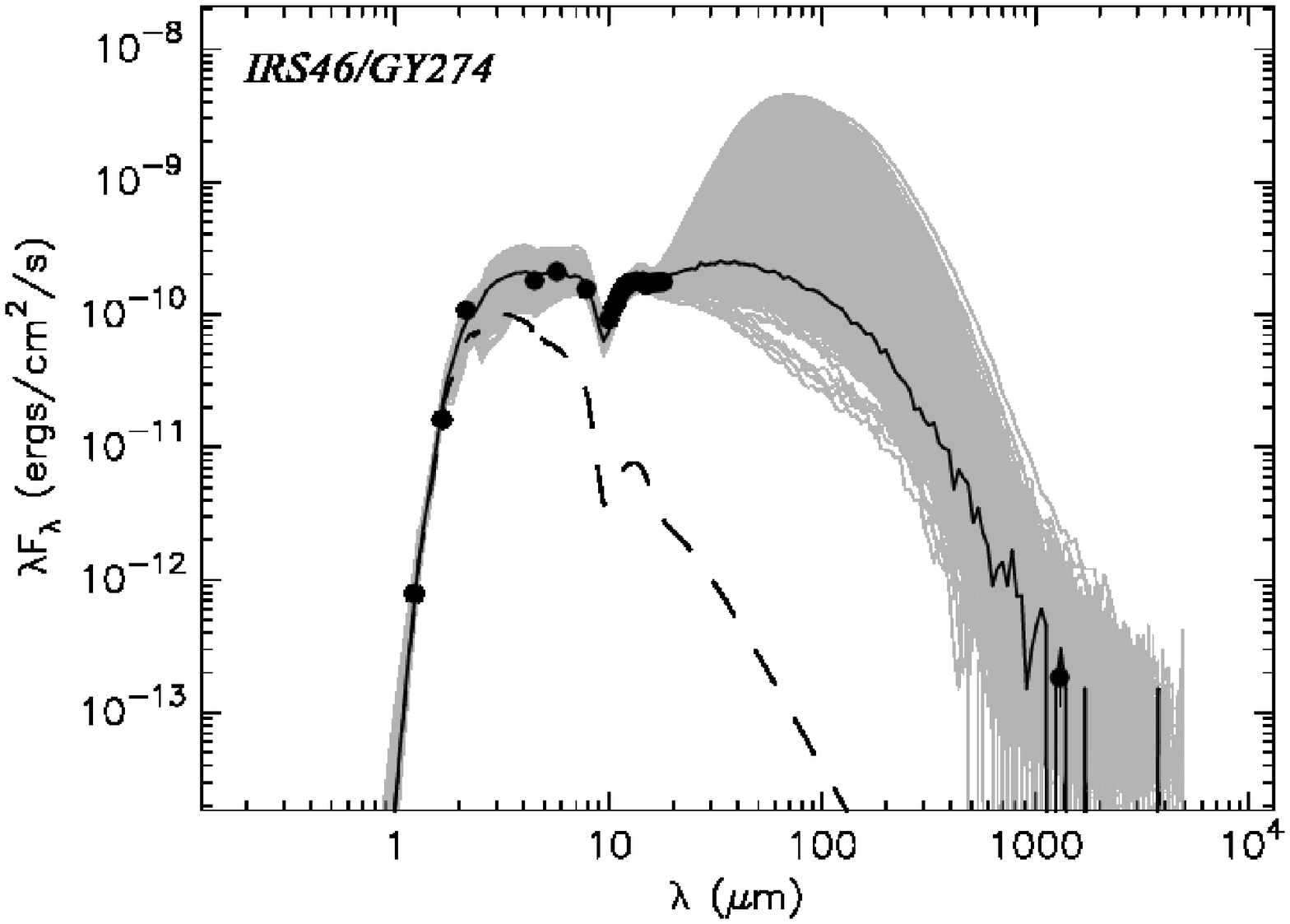, width=4.60cm} 
\epsfig{file=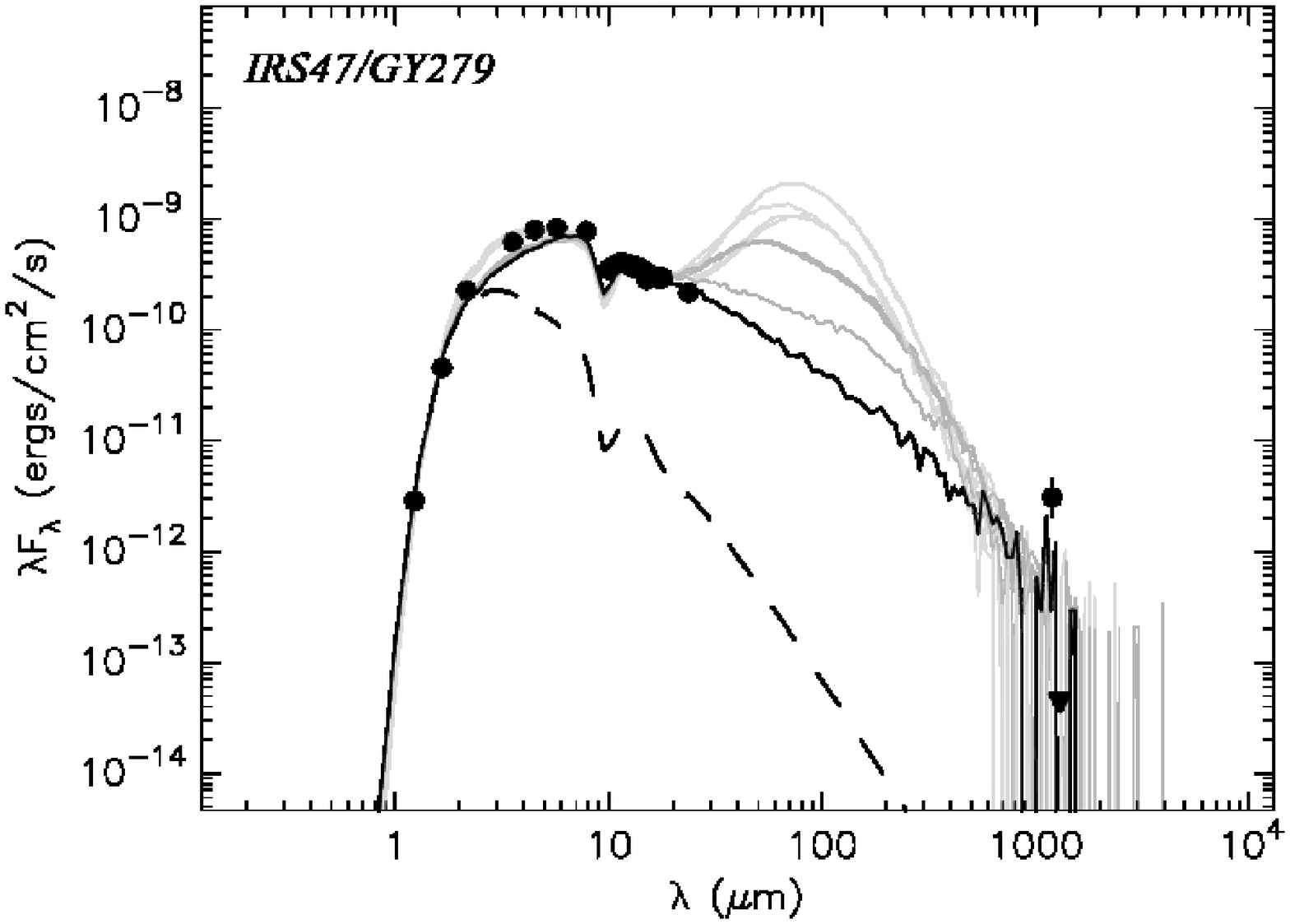, width=4.60cm} 
\epsfig{file=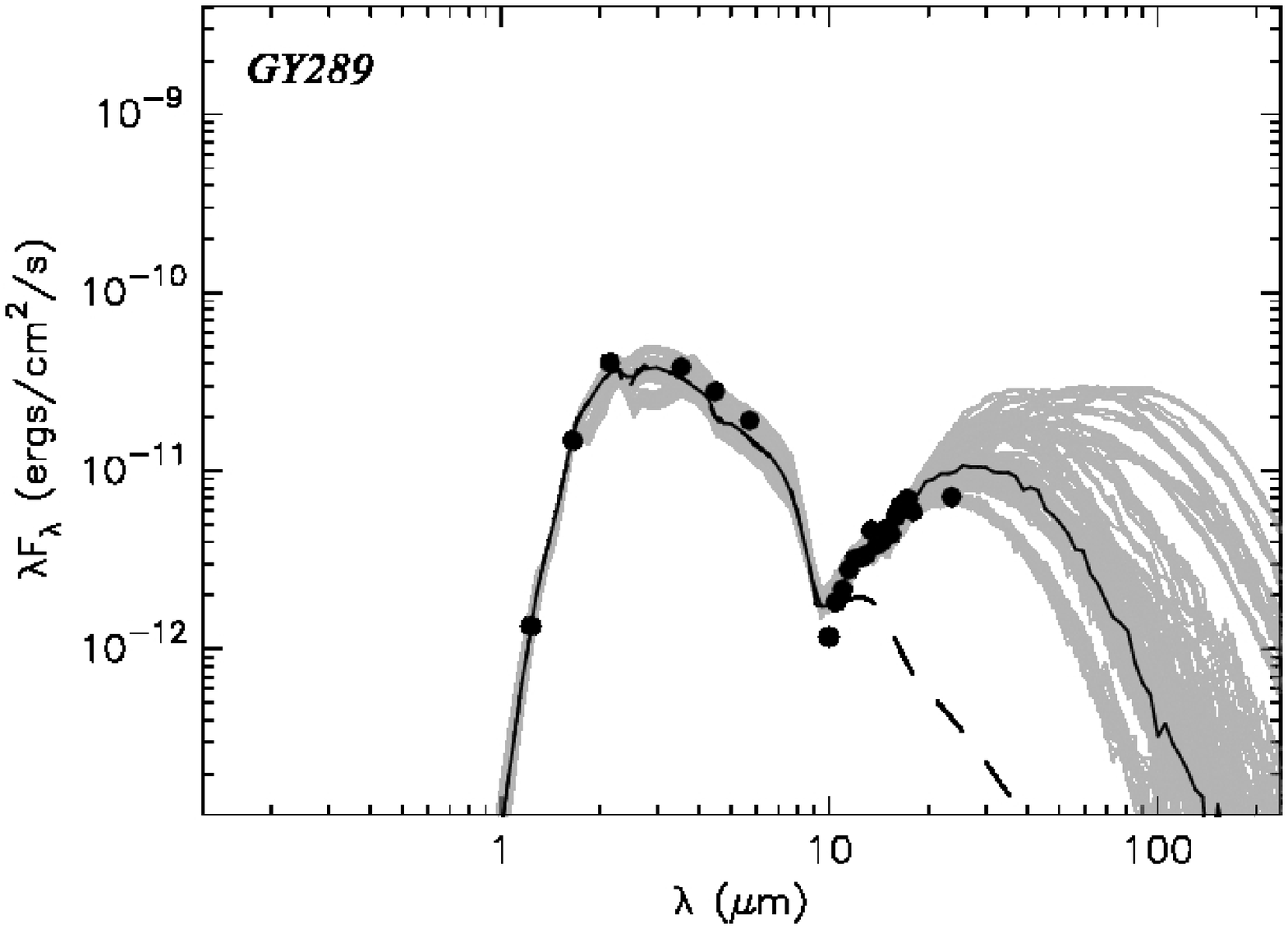, width=4.60cm} 
}\centerline{
\epsfig{file=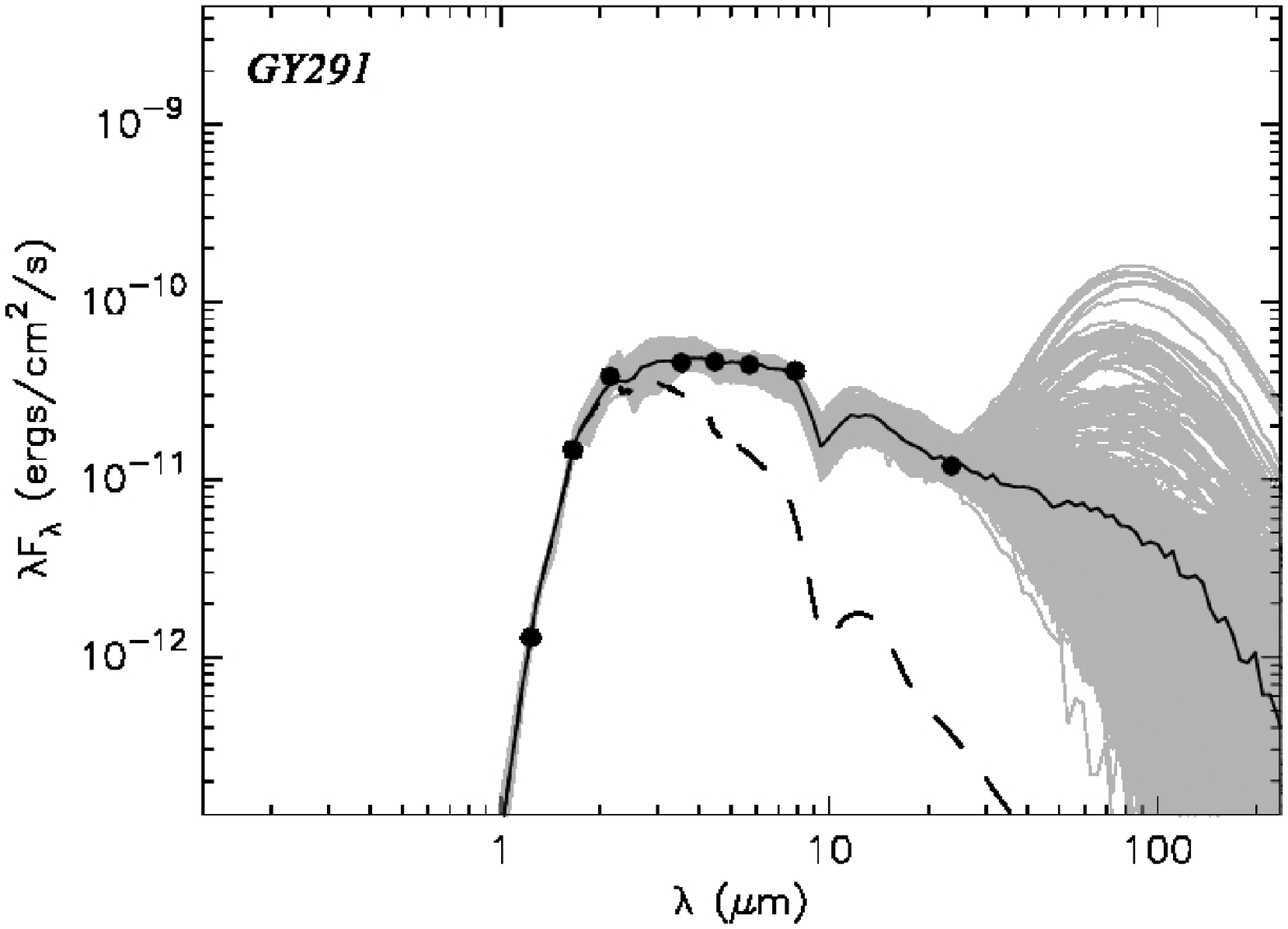, width=4.60cm} 
\epsfig{file=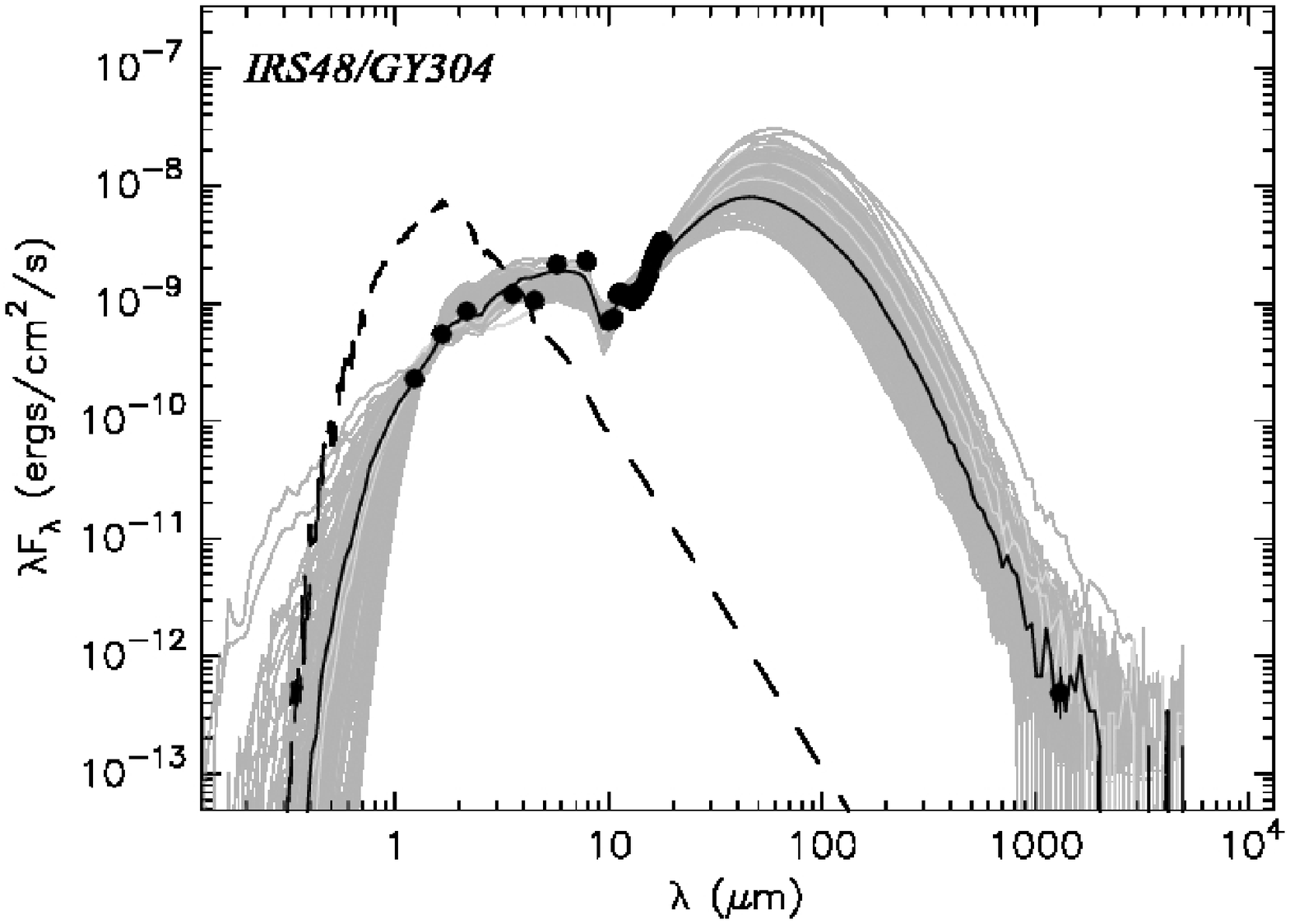, width=4.60cm} 
\epsfig{file=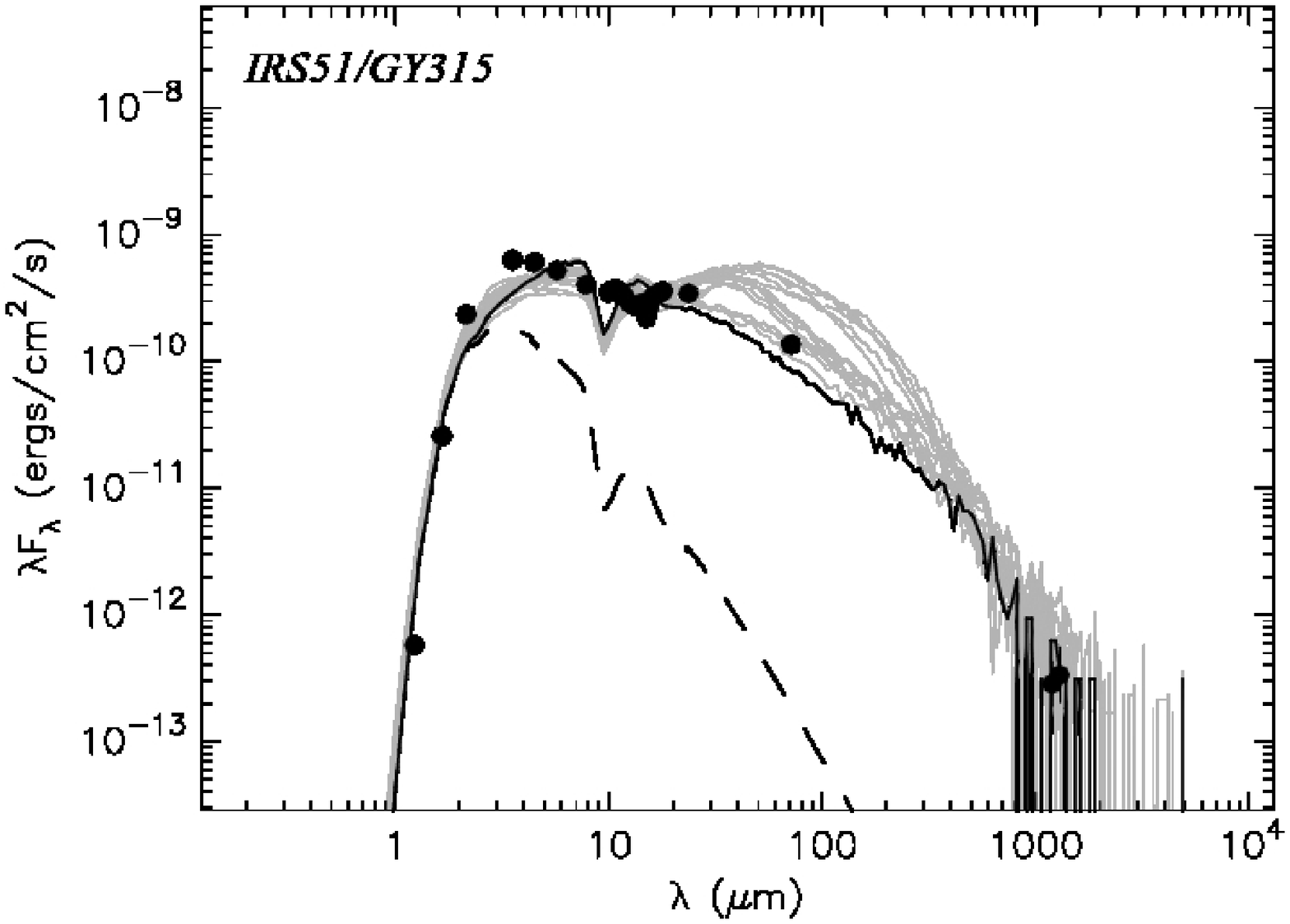, width=4.60cm}
\epsfig{file=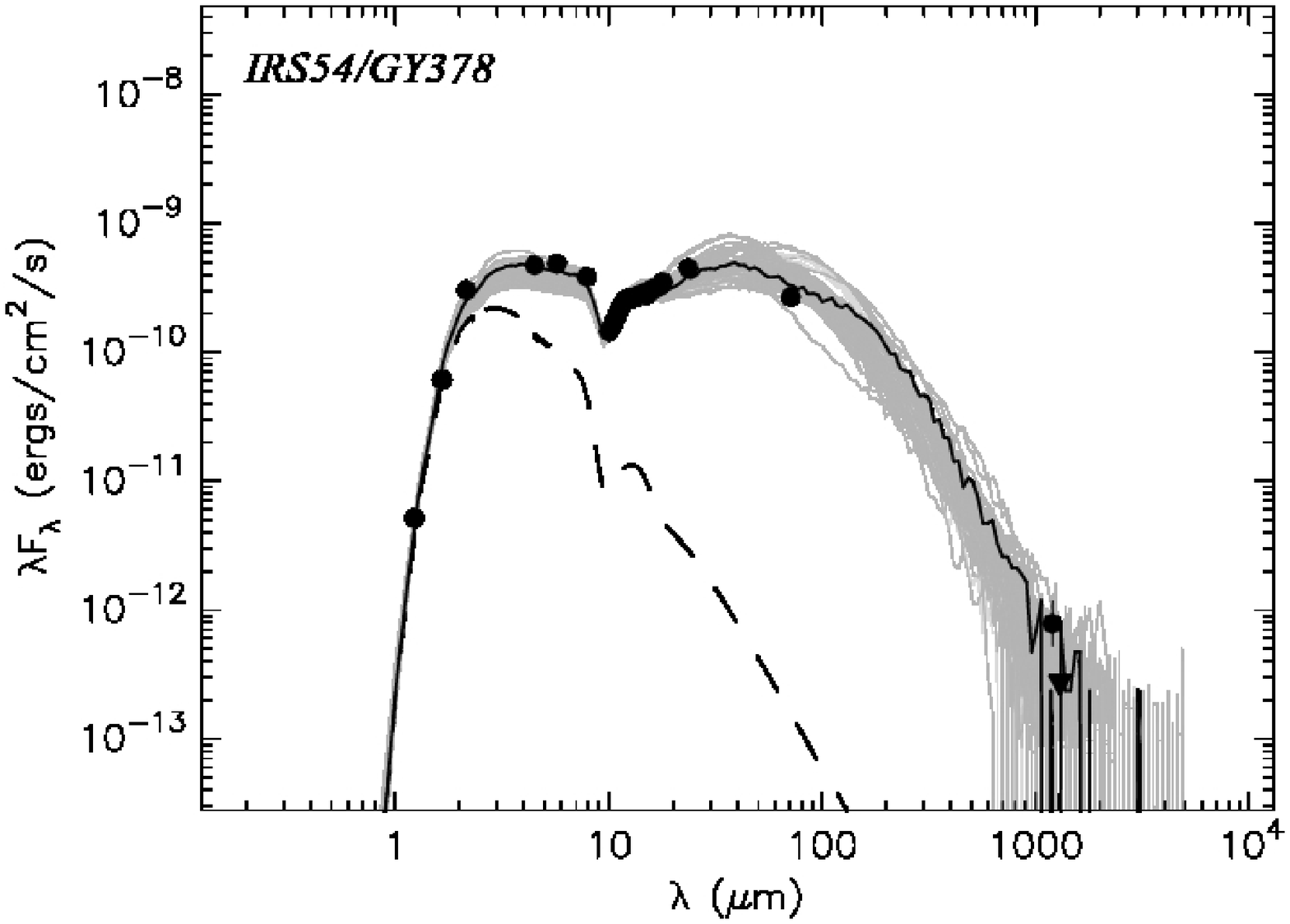, width=4.60cm}
}
\caption{SEDs and best fit models, as produced by the Web interface
provided by \citet{rob06}, for all the 28 YSOs in our sample.}
\label{fig:sed_all}
\end{figure*}

Figure \ref{fig:sed_examples} exemplifies the ``fitting'' procedure
described in \S\,\ref{ap:sed_method} for three of our YSOs. It shows
the SEDs with the best fit models and the distributions of two fit
parameters, $A_{\rm V}$  and $\dot{M}_{\rm disk}$, both for the 1000
models with lowest $\chi^2$ and for the {\em statistically reasonable}
ones (cf. \ref{ap:sed_method}). SEDs and best fit models for all
the 28 YSOs in our sample are shown in Fig.\,\ref{fig:sed_all}.

Following visual examination of the SED fits and of the distributions
of model parameters used to define the confidence intervals, we have
decided to modify the input datapoints for two objects: for
IRS45/GY273 we have excluded the 1.2 and 1.3\,mm datapoints from
\citet{stan06} and \citet{and94}; in both cases including these points
significantly worsened the quality of the fit and had a significant
effect on the values of the parameters. The 1.2\,mm flux is $>$20
times higher than the 1.3\,mm flux (an upper limit) and probably
refers to an extended source that {\em includes} our YSO. For GY289, a
source with average IRS flux $<$0.5\,Jy, we have decided to include
the IRS datapoints because: $i)$ they agree quite well with the MIPS
fluxes at similar wavelengths, $ii)$ the quality of the model fit is
reasonable ($\chi^2_{\rm best}\sim$2) and, $iii)$ the confidence
intervals of the model parameters are narrower but compatible with
those from the fit performed without these points. 

Finally for one object, WL5/GY246, we could not obtain a unique fit
with the above procedure. The object was previously classified as a
deeply absorbed Class~III star with an F7 spectral type \citep{gre95},
and our SED was defined by $J$,$H$,$K$, {\em Spitzer} IRAC 1-4 and
1.2/1.3mm fluxes. Fits both with and without the mm fluxes,
likely contaminated by nearby sources \citep[c.f.][]{and94,stan06},
consistently yield high envelope and/or disk accretion rates, typical
of a Class~I object, but having  little effect on the NIR/MIR part of
the SED due to the associated large inner disk radii. The NIR/MIR SED
can however be fit equally well by purely photospheric `Phoenix'
models as suggested by the same \citet{rob07} web interface used to
fit the star/disk/envelope models. We thus decided to assume that
WL5/GY246 is a Class~III object and to derive its extinction,
effective temperature, and stellar mass using the $J$, $H$, and $K$
photometry, the spectral type, and the calibrations tabulated by
\citet{kh95}. Uncertainties were estimated from the assumed
uncertainty on the spectral type, one subclass, and the range of
values obtained by estimating the absorption from the J-H, H-K, and
J-K colors.

Table \ref{tab:sed_results}, introduced in the main text (\S
\ref{sect:anc_SED}) lists the outcome of the SED-fit process: the
quality of the fit (the $\chi^2$ of the ``best-fit'' model), the
object extinction (the sum of interstellar and envelope extinction),
the stellar effective temperature and mass, the disk mass, the disk
and envelope accretion rates, the evolutionary Stage. This latter
quantity was assigned following \citet{rob07}. Stage I: $\dot{M}_{\rm
env}/M_* > 10^{-6}$; Stage II: $\dot{M}_{\rm env}/M_* \le 10^{-6}$ and
$M_{\rm disk}/M_* > 10^{-6}$; Stage III: $\dot{M}_{\rm env}/M_* \le
10^{-6}$ and $M_{\rm disk}/M_* \le 10^{-6}$. As indicated in the main
text, in order to use a  more familiar designation to researchers in
the field, we also refer to the `Stages' as `Classes'. 

\begin{figure}
\centerline{\epsfig{file=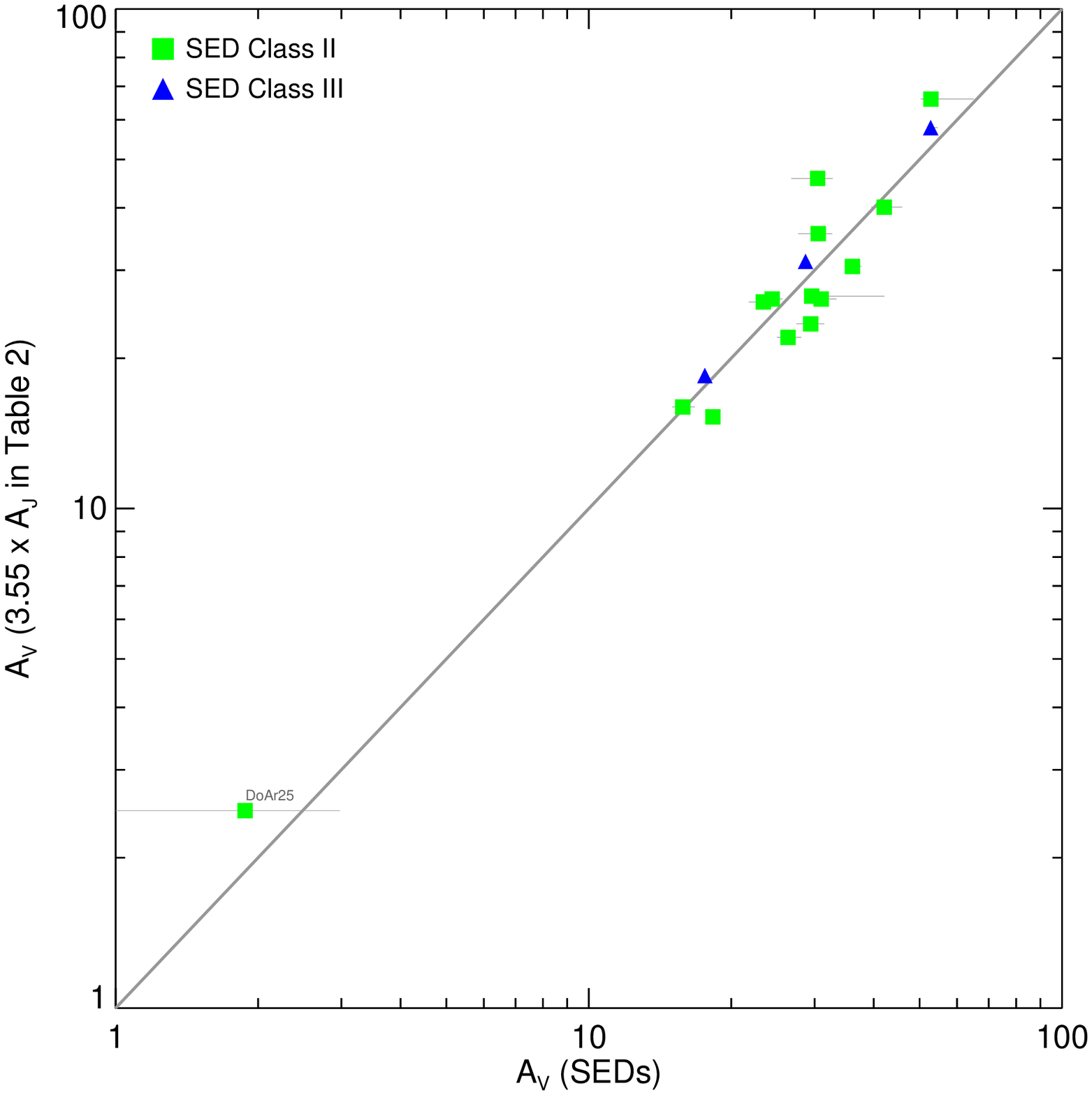, width=8.8cm}}
\caption{Comparison of the $A_{\rm V}$
values derived from fitting the SEDs with the \citet{rob06} models,
with values derived from 2MASS photometry (cf. Table \ref{tab:target_litdata}).
Objects of different SED Class are indicated by different symbols as
shown in the legend.}
\label{fig:sedAv_vs_lit} 
\end{figure}

\begin{figure}
\centerline{\epsfig{file=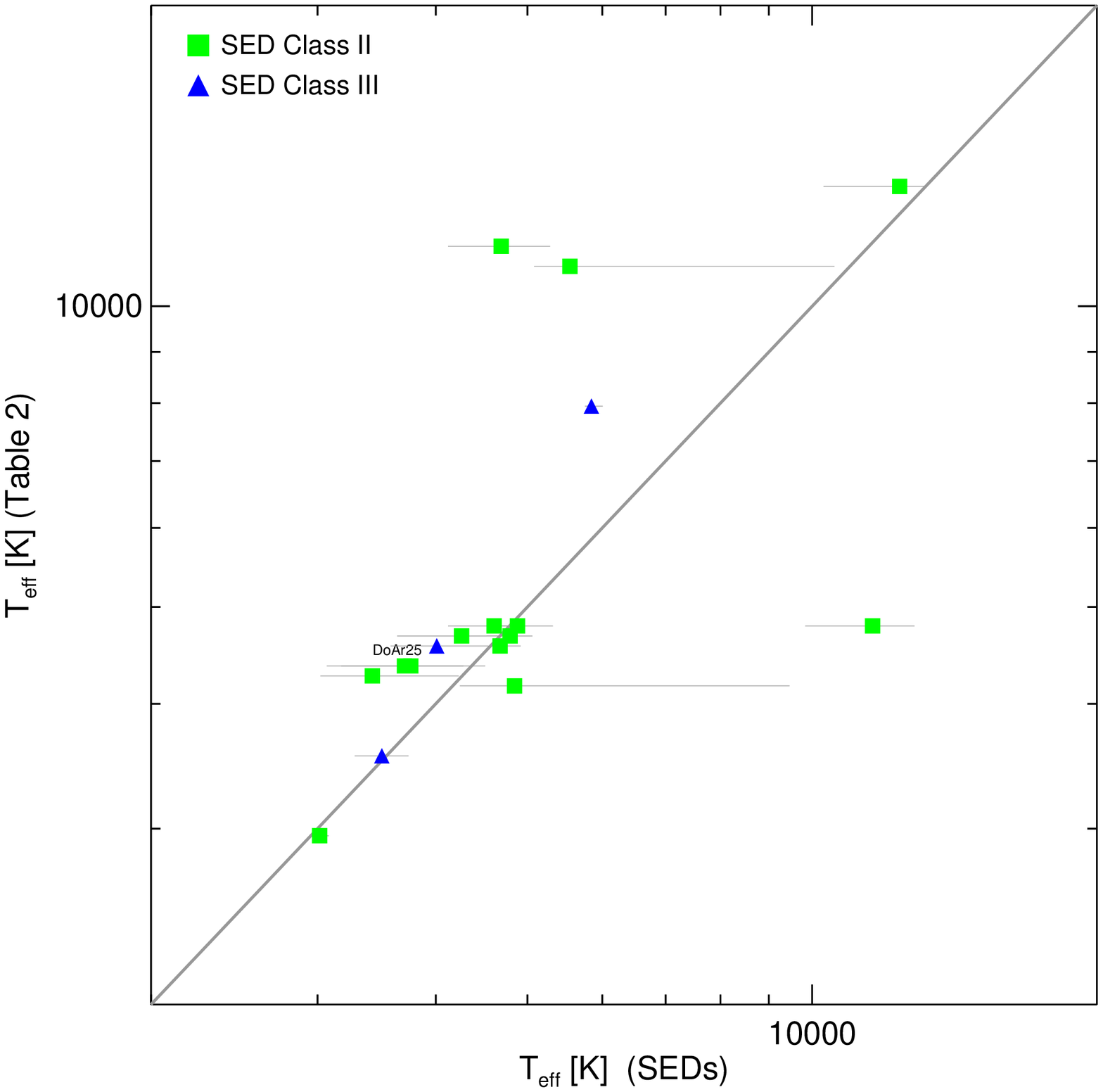, width=8.8cm}}
\caption{Same as Fig. \ref{fig:sedAv_vs_lit} for the effective
temperatures.}
\label{fig:sedTe_vs_lit} 
\end{figure}

Figures \ref{fig:sedAv_vs_lit} and \ref{fig:sedTe_vs_lit} compare the
extinction values ($A_{\rm V}$) and stellar $T_{\rm eff}$ obtained
from the SED fits with the same parameters listed in
Table\,\ref{tab:target_litdata} for Class\,II and Class\,III stars.
Given the considerable uncertainties of both determinations, the SED
fits yield results similar to those obtained with the method of
\citet{nat06}. A similar comparison with the accretion rates derived
from the Pa$\beta$ and Br$\gamma$ NIR line fluxes (in Table
\,\ref{tab:target_litdata}), is less conclusive due to the large
number of upper limits and to the large uncertainties that affect the
spectroscopic measurements as well as the SED fits. Seven objects can
be used for the comparison having accretion rate estimates or upper
limits from both methods. For only two stars both methods yield
estimates: those for IRS\,54 are in good agreement; for WL\,16 the
spectroscopic estimate is 2.6\,dex higher than the value from the SED
fits, $\dot{M}\sim 10^{-8}$\,M$_{\odot}$yr$^{-1}$. The discrepancy is
however reduced to 1.2\,dex when comparing the result of the SED fit
with the \citet{nat06} value. Moreover the derivation of $\dot{M}$
from the Pa$_\beta$ line with the method of \citet[][see also \S
\ref{sect:anc_SED}]{nat06} is better suited for cool stars and likely
to yield inaccurate results for WL\,16 ($T_{\rm eff}\sim10^4$\,K).  An
independent estimate by \citet{naj96} yielded an upper limit
compatible with the SED value: $\dot{M}\lesssim
2\times10^{-7}$\,M$_{\odot}$yr$^{-1}$. Other three stars have
$\dot{M}$ estimates from the SED fits and upper limits from
Table\,\ref{tab:target_litdata}: in two cases, IRS\,51 and IRS\,47,
the confidence intervals from the SED fits are consistent with the
upper limits; for DoAr\,25/GY17, the only star with optical
magnitudes, the SED fit yields an accretion rate that is 1.6\,dex
higher than the upper limit from the Pa$_\beta$ line. Finally, for two
stars, WL\,10 and WL\,11, the spectroscopic estimates are 0.4\,dex and
0.1\,dex larger than the upper limits from the SED fits. The
discrepancy is however reduced to 0.24 dex for WL\,10 and disappears
for WL\,11 if the slightly larger $\dot{M}$ values from \citet{nat06}
are considered instead of those in Table \ref{tab:target_litdata}.

\subsection{Summary}

In this Appendix we have shown that the SED models of \citet{rob06},
although undeniably approximate, can be useful to constrain parameters
such as the line-of-sight absorption and the disk accretion  rate,
even in the absence of optical photometry. Although resulting
uncertainties in these parameters are often large, the constraints are
by and large compatible with independent determinations obtained with
more direct methods.

\begin{sidewaystable*}[!h]
\caption{Flux densities, in mJy, collected from the literature (cf. \S
\ref{ap:sed_rhoOph}) and used for the SED fits.}
\label{tab:sed_input}
\begin{tabular}{lrrrrrrrrrrr}
\hline
\scriptsize
Name  &                    J&                    H&                K$_s$&             IRAC [1]&             IRAC [2]&             IRAC [3]&             IRAC [4]&             MIPS [1]&             MIPS [2]&                1.2mm&                1.3mm\\
\hline
DoAr25/GY17
&                   279.00&                   448.00&                   484.00&                   367.00&                   292.00&                   299.00&                   258.00&                   399.00&                         &                   153.00&                   280.00 \\
IRS14/GY54 
&                     1.66&                    12.10&                    30.20&                    35.10&                    29.20&                    23.40&                    14.00&                         &                         &                         &$<$                 30.00 \\
WL12/GY111 
&                     0.32&                     3.21&                    15.20&                   239.00&                   744.00&                  1610.00&                  2240.00&                         &                  8120.00&                   415.00&                   130.00 \\
WL22/GY174 
&$<$                  0.20&$<$                  0.08&                     0.59&                         &                         &                         &                  3230.00&                         &                         &                         &                   400.00 \\
WL16/GY182 
&                     3.44&                    65.90&                   397.00&                  1400.00&                  1970.00&                  5030.00&                         &                         &                 29100.00&                         &$<$                 10.00 \\
WL17/GY205 
&$<$                  0.19&                     1.96&                    27.30&                   240.00&                   416.00&                   553.00&                   695.00&                  2790.00&                  6070.00&                   144.00&                    70.00 \\
WL10/GY211 
&                    15.30&                    86.10&                   181.00&                   259.00&                   310.00&                   272.00&                   222.00&                   339.00&                   784.00&                         &$<$                 30.00 \\
EL29/GY214 
&                     0.31&                    39.00&                   929.00&                         &                         &                 12800.00&                         &                         &                         &                   316.00&                   300.00 \\
GY224      
&$<$                  0.20&                     7.21&                    55.70&                   203.00&                         &                   358.00&                   367.00&                   907.00&                   908.00&                         &                          \\
WL19/GY227 
&$<$                  0.06&                     0.97&                    25.20&                   215.00&                         &                   406.00&                   328.00&                   223.00&                         &                         &$<$                 20.00 \\
WL11/GY229 
&                     0.90&                     5.83&                    17.00&                    33.70&                    35.80&                    34.20&                    31.00&                    43.00&                         &                         &$<$                 20.00 \\
WL20/GY240 
&                     4.41&                    32.00&                    97.30&                   127.00&                   143.00&                   140.00&                         &                         &                         &                         &                    90.00 \\
IRS37/GY244
&$<$                  0.21&                     1.75&                    15.50&                   127.00&                   206.00&                   286.00&                   268.00&                   780.00&                         &                         &                   300.00 \\
WL5/GY246  
&               {\em 0.02}&               {\em 1.37}&              {\em 39.90}&             {\em 209.00}&             {\em 297.00}&             {\em 298.00}&             {\em 163.00}&                         &                         &$<$          {\em 375.00}&              {\em 35.00} \\
IRS42/GY252
&                     1.31&                    32.50&                   270.00&                  1060.00&                  1630.00&                  2100.00&                  2980.00&                  3450.00&                  2940.00&                         &$<$                 35.00 \\
GY253      
&                     0.20&                     5.39&                    32.70&                    64.70&                    60.40&                    53.30&                    30.90&                     3.18&                         &                         &                          \\
WL6/GY254  
&$<$                  0.06&                     0.72&                    31.10&                         &                   925.00&                  1440.00&                  1730.00&                  4360.00&                  5110.00&                         &$<$                 75.00 \\
CRBR85     
&$<$                  0.01&                     0.06&                     2.26&                    54.00&                   131.00&                   191.00&                   198.00&                  1340.00&                         &$<$                921.00&                          \\
IRS43/GY265
&$<$                  0.06&                     4.01&                    84.30&                   629.00&                  1240.00&                  1790.00&                  2190.00&                         &                         &                   967.00&                   190.00 \\
IRS44/GY269
&$<$                  0.38&                     3.46&                    47.00&                   731.00&                  1830.00&                         &                         &                         &                 34700.00&$<$                576.00&                   180.00 \\
IRS45/GY273
&                     0.81&                    12.20&                    60.70&                   187.00&                   272.00&                   382.00&                   481.00&                   712.00&                         &            {\em 1058.00}&$<$           {\em 40.00} \\
IRS46/GY274
&                     0.33&                     9.03&                    77.50&                         &                   271.00&                   402.00&                   411.00&                         &                         &                         &                    90.00 \\
IRS47/GY279
&                     1.19&                    25.40&                   164.00&                   740.00&                  1190.00&                  1580.00&                  2040.00&                  1720.00&                         &                  1336.00&$<$                 20.00 \\
GY289      
&                     0.56&                     8.24&                    29.10&                    45.60&                    41.90&                    36.80&                         &                    56.50&                         &                         &                          \\
GY291      
&                     0.53&                     8.18&                    27.50&                    54.20&                    69.50&                    84.80&                   108.00&                    94.50&                         &                         &                          \\
IRS48/GY304
&                    94.60&                   305.00&                   618.00&                  1410.00&                  1600.00&                  4060.00&                  6000.00&                         &                         &                         &                   240.00 \\
IRS51/GY315
&                     0.24&                    14.40&                   169.00&                   752.00&                   916.00&                  1000.00&                  1070.00&                  2730.00&                  3260.00&                   124.00&                   165.00 \\
IRS54/GY378
&                     2.14&                    34.30&                   218.00&                         &                   712.00&                   931.00&                  1010.00&                  3560.00&                  6500.00&                   340.00&$<$                120.00 \\
\hline
\end{tabular}
\\Note: Values in italic were not used for the SED fits.
\end{sidewaystable*}

\begin{sidewaystable*}[!h]
\scriptsize
\begin{center}
\caption{Flux densities, in Jy, obtained from the IRS spectra for the
SED fits.}
\label{tab:sed_input_IRS}
\begin{tabular}{lrrrrrrrrrrrrrrrrr}
\hline
\footnotesize
Name / $\lambda$[$\mu$m] &     10.0 &     10.5 &     11.0 &     11.5 &     12.0 &     12.5 &     13.0 &     13.5 &     14.0 &     14.5 &     15.0 &     15.5 &     16.0 &     16.5 &     17.0 &     17.5 &     18.0\\
\hline
{\em DoAr25}
 & {\em     0.23} & {\em     0.23} & {\em     0.24} & {\em     0.27} & {\em     0.25} & {\em     0.25} & {\em     0.23} & {\em     0.23} & {\em     0.24} & {\em     0.22} & {\em     0.24} & {\em     0.25} & {\em     0.28} & {\em     0.30} & {\em     0.33} & {\em     0.36} & {\em     0.37}\\
{\em IRS14}
 & {\em    0.022} & {\em    0.022} & {\em    0.040} & {\em    0.094} & {\em    0.096} & {\em     0.11} & {\em    0.098} & {\em    0.081} & {\em    0.066} & {\em    0.061} & {\em    0.062} & {\em    0.066} & {\em    0.073} & {\em     0.11} & {\em     0.11} & {\em    0.095} & {\em    0.067}\\
WL12
 &     0.93 &      1.3 &      1.7 &      2.3 &      3.0 &      3.7 &      4.1 &      4.3 &      4.5 &      4.7 &      4.1 &      4.1 &      4.4 &      4.2 &      4.3 &      4.4 &      4.5\\
WL22
 &     0.16 &     0.25 &     0.62 &     0.87 &      1.0 &      1.4 &      1.1 &      1.1 &     0.96 &     0.96 &     0.44 &     0.48 &     0.96 &      1.5 &      1.3 &      1.4 &      1.1\\
WL16
 &      1.3 &      1.5 &      4.4 &      5.1 &      4.8 &      6.3 &      4.4 &      3.8 &      3.2 &      2.9 &      2.6 &      2.4 &      2.9 &      4.4 &      4.0 &      3.9 &      2.7\\
WL17
 &     0.18 &     0.21 &     0.27 &     0.31 &     0.34 &     0.37 &     0.42 &     0.48 &     0.48 &     0.52 &     0.51 &     0.54 &     0.64 &     0.64 &     0.70 &     0.74 &     0.76\\
{\em WL10}
 & {\em     0.17} & {\em     0.18} & {\em     0.19} & {\em     0.20} & {\em     0.21} & {\em     0.21} & {\em     0.21} & {\em     0.22} & {\em     0.22} & {\em     0.22} & {\em     0.23} & {\em     0.24} & {\em     0.25} & {\em     0.26} & {\em     0.27} & {\em     0.27} & {\em     0.28}\\
EL29
 &      7.2 &      9.5 &      12. &      16. &      19. &      22. &      23. &      24. &      24. &      26. &      22. &      22. &      24. &      24. &      24. &      25. &      25.\\
GY224      
 &     0.27 &     0.30 &     0.34 &     0.37 &     0.38 &     0.39 &     0.40 &     0.43 &     0.45 &     0.47 &     0.48 &     0.48 &     0.53 &     0.56 &     0.60 &     0.62 &     0.63\\
{\em WL19}
 & {\em    0.053} & {\em    0.071} & {\em    0.077} & {\em    0.096} & {\em     0.13} & {\em     0.15} & {\em     0.15} & {\em     0.16} & {\em     0.17} & {\em     0.18} & {\em     0.16} & {\em     0.15} & {\em     0.16} & {\em     0.14} & {\em     0.14} & {\em     0.15} & {\em     0.14}\\
{\em WL11}
 & {\em    0.012} & {\em    0.014} & {\em    0.011} & {\em    0.010} & {\em    0.014} & {\em    0.014} & {\em    0.015} & {\em    0.019} & {\em    0.017} & {\em    0.017} & {\em    0.022} & {\em    0.021} & {\em    0.027} & {\em    0.029} & {\em    0.034} & {\em    0.037} & {\em    0.033}\\
WL20
 &     0.34 &     0.45 &     0.59 &     0.73 &     0.83 &     0.89 &      1.0 &      1.2 &      1.4 &      1.6 &      1.6 &      1.8 &      2.3 &      2.5 &      2.8 &      3.1 &      3.4\\
{\em IRS37}
 & {\em     0.13} & {\em     0.17} & {\em     0.22} & {\em     0.26} & {\em     0.29} & {\em     0.33} & {\em     0.35} & {\em     0.37} & {\em     0.39} & {\em     0.43} & {\em     0.40} & {\em     0.42} & {\em     0.49} & {\em     0.53} & {\em     0.57} & {\em     0.63} & {\em     0.66}\\
{\em WL5}
 & {\em    0.015} & {\em    0.021} & {\em    0.029} & {\em    0.037} & {\em    0.052} & {\em    0.064} & {\em    0.062} & {\em    0.066} & {\em    0.065} & {\em    0.066} & {\em    0.047} & {\em    0.047} & {\em    0.068} & {\em    0.079} & {\em    0.088} & {\em     0.11} & {\em    0.098}\\
IRS42
 &      1.8 &      2.0 &      2.3 &      2.4 &      2.1 &      2.1 &      2.1 &      2.1 &      2.1 &      2.2 &      2.1 &      2.1 &      2.4 &      2.4 &      2.4 &      2.6 &      2.7\\
{\em GY253      }
 & {\em  9.4e-05} & {\em  0.00031} & {\em  -0.0015} & {\em  -0.0021} & {\em   0.0024} & {\em  0.00084} & {\em   0.0028} & {\em   0.0059} & {\em   0.0035} & {\em   0.0023} & {\em   0.0051} & {\em   0.0024} & {\em   0.0047} & {\em   0.0048} & {\em   0.0080} & {\em   0.0092} & {\em   0.0058}\\
WL6
 &     0.73 &     0.92 &      1.1 &      1.4 &      1.6 &      1.8 &      2.0 &      2.1 &      2.2 &      2.4 &      1.9 &      2.0 &      2.4 &      2.5 &      2.6 &      2.7 &      2.8\\
{\em CRBR85     }
 & {\em    0.066} & {\em     0.10} & {\em     0.14} & {\em     0.18} & {\em     0.23} & {\em     0.27} & {\em     0.31} & {\em     0.35} & {\em     0.39} & {\em     0.43} & {\em     0.35} & {\em     0.37} & {\em     0.49} & {\em     0.51} & {\em     0.55} & {\em     0.62} & {\em     0.65}\\
IRS43
 &     0.87 &      1.2 &      1.7 &      2.3 &      2.9 &      3.5 &      4.0 &      4.4 &      4.9 &      5.4 &      4.0 &      4.4 &      5.6 &      5.8 &      6.3 &      6.8 &      7.3\\
IRS44
 &      1.5 &      2.4 &      3.5 &      5.3 &      7.2 &      8.7 &      10. &      12. &      13. &      15. &      13. &      14. &      17. &      18. &      20. &      22. &      24.\\
IRS45
 &     0.34 &     0.40 &     0.44 &     0.48 &     0.46 &     0.48 &     0.48 &     0.50 &     0.52 &     0.53 &     0.49 &     0.52 &     0.59 &     0.57 &     0.60 &     0.64 &     0.63\\
IRS46
 &     0.31 &     0.39 &     0.48 &     0.57 &     0.66 &     0.72 &     0.76 &     0.81 &     0.83 &     0.87 &     0.85 &     0.88 &     0.93 &     0.96 &      1.0 &      1.0 &      1.1\\
IRS47
 &      1.2 &      1.3 &      1.5 &      1.6 &      1.6 &      1.6 &      1.6 &      1.6 &      1.7 &      1.7 &      1.4 &      1.5 &      1.7 &      1.6 &      1.7 &      1.7 &      1.8\\
{\em GY289      }
 & {\em   0.0039} & {\em   0.0065} & {\em   0.0080} & {\em    0.011} & {\em    0.013} & {\em    0.014} & {\em    0.015} & {\em    0.021} & {\em    0.018} & {\em    0.019} & {\em    0.024} & {\em    0.023} & {\em    0.030} & {\em    0.035} & {\em    0.037} & {\em    0.041} & {\em    0.036}\\
{\em GY291      }
 & {\em    0.055} & {\em    0.060} & {\em    0.063} & {\em    0.064} & {\em    0.062} & {\em    0.062} & {\em    0.063} & {\em    0.065} & {\em    0.063} & {\em    0.064} & {\em    0.064} & {\em    0.063} & {\em    0.075} & {\em    0.085} & {\em    0.091} & {\em    0.095} & {\em    0.091}\\
IRS48
 &      2.4 &      2.6 &      4.3 &      4.7 &      4.5 &      4.9 &      4.6 &      5.1 &      5.8 &      6.7 &      8.0 &      9.2 &      12. &      14. &      16. &      18. &      20.\\
IRS51
 &      1.2 &      1.3 &      1.3 &      1.3 &      1.3 &      1.2 &      1.2 &      1.3 &      1.3 &      1.4 &      1.1 &      1.3 &      1.7 &      1.8 &      2.0 &      2.1 &      2.2\\
IRS54
 &     0.48 &     0.59 &     0.73 &     0.89 &      1.0 &      1.1 &      1.2 &      1.2 &      1.3 &      1.4 &      1.4 &      1.5 &      1.7 &      1.7 &      1.8 &      1.9 &      2.1\\
\hline
\end{tabular}
\\
Note: Values in italic were considered uncertain and were not used for the SED fits.
\end{center}
\end{sidewaystable*}

\end{document}